\documentclass[10pt,twocolumn,english,aps,manuscript,pra,showpacs]{revtex4}
\usepackage[T1]{fontenc}
\usepackage[latin9]{inputenc}
\usepackage{babel}
\usepackage{textcomp}
\usepackage{mathrsfs}
\usepackage{mathtools}
\usepackage{amsmath}
\usepackage{amssymb}
\usepackage{cancel}
\usepackage{graphicx}
\usepackage{esint}
\usepackage[unicode=true,pdfusetitle,
 bookmarks=true,bookmarksnumbered=false,bookmarksopen=false,
 breaklinks=false,pdfborder={0 0 1},backref=false,colorlinks=false]
 {hyperref}

\makeatletter
\@ifundefined{textcolor}{}
{%
 \definecolor{BLACK}{gray}{0}
 \definecolor{WHITE}{gray}{1}
 \definecolor{RED}{rgb}{1,0,0}
 \definecolor{GREEN}{rgb}{0,1,0}
 \definecolor{BLUE}{rgb}{0,0,1}
 \definecolor{CYAN}{cmyk}{1,0,0,0}
 \definecolor{MAGENTA}{cmyk}{0,1,0,0}
 \definecolor{YELLOW}{cmyk}{0,0,1,0}
}

\usepackage{ bbm }
\DeclareMathAlphabet{\mathbbold}{U}{bbold}{m}{n} 

\makeatother

\begin{document}

\title{Time Averaged Density Matrix as an Optimization Problem}

\author{V. Nebendahl}

\affiliation{Institut für Theoretische Physik, Universität Innsbruck, Technikerstr.
25, A-6020 Innsbruck, Austria}

\email{Volckmar.Nebendahl@uibk.ac.at}

\begin{abstract}
A new method is presented which allows time averaged density matrices
of closed quantum systems to be computed via a constraint overlap
maximization. Due to its simplicity, this method can be combined with
algorithms based on tensor networks, as, e.g., matrix product operators
(MPO). An algorithm is explained and several results for non-integrable
Ising chains are given. Among them are scaling examples, time averaged
expectation values, their variances and operator space entanglement
entropies.
\end{abstract}


\maketitle

\section{Introduction\label{sec:Inroduction TADM}}

An isolated classical system approaches its thermal equilibrium by
maximizing its entropy. A closed quantum system on the other hand
can not thermalize in this way, since it is subjected to a unitary
time evolution, which does not change the entropy. So, by which mechanism
do closed quantum systems equilibrate, if they equilibrate at all?
This question was already investigated 1929 by John von Neumann in
the early days of quantum mechanics \cite{Neumann1929}, see also
the comments in Ref.~\cite{Goldstein2010_Neumann}. While at that
time, the thermalization of a pure quantum state might have seem solely
as an academic question, the interest in this subject has recently
rekindled with the advent of new experimental techniques which allow
to study almost undisturbed long-time evolution of ultracold atoms
and trapped ions \cite{Kinoshita2006,Yukalov2011_IonThermo}.

To start with, we like to remark that even in classical physics, the
definition of thermal equilibrium is far from trivial if we look at
the deterministic evolution of a specific microstate. Therefore, one
resorts to macrostates, which implies an averaging over a vast number
of microstates. In statistical mechanics, an isolated system with
known total energy is described by the a microcanonical ensemble,
which assigns to each microstate with suitable energy the same probability.
For ergodic systems, the microcanonical ensemble coincides with the
long-time average of the system. These two definitions can also be
used for closed quantum systems. But here, the density matrices obtained
from the microcanonical ensemble $\varrho_{{\rm m.c.}}$ and the long-time
average 
\begin{equation}
\bar{\varrho}=\lim_{T\rightarrow\infty}\left(\frac{1}{T}\int_{0}^{T}\varrho(t)dt\right)\label{eq:First time TADM}
\end{equation}
do generally not coincide \cite{Neumann1929}. Both density matrices
$\varrho_{{\rm m.c.}}$ and $\bar{\varrho}$ are diagonal in the energy
eigenstates basis. Yet, the diagonal elements of $\varrho_{{\rm m.c.}}$
consist only of zeros and ones (disregarding normalization), while
the diagonal elements of $\bar{\varrho}$ depend on the initial state
(see also Eq.~(\ref{eq:Diagonale TADM}), below).

Instead of considering the entire system, one can also focus on a
small subsystem. Then, the (much larger) rest of the system can be
seen as thermal bath of the subsystem. Interestingly, in this case
thermalization might even be obtained without any ensemble or time
averaging. Due to entanglement, the reduced density matrix of a pure
state is a mixed state and therefore it is possible that after sufficiently
long time, the reduced density matrix of a subsystem describes a thermal
state. Various publications have shown that this is indeed the case
for many systems, see e.g.\ Ref.~\cite{Deutsch1991_ETH,Srednicki1994,Rigol2007_GeneralizedGibbs,Reimann2008,Rigol2008,EisertCramer2008_Therm0,Linden2009_Thermo,Banuls2011_Thermo},
but also Ref.~\cite{Rigol2009,Gogolin2011_Thermo} as counter examples. 

One of the leading theories to explain thermalization is the so called
eigenstate thermalization hypothesis (ETH) \cite{Deutsch1991_ETH,Srednicki1994}.
Here, it is assumed that for energy eigenstates, the reduced density
matrices of local subsystems are thermal. That is, an initially out
of equilibrium state relaxes due to a dephasing of the different energy
eigenstates such that the coherences average out. More background
information can also be found in the reviews \cite{Cazalilla2010_OverviewThermo,Polkovnikov2011_ThermoOverview,Eisert2014_Thermo}.

In the context of thermalization, it is important to distinguish between
integrable and non-integrable systems. While for non-integrable systems
it is widely assumed that their reduced density matrices relax to
a standard Gibbs ensemble, this is generally not possible for integrable
systems, where an extensive number of local integrals of motions $\mathcal{I}_{j}$
conserves a memory of the initial state. In such a case, the system
can still equilibrate to a generalized Gibbs ensemble \cite{Rigol2007_GeneralizedGibbs}\emph{.}

However, integrable systems have the undoubted advantage that closed
analytical solution might be found \cite{Calabrese2011_QuenschIsing,KormosCalabrese2014},
while the numerically accessible timespan of non-integral systems
might not suffice to observe equilibration \cite{EntropyTimeSchuch2008,Banuls2011_Thermo}.
The situation might even be worse, if the system is non-integrable
but has quasi-conserved local integrals of motions, which can result
in very long relaxation times \cite{Kim2014_Banuls}.

In any case, if a closed quantum system respectively its subsystems
equilibrate, the equilibrium states must coincide with the time averages
of these states and hence, can all be obtained from the time averaged
density matrix (TADM) $\bar{\varrho}$ defined in Eq.~(\ref{eq:First time TADM}).
In order to actually compute the TADM $\bar{\varrho}$, two methods
come immediately into mind: One can take Eq.~(\ref{eq:First time TADM})
literally and calculate various time evolved states $\Psi(t)$ and
sum up their $\varrho(t)$. Alternatively, one can recall that the
TADM $\bar{\varrho}$ consists of the diagonal elements of the initial
density matrix $\varrho_{0}$ expressed in energy eigenstates $|E_{j}\rangle$.
Hence, one can use diagonalization techniques to find the relevant
eigenstates of the Hamiltonian to reconstruct $\bar{\varrho}$. Unfortunately,
for many systems of interest, both methods are of very limited applicability.

Here, we introduce an alternative strategy based on a simple constraint
overlap optimization procedure. Due to its simple structure, the optimization
can be easily carried out with matrix product operators (MPO) or other
tensor network operators \cite{Verstraete2008Review,Schollwoeck2011,Eisert2013OverviewTNS,Orus2014_TensorNetworks}.
This is demonstrated by various applications in Sec.~\ref{sec:Results TADM},
where among others expectation values, variances and operator space
entanglement entropies for non-integrable Ising spin chains are presented.
Further, we compare spin chains of different lengths and also study
the influence of the so-called bond dimension of the MPO on the obtainable
results.

The key idea of tensor networks as matrix product states (MPS) and
MPO is to express high-dimensional quantum states and operators as
products of low-dimensional matrices respectively tensors. Such an
ansatz works fine as long as the system's entanglement is limited.
Unfortunately, this is not a situation we can expect to find in a
generic TADM $\bar{\varrho}$. Hence, in many cases of interest, a
tensor network can only represent an approximation of the TADM which
is more or less rough. However, since our approach is based on an
optimization principle, we can at least hope to get the best out of
our limited resources within the chosen ansatz class. 

The quality of the solution also depends on the chosen type of tensor
network. In this paper, we will mainly deal with MPO. That is, we
aim directly for the density \emph{operator, }while e.g.\ the methods
based on time evolution or diagonalization primary calculate the \emph{states
}$|\Psi(t)\rangle$ or $|E_{j}\rangle$. In appendix~\ref{sec:Double-MPS tadm},
we will also give brief account how the algorithm can be used with
MPS.

Not surprisingly, the decision to calculate states or to aim directly
for the operator entails certain advantages and disadvantages. For
example, if the TADM is mainly described by one dominant energy eigenstate
$\bar{\varrho}\approx|E\rangle\langle E|$, it is generally easier
to compute this state $|E\rangle$ instead of the operator $|E\rangle\langle E|$.
Here, the entropy of $\bar{\varrho}\approx|E\rangle\langle E|$ is
close to zero. Then again, if the entropy of the TADM is high, i.e.,
if it is composed of many similarly weighted energy eigenstates $\bar{\varrho}=\sum_{j}p_{jj}|E_{j}\rangle\langle E_{j}|$,
targeting the operator directly seems favorable, since a MPO can handle
arbitrary amounts of entropy. For the same reason, MPO are better
suited for the time average of a (local) measurement operator $\bar{O}=\sum_{j}o_{jj}|E_{j}\rangle\langle E_{j}|$.

\subsection{Structure of this paper}

In writing this paper, we had two different types of readers in mind.
On the one hand, the reader who likes to understand the basic ideas,
but has no need for all algorithmic detail. On the other hand, the
reader who likes to reproduce our algorithm and hence, needs all the
details (s)he can get. The main paper should fit the first type of
readers, while readers of the second type find all the information
they need in a vast appendix, where various special topics are treated. 

The key insight of our method is that the search for the TADM can
be phrased as simple optimization problem. This is explained in Sec.~\ref{sec:Time-averaged-density as optimizatio problem},
while a detailed presentation of an algorithm solving the optimization
problem is outsourced into the appendix: In appendix~\ref{sec:Solving-the-optimization problem general approach tadm},
a general strategy for solving the optimization problem is explained,
yet without any references to tensor networks. The modifications needed
to incorporate tensor networks are discussed in appendix~\ref{sec:Time-averaged-density as MPO}.
Further improvements are presented in the appendices~\ref{sub:Gauging-the-MPO tadm},
\ref{sub:Speeding-up-convergence tadm}, and \ref{sec:Mapping-hermitian-matrices onto real tADM}. 

For readers who do not intent to study the appendix, Sec.~\ref{sub:Solving-the-optimization overview}
provides a quick overview of the crucial ideas used for the numerical
solution, but here, explanations are sparse. In Sec.~\ref{sec:Results TADM},
we present our numerical results. Finally, the main paper is concluded
with a discussion and outlook in Sec.~\ref{sec:Discussion-and-outlook TADM}.

\section{Formal solutions for the time averaged density matrix\label{sec:Time-averaged-density as optimizatio problem}}

In this section, we look at the general structure of the time averaged
density matrix (TADM) $\bar{\varrho}$ and show in the following subsections
how the calculation of $\bar{\varrho}$ can be cast into the alternative
form of a simple optimization problem. 

For the theoretical considerations, we always choose the energy eigenstates
as preferred basis. In this basis, we express the initial density
matrix $\varrho_{0}$ at time $t=0$ as
\begin{equation}
\varrho_{0}=\sum_{j,k}p_{jk}|E_{j}\rangle\langle E_{k}|\quad\textrm{with}\quad p_{jk}=\langle E_{j}|\varrho_{0}|E_{k}\rangle,\label{eq:Rho_Null}
\end{equation}
while at any other time $t$, the time evolved density matrix $\varrho(t)$
is given as
\begin{equation}
\varrho(t)=\sum_{j,k}\exp\left(-\frac{i}{\hbar}(E_{j}-E_{k})t\right)p_{jk}|E_{j}\rangle\langle E_{k}|.
\end{equation}
Inserting this notation into the definition of the TADM~$\bar{\varrho}$
\begin{equation}
\bar{\varrho}\coloneqq\lim_{T\rightarrow\infty}\left(\frac{1}{T}\int_{0}^{T}\varrho(t)dt\right),\label{eq:Def time averaged desity matrix}
\end{equation}
we obtain 
\begin{eqnarray}
\bar{\varrho} & = & \sum_{j,k}\bar{p}_{jk}|E_{j}\rangle\langle E_{k}|,\nonumber \\
 & = & \sum_{j,k}\delta_{E_{j},E_{k}}p_{jk}|E_{j}\rangle\langle E_{k}|,\label{eq:Diagonale TADM}
\end{eqnarray}
where we used the symbol $\delta_{E_{j},E_{k}}$ as abbreviation for
\begin{eqnarray}
\delta_{E_{j},E_{k}} & = & \lim_{T\rightarrow\infty}\left(\frac{1}{T}\int_{0}^{T}\exp\left(-\frac{i}{\hbar}(E_{j}-E_{k})t\right)dt\right)\nonumber \\
 & = & \begin{cases}
1 & \textrm{for }E_{j}=E_{k}\\
0 & \textrm{for }E_{j}\neq E_{k}
\end{cases}.\label{eq:Delta-Integral time averaged density matrix}
\end{eqnarray}
For a non-degenerate energy spectrum, the TADM $\bar{\varrho}$ consists
of the diagonal elements of $\varrho_{0}$. In case of a degenerate
energy spectrum, $\bar{\varrho}$ is made up of the corresponding
block diagonal elements of $\varrho_{0}$. To keep the notation simple,
we will still refer to these elements as $\varrho_{\textrm{diag}}$\label{Diag gleich Blocj Diag TADM},
i.e. 
\begin{align}
\varrho_{\textrm{diag}}=\bar{\varrho} & =\sum_{j,k}\delta_{E_{j},E_{k}}p_{jk}|E_{j}\rangle\langle E_{k}|\nonumber \\
 & =\sum_{E_{j}=E_{k}}p_{jk}|E_{j}\rangle\langle E_{k}|.\label{eq:Del Rho diag tadm}
\end{align}
Correspondingly, we define $\varrho_{\textrm{off-diag}}$ as
\begin{eqnarray}
\varrho_{\textrm{off-diag}} & = & \sum_{j,k}\left(1-\delta_{E_{j},E_{k}}\right)p_{jk}|E_{j}\rangle\langle E_{k}|\nonumber \\
 & = & \sum_{E_{j}\neq E_{k}}p_{jk}|E_{j}\rangle\langle E_{k}|.\label{eq:Rho Off Diag}
\end{eqnarray}

\subsection{Structure of the solution\label{sub:Structure-of-the prolem TADM}}

The idea we pursue to obtain the time averaged density matrix (TADM)
$\bar{\varrho}$ is to calculate (respectively approximate) $\varrho_{\textrm{off-diag}}$
(\ref{eq:Rho Off Diag}) and subtract it from $\varrho_{0}$ 
\begin{equation}
\bar{\varrho}\overset{\eqref{eq:Del Rho diag tadm}}{=}\varrho_{\textrm{diag}}=\varrho_{0}-\varrho_{\textrm{off-diag}}.\label{eq:Ansatz Struktur tadm}
\end{equation}
To this end, we express $\varrho_{\textrm{off-diag}}$ as commutator
of the Hamiltonian and an unknown matrix $M$, which still has to
be determined, i.e.\  
\begin{equation}
\varrho_{\textrm{off-diag}}=[H,M].\label{eq:Rho off as HM tadm}
\end{equation}
The purpose of the commutator will become clear in the following.
In a first step, we write the matrix $M=\sum_{jk}m_{jk}|E_{j}\rangle\langle E_{k}|$
as 
\begin{equation}
M=\underbrace{\sum_{E_{j}=E_{k}}m_{jk}|E_{j}\rangle\langle E_{k}|}_{=M_{\textrm{diag}}}+\underbrace{\sum_{E_{j}\neq E_{k}}m_{jk}|E_{j}\rangle\langle E_{k}|}_{=M_{\textrm{off-diag}}}.
\end{equation}
With that, we obtain for the commutator of \emph{any} matrix $M$
with the Hamiltonian $H$
\begin{equation}
[H,M]=\underbrace{[H,M_{\textrm{diag}}]}_{=0}+[H,M_{\textrm{off-diag}}],
\end{equation}
where 
\begin{eqnarray}
[H,M_{\textrm{off-diag}}] & = & \sum_{E_{j}\neq E_{k}}\left(E_{j}-E_{k}\right)m_{jk}|E_{j}\rangle\langle E_{k}|\label{eq:H M_off tadm}\\
{}[H,M_{\textrm{diag}}] & = & 0.\label{eq:C vernichtet rho_diag}
\end{eqnarray}
Obviously, $[H,M]=[H,M_{\textrm{off-diag}}]$ is always an off-diagonal
matrix for any matrix $M$. Further, Eq.~(\ref{eq:H M_off tadm})
can be inverted to solve $\varrho_{\textrm{off-diag}}=[H,M]$~(\ref{eq:Rho off as HM tadm}),
since the term $E_{j}-E_{k}$ never becomes zero for $E_{j}\neq E_{k}$.
That is, 
\begin{equation}
\forall\varrho_{\textrm{off-diag}},\;\exists M\quad\textrm{with}\quad\varrho_{\textrm{off-diag}}=[H,M].\label{eq:Existenz vom M tadm}
\end{equation}
In this equation, only the off-diagonal part $M_{\textrm{off-diag}}$
of the matrix $M$ is unique, while the diagonal part $M_{\textrm{diag}}$
is arbitrary due to $[H,M_{\textrm{diag}}]=0$ (\ref{eq:C vernichtet rho_diag}).
Inserting Eq.~(\ref{eq:Existenz vom M tadm}) into Eq.~(\ref{eq:Ansatz Struktur tadm}),
we find
\begin{equation}
\forall\varrho_{0},\;\exists M\quad\textrm{with}\quad\bar{\varrho}=\varrho_{0}-[H,M].\label{eq:Vorank=0000FCndigung der Formel TADM}
\end{equation}

\subsection{Commutator operator $\mathfrak{C}$\label{sub:Commutator-operatorTADM}}

Before we explain the advantage of expressing the TADM as $\bar{\varrho}=\varrho_{0}-[H,M]$
(\ref{eq:Vorank=0000FCndigung der Formel TADM}), it will be convenient
to introduce the superoperator $\mathfrak{C}=\left[H,\ldots\right]$
which acts on a matrix $A$ as 
\begin{equation}
\mathfrak{C}A:=\left[H,A\right].\label{eq:Kommutatorzeiche C def t.a.d.m.}
\end{equation}
Formally, $\mathfrak{C}$ can be seen as a tensor of fourth order
$\left(\left[H,A\right]\right)^{lm}=\mathbf{\mathfrak{C}}_{jk}^{lm}\cdot A^{jk}$.
In other contexts, the superoperator $\mathfrak{C}$ is often called
Liouvillian. Unfortunately, the term Liouvillian is not unique and
also understood in other ways. Therefore, we will simply refer to
$\mathfrak{C}$ as ``commutator operator'', which should be free
of any ambiguity. 

In the following, we will not distinguish between operator and superoperator
and also vectorize matrices as $M$ writing $|M\rangle$. Here, we
take advantage of the Choi-Jamiolkowski isomorphism $M_{jk}\cdot|j\rangle\langle k|\Leftrightarrow M_{jk}\cdot|j\rangle\otimes|k\rangle.$

For the Hilbert-Schmidt inner product (which we use throughout this
paper) of matrices $A$, $B$, we find that the commutator operator
$\mathfrak{C}$ behaves self-adjoint $\left\langle \mathfrak{C}A|B\right\rangle =\langle A|\mathfrak{C}B\rangle$,
despite the anti-hermiticity $\bigl(\mathfrak{C}A\bigr)^{\dagger}=-\mathfrak{C}A$
for Hermitian matrices $A^{\dagger}=A$. Hence, we can use notations
like e.g.\ $\left\Vert \left[H,M\right]\right\Vert {}^{2}=\left\langle M|\mathfrak{C}^{2}|M\right\rangle .$
Since we will need this property quite often, we derive it explicitly
\begin{eqnarray}
\left\langle \mathfrak{C}A|B\right\rangle  & := & \textrm{tr}\Bigl((HA-AH)^{\dagger}B\Bigr)\nonumber \\
 & = & \textrm{tr}\Bigl(A^{\dagger}HB-HA^{\dagger}B\Bigr)\nonumber \\
 & = & \textrm{tr}\Bigl(A^{\dagger}HB-A^{\dagger}BH\Bigr)\nonumber \\
 & = & \textrm{tr}\Bigl(A^{\dagger}[H,B]\Bigr)\nonumber \\
 & = & \langle A|\mathfrak{C}B\rangle.\label{eq:Selbstadjungierter kommutator t.a.d.m.}
\end{eqnarray}

\subsection{Optimization problem\label{sub:Optimization-problem TADM}}

In this subsection, we show how the TADM $\bar{\varrho}$ can be solved
as optimization problem. We start with Eq.~(\ref{eq:Rho off as HM tadm}),
which expresses the off-diagonal elements of $\varrho_{0}$ as commutator
\begin{equation}
\varrho_{\textrm{off-diag}}\overset{\eqref{eq:Rho off as HM tadm}}{=}[H,M]\overset{\eqref{eq:Kommutatorzeiche C def t.a.d.m.}}{=}\mathfrak{C}M,\label{eq:Rho Off gleich HM}
\end{equation}
with a yet unknown matrix $M$. Formally, this can be solved as 
\begin{equation}
M=\mathfrak{C}^{-1}\varrho_{\textrm{off-diag}}.
\end{equation}
In appendix~\ref{sub: Inverse Problem tadm}, we have a closer look
at this strategy and also comment on problems arising due to quasi-degenerate
eigenstates, but we will not use these findings in the rest of the
paper. Here, we follow a different approach.

In a first step, we note that the TADM $\bar{\varrho}=\varrho_{\textrm{diag}}$
has a vanishing overlap with the commutator $\mathfrak{C}M$ for \emph{any
}matrix $M$ with finite norm, since $\mathfrak{C}M$ is a purely
off-diagonal matrix (in energy eigenstates) 
\begin{equation}
\left\langle \varrho_{\textrm{diag}}|\mathfrak{C}M\right\rangle \overset{\eqref{eq:Selbstadjungierter kommutator t.a.d.m.}}{=}\langle\underbrace{\mathfrak{C}\varrho_{\textrm{diag}}}_{=0\;\eqref{eq:C vernichtet rho_diag}}|M\rangle=0.\label{eq:Null Overlap Rho diag CM TADM}
\end{equation}
So far, neither $\varrho_{\textrm{diag}}$ nor $\varrho_{\textrm{off-diag}}$
nor $M$ are known objects. What we know is $\varrho_{0}$. Hence,
let us look at the overlap of $\varrho_{0}$ with the commutator $\mathfrak{C}M$
\begin{eqnarray}
\langle\varrho_{0}|\mathfrak{C}M\rangle & = & \left\langle \varrho_{\textrm{diag}}+\varrho_{\textrm{off-diag}}|\mathfrak{C}M\right\rangle \nonumber \\
 & = & \underbrace{\xcancel{\left\langle \varrho_{\textrm{diag}}|\mathfrak{C}M\right\rangle }}_{=0\:\eqref{eq:Null Overlap Rho diag CM TADM}}+\left\langle \varrho_{\textrm{off-diag}}|\mathfrak{C}M\right\rangle \nonumber \\
 & = & \langle\varrho_{\textrm{off-diag}}|\mathfrak{C}M\rangle.\label{eq:Gleicher Overlar rho null Rho off}
\end{eqnarray}
This simple result is one of the cornerstones for the time averaged
density matrix algorithms we are going to derive in this paper: For
any matrix $M$, the inner product of the commutator $\mathfrak{C}M$
with the unknown matrix $\varrho_{\textrm{off-diag}}$ equals the
inner product with the known matrix $\varrho_{0}$. 

With the identity $\left\langle \varrho_{0}|\mathfrak{C}M\right\rangle =\left\langle \varrho_{\textrm{off-diag}}|\mathfrak{C}M\right\rangle $
(\ref{eq:Gleicher Overlar rho null Rho off}), we also find that the
matrices $M$ which maximize the two inner products are the same 
\begin{equation}
\underset{\left\Vert \mathfrak{C}M\right\Vert {}^{2}=1}{\textrm{arg max}}\Bigl(\left\langle \varrho_{0}|\mathfrak{C}M\right\rangle \Bigr)=\underset{\left\Vert \mathfrak{C}M\right\Vert ^{2}=1}{\textrm{arg max}}\Bigl(\left\langle \varrho_{\textrm{off-diag}}|\mathfrak{C}M\right\rangle \Bigr),\label{eq:Gleiches Maximum dichte matrix}
\end{equation}
where we use 
\begin{equation}
\left\Vert \mathfrak{C}M\right\Vert {}^{2}=\left\langle \mathfrak{C}M|\mathfrak{C}M\right\rangle =1\label{eq:Norm Dichte Matrix}
\end{equation}
as normalization condition and not $\left\langle M|M\right\rangle =1$.
Eq.~(\ref{eq:Gleiches Maximum dichte matrix}) is maximized for 

\begin{equation}
\mathfrak{C}M=\frac{1}{c}\varrho_{\textrm{off-diag}},\label{eq:Existence M_bar}
\end{equation}
with $c=\left\Vert \varrho_{\textrm{off-diag}}\right\Vert =\langle\mathfrak{C}M|\varrho_{\textrm{off-diag}}\rangle=\langle\mathfrak{C}M|\varrho_{0}\rangle$.
Actually, we also have to ensure the existence of matrices $M$ which
satisfy Eq.~(\ref{eq:Existence M_bar}). But this we have already
done in Eq.~(\ref{eq:Existenz vom M tadm}).

Putting all together, we find that any matrix $M$ which maximizes
the inner product $\langle\varrho_{0}|\mathfrak{C}M\rangle$ under
the condition $\Vert\mathfrak{C}M\Vert^{2}=1$
\begin{equation}
M=\underset{\Vert\mathfrak{C}M'\Vert^{2}=1}{\textrm{arg max}}\Bigl(\left\langle \varrho_{0}|\mathfrak{C}M'\right\rangle \Bigr)\label{eq:Optimization Objective tadm}
\end{equation}
 also satisfies
\begin{eqnarray}
\bar{\varrho} & \overset{\eqref{eq:Ansatz Struktur tadm}}{=} & \varrho_{0}-\varrho_{\textrm{off-diag}}\nonumber \\
 & \overset{\eqref{eq:Existence M_bar}}{=} & \varrho_{0}-\left\langle \mathfrak{C}M|\varrho_{0}\right\rangle \mathfrak{C}M\nonumber \\
 & = & \varrho_{0}-\mathfrak{C}\bar{M},\label{eq:=0000DCbersichtsformel Time averaged d.m.}
\end{eqnarray}
with $\bar{M}=\left\langle \mathfrak{C}M|\varrho_{0}\right\rangle M$.
That is, the TADM $\bar{\varrho}$ can be solved as an optimization
problem, which only involves basic matrix operations as addition,
multiplication and inner product.

\subsubsection{Alternative optimization\label{sub:Vorank=0000FCndigung Alternative-optimization-ansatz}
ansatz}

It is relative straight forward to see that the conditioned maximization
of Eq.~(\ref{eq:Optimization Objective tadm}) is equivalent to the
unconditioned minimization of 
\begin{align}
M & =\underset{}{\textrm{arg min}}\Bigl(\left\Vert \varrho_{0}-\mathfrak{C}M'\right\Vert ^{2}\Bigr)\nonumber \\
 & =\textrm{arg min}\Bigl(\underbrace{\xcancel{\langle\varrho_{0}|\varrho_{0}\rangle}}_{={\rm const.}}-2{\rm Re}\bigl(\langle\varrho_{0}|\mathfrak{C}|M'\rangle\bigr)+\langle M'|\mathfrak{C}^{2}|M'\rangle\Bigr).\label{eq: Minimierung Rho_Approx TADM}
\end{align}
While Eq.~(\ref{eq:Optimization Objective tadm}) and Eq.~(\ref{eq: Minimierung Rho_Approx TADM})
are equivalent, one might also find alternative approaches for $M$
which would yield the same $\bar{\varrho}$ for optimal $M$, but
result in qualitatively different approximations $\bar{\varrho}_{{\rm approx}}$
for imperfect $M$. In appendix~\ref{sub:Alternative-optimization-ansatz tadm},
we discuss such an approach given by
\begin{equation}
M=\textrm{arg min}\Bigl(\left\Vert \mathfrak{C}\left(\varrho_{0}-\mathfrak{C}M'\right)\right\Vert ^{2}\Bigr).\label{eq:Vorank=0000FCndigung T+Methode}
\end{equation}
This method minimizes the residual time dependence of $\bar{\varrho}_{{\rm approx}}$,
while for most other physical properties, the standard method described
by Eqs.~(\ref{eq:Optimization Objective tadm}) and (\ref{eq: Minimierung Rho_Approx TADM})
seems more promising; see appendix~\ref{sub:Which-method-is better T+ T-}.

\subsubsection{General eigenvector problem}

Instead of maximizing $\langle\varrho_{0}|\mathfrak{C}M\rangle$ as
in Eq.~(\ref{eq:Optimization Objective tadm}), one can also maximize
$\left\langle \mathfrak{C}M|\varrho_{0}\right\rangle \left\langle \varrho_{0}|\mathfrak{C}M\right\rangle $.
The advantage of this bilinear form is that the search for an optimal
$M$ can now be phrased as 
\begin{equation}
M=\textrm{arg max}\left(\frac{\left\langle \mathfrak{C}M'|\varrho_{0}\right\rangle \left\langle \varrho_{0}|\mathfrak{C}M'\right\rangle }{\left\langle \mathfrak{C}M'|\mathfrak{C}M'\right\rangle }\right),
\end{equation}
which can be solved as a general eigenvector problem for the maximal
eigenvalue~$\lambda$
\begin{equation}
\left(\mathfrak{C}|\varrho_{0}\rangle\langle\varrho_{0}|\mathfrak{C}\right)|M\rangle=\lambda\cdot\mathfrak{C}\mathfrak{C}|M\rangle\label{eq:General Eigenvector problem TADM}
\end{equation}
Unfortunately, both sides of the eigenvector equation can become zero
at the same time for $\left\Vert M\right\Vert >0$, which is a notorious
source of trouble for the numerical treatment of generalized eigenvector
problems. Therefore, we do not consider this approach as ideal and
present an alternative strategy in the appendices~\ref{sec:Solving-the-optimization problem general approach tadm}
and \ref{sec:Time-averaged-density as MPO}. Still, the reader who
has already a good and stable software solution for this problem at
his or her disposal might give it a try, anyway.

\subsection{Solving the optimization problem\label{sub:Solving-the-optimization overview}}

In the following, we just give a short (an hence incomplete) overview
of the method used for solving the optimization problem. Further information
and omitted explanations can be found in the appendices~\ref{sec:Solving-the-optimization problem general approach tadm},
\ref{sec:Time-averaged-density as MPO}, and beyond, where an in-detail
description is provided.

We need to find a matrix $M$ (\ref{eq:=0000DCbersichtsformel Time averaged d.m.})
, with $\varrho_{\textrm{off-diag}}=\mathfrak{C}M$ (\ref{eq:Rho Off gleich HM}).
To this end, the matrix $M$ is expressed as a linear combination
\begin{equation}
M=\sum_{j}\alpha_{j}\mathcal{M}_{j},\label{eq:Vorweg summenansatz shortpaper}
\end{equation}
i.e., $\varrho_{\textrm{off-diag}}=\sum_{j}\alpha_{j}\mathfrak{C}\mathcal{M}_{j}$.
For the matrices $\mathcal{M}$, we demand $\langle\mathfrak{C}\mathcal{M}_{j}|\mathfrak{C}\mathcal{M}_{k}\rangle=\delta_{jk}$,
which allows us to obtain the optimal coefficients $\alpha_{j}$ as
\begin{equation}
\alpha_{j}=\langle\mathfrak{C}\mathcal{M}_{j}|\varrho_{{\rm off-diag}}\rangle\overset{\eqref{eq:Gleicher Overlar rho null Rho off}}{=}\langle\mathfrak{C}\mathcal{M}_{j}|\varrho_{0}\rangle.
\end{equation}
We like to find matrices $\mathcal{M}_{j}$ with high $\alpha_{j}$,
respectively with high overlap $\langle\mathfrak{C}\mathcal{M}_{j}|\varrho_{0}\rangle$.
A suitable way is to generate the matrices $\mathcal{M}_{j}$ iteratively
as elements of a Krylov subspace $\mathcal{K}$
\begin{equation}
\mathcal{K}=\textrm{span}\left\{ \mathfrak{C}\varrho_{0},\mathfrak{C}^{3}\varrho_{0},\mathfrak{C}^{5}\varrho_{0},\ldots,\mathfrak{C}^{2j-1}\varrho_{0}\right\} .\label{eq:Krylov Shortoverview}
\end{equation}
But in this Krylov subspace approach, we still ignore the fact that
the matrices $\mathcal{M}_{j}$ exhibit the same exponential scaling
with the system size as $\varrho_{0}$ itself, which generally foils
an explicit calculation of these matrices. 

To master this exponential scaling, we resort to a tensor network
representation \cite{Verstraete2008Review,Schollwoeck2011,Eisert2013OverviewTNS,Orus2014_TensorNetworks},
i.e., we use MPO (appendix~\ref{sub:Short-introduction-to MPO in TADM})
and double MPS, a tensor network explained in appendix~\ref{sec:Double-MPS tadm}.
The basic idea of a tensor network is to express (or approximate)
a high-dimensional object $M$ as a product of low-dimensional tensors
$\mathsf{M}_{[k]}$
\begin{equation}
M=\prod_{k}\mathsf{M}_{[k]}.\label{eq:Vorweg tensor network}
\end{equation}
This is a short hand notation, where we omitted the indices used for
the multiplications of the tensors $\mathsf{M}_{[k]}$; see also appendix
\ref{sub:Short-introduction-to MPO in TADM}. The dimension of these
indices is commonly referred to as bond dimension and has to be limited
for a successful numerical handling. This also imposes limitations
on the maximal amount of entanglement which can be represented faithfully.

In case of a tensor network, we need an optimization procedure for
the network tensors $\mathsf{M}_{[k]}$ (\ref{eq:Vorweg tensor network}).
Here, we can use the same idea as before and express each $\mathsf{M}_{[k]}$
as a linear combination of iteratively generated tensors $\mathsf{M}_{[k]}^{(j)}$
\begin{equation}
\mathsf{M}_{[k]}=\sum\alpha'_{j}\mathsf{M}_{[k]}^{(j)}.
\end{equation}
But it contrast to Eq.~(\ref{eq:Krylov Shortoverview}), it is no
longer advisable to generate the tensors $\mathsf{M}_{[k]}^{(j)}$
as elements of a simple Krylov subspace. Here, a more elaborated iteration
rule is needed (\ref{eq:Overarching iteration tadm}), which also
takes information from previous optimizations into account. This is
explained in appendix~\ref{sub:Advisable-modification-of the algorithm tadm}
and further improved in appendix~\ref{sub:Speeding-up-convergence tadm}.

\section{Results\label{sec:Results TADM}}

In this paper, we have presented a new numerical method, which naturally
raises lots of questions concerning its performance. In case of highly
entangled time averaged density matrices or operators, the probably
most urgent question is how much insight we can really gain if the
chosen tensor network ansatz only supports a limited amount of entanglement.
We strongly focus on this question comparing results for different
bond dimensions $D=2^{n}$ ranging from $D=4$ to $D=512$. Hereby,
$D$ always refers to the bond dimension used for the ansatz $M$
in $\bar{\varrho}=\varrho_{0}-\mathfrak{C}M$ (\ref{eq:=0000DCbersichtsformel Time averaged d.m.}).
Other interesting aspects as convergence properties and the achievable
precision are addressed in appendix~\ref{sec:Numerical-aspects TADM}.

As already mentioned in the introduction (Sec.~\ref{sec:Inroduction TADM}),
integrable and non integrable systems are expected to thermalize differently.
Further, for integrable systems, the tensor network based simulation
of time evolution can often be done with less computational resources,
i.e., with lower bond dimensions \cite{Prosen2007_Integrabel}. As
an example for an integrable system, we look at the Ising Hamiltonian
$H$ of a spin chain of length $L$
\begin{equation}
H=-\sum_{j=1}^{L-1}\sigma_{z}^{(j)}\sigma_{z}^{(j+1)}-\sum_{j=1}^{L}\sigma_{x}^{(j)},
\end{equation}
where $\sigma_{x}^{(j)}$ and $\sigma_{z}^{(j)}$ denote the Pauli
matrices applied to the $j$th spin. This Hamiltonian can be mapped
onto a system of free fermions by a Jordan-Wigner transformation.
Also numerically, one quickly finds that the time average of a single
$\sigma_{x}^{(k)}$ operator (Heisenberg picture, see appendix~\ref{sec:Time-averaged-operator variance and error reduction tadm})
can be described by a MPO with bond dimension $D\leqslant L+2$ and
for the time average of the operator $S_{x}=\sum_{j=1}^{L}\sigma_{x}^{(j)}$,
even $D=4$ is sufficient. For non-integrable Ising models on the
other hand, such simplifications cannot be found.

For the rest of this paper, we consider the non-integrable Ising Hamiltonian
$H$ 
\begin{equation}
H=-\sum_{j=1}^{L-1}\sigma_{z}^{(j)}\sigma_{z}^{(j+1)}-\sum_{j=1}^{L}\frac{\sigma_{x}^{(j)}+\sigma_{z}^{(j)}}{\sqrt{2}},\label{eq:Hamilton Nichtintegrabel TADM}
\end{equation}
for which we compare spin chains of different length, $L=13,25,51$.
As examples for time averaged operators, we look at the polarization
of the central spin $\sigma_{{\rm field}}^{{\rm central}}$ and the
average polarization $S_{{\rm field}}$ in direction of the applied
field 
\begin{eqnarray}
\sigma_{{\rm field}}^{{\rm central}} & = & \tfrac{\sigma_{x}^{(c)}+\sigma_{z}^{(c)}}{\sqrt{2}},\quad{\rm with}\quad c=\lceil\tfrac{L}{2}\rceil\label{eq:HeisenbergMitte TADM}\\
S_{{\rm field}} & = & \sum_{j=1}^{L}\frac{\sigma_{x}^{(j)}+\sigma_{z}^{(j)}}{\sqrt{2}\cdot L}.\label{eq:HeisenbergAll TADM}
\end{eqnarray}
As states, we consider the two initial state
\begin{alignat}{3}
|\Psi_{+}\rangle & = & |+\rangle^{\otimes L} & = & 2^{-\frac{L}{2}}\binom{1}{1}^{\otimes L}\label{eq:RhoAllMixed TADM}\\
|\Psi_{\uparrow}\rangle & = & |0\rangle^{\otimes L} & = & \binom{1}{0}^{\otimes L}.\label{eq:RhoAllUp TADM}
\end{alignat}

Further, we look at the ground state $|E_{0}\rangle$ of the Hamiltonian
(\ref{eq:Hamilton Nichtintegrabel TADM}) where either the central
spin is flipped or the left and right outer spins together
\begin{eqnarray}
|\Psi_{\textrm{central flip}}\rangle & = & \sigma_{x}^{(c)}|E_{0}\rangle,\quad{\rm with}\quad c=\lceil\tfrac{L}{2}\rceil\label{eq:EigenRhoMitte}\\
|\Psi_{\textrm{outer flip}}\rangle & = & \sigma_{x}^{(1)}\otimes\sigma_{x}^{(L)}|E_{0}\rangle.\label{eq:EigenRhoAussen}
\end{eqnarray}
In the ground state, the spins are mostly in the ``up'' position
$\tbinom{1}{0}$ such that we can e.g.\ expect much higher precisions
for $|\Psi_{\uparrow}\rangle$~(\ref{eq:RhoAllUp TADM}) than for
$|\Psi_{+}\rangle$~(\ref{eq:RhoAllMixed TADM}). For $|\Psi_{\textrm{outer flip}}\rangle$,
the precisions is even high enough to calculate reliable results for
the variances of $\overline{\langle\sigma_{z}^{(j)}\rangle}$ with
the help of the method explained in appendix~\ref{sec:Variance-of-expectation values TADM}.

\begin{figure*}
\includegraphics[scale=0.2]{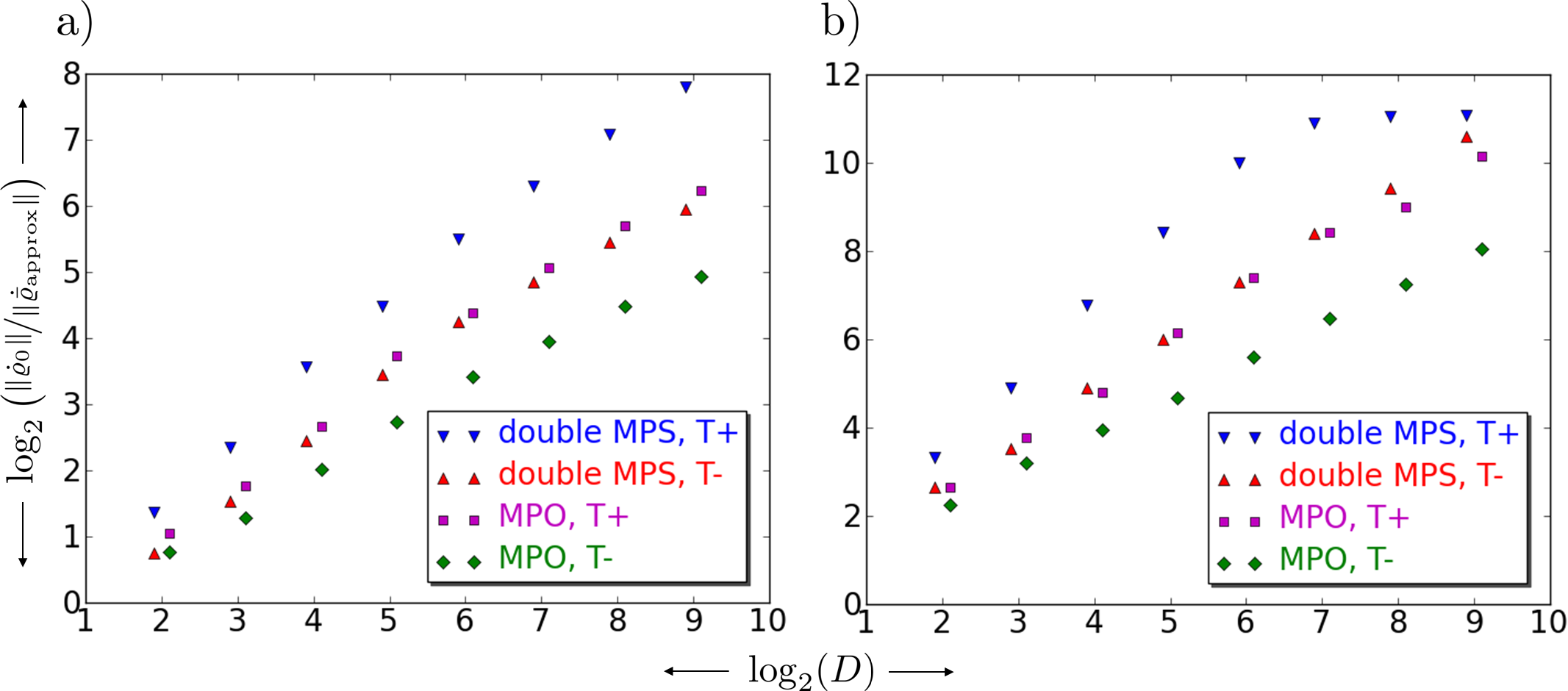}

\protect\caption{\label{fig:AbleitungRhoAllTyps TADM}Comparison of the $q$ value
(\ref{eq:Guete again TADM}) for different methods. The logarithm
of the $q$ value is plotted over the logarithm of the bond dimension
$D$, for a) $|\Psi_{+}\rangle$ (\ref{eq:RhoAllMixed TADM}) and
b) $|\Psi_{\uparrow}\rangle$ (\ref{eq:RhoAllUp TADM}). In both cases,
the system size is $L=25$. The different methods are explained in
the main text (Sec.~\ref{sub:Performance-of-the different methods TADM}).
For better visibility, the data points are slightly shifted \---
$\log_{2}(D)$ is always an integer.}
\end{figure*}

\subsection{Performance of the different methods\label{sub:Performance-of-the different methods TADM} }

We start by comparing the performance of four different ways to calculate
the TADM of the initial states $|\Psi_{\uparrow}\rangle$\ (\ref{eq:RhoAllUp TADM})
and $|\Psi_{+}\rangle$\ (\ref{eq:RhoAllMixed TADM}).

\subsubsection{$q$ value}

To estimate the quality of the approximated TADM $\bar{\varrho}_{{\rm approx}}$
without knowledge of the exact TADM $\bar{\varrho}$, we look at the
residual time dependence of $\mathfrak{\bar{\varrho}}_{{\rm approx}}$
and set it in relation to the time dependence of the initial state
$\mathfrak{\varrho_{{\rm 0}}}$
\begin{equation}
q=\frac{\left\Vert [H,\mathfrak{\varrho_{{\rm 0}}}]\right\Vert }{\left\Vert [H,\mathfrak{\bar{\varrho}_{{\rm approx}}}]\right\Vert }=\frac{\left\Vert \mathfrak{\dot{\varrho}_{{\rm 0}}}\right\Vert }{\left\Vert \mathfrak{\dot{\bar{\varrho}}}_{{\rm approx}}\right\Vert }.\label{eq:Guete again TADM}
\end{equation}
That is, $q$ is the factor by which the time dependence of $\bar{\varrho}_{{\rm approx}}$
was reduced compared to $\varrho_{0}.$ In Sec.~\ref{sub:Vorank=0000FCndigung Alternative-optimization-ansatz},
we mentioned an alternative optimization ansatz which minimizes the
time derivative $\|\mathfrak{\dot{\bar{\varrho}}}_{{\rm approx}}\|$
(\ref{eq:Vorank=0000FCndigung T+Methode}) (and with that maximizes
$q)$, while the standard method used in the rest of the paper minimizes
$\|\bar{\varrho}_{{\rm approx}}\|$ (\ref{eq: Minimierung Rho_Approx TADM})
without time derivative. This alternative optimization ansatz is discussed
in more detail in appendix~\ref{sub:Alternative-optimization-ansatz tadm}.
To distinguish these two methods, we denote them by $T+$ and $T-$

\begin{equation}
\begin{array}{ccl}
T+ &  & \textrm{method which minimizes }\left\Vert \mathfrak{\dot{\bar{\varrho}}}_{{\rm approx}}\right\Vert \\
T- &  & \textrm{standard method, which minimizes }\left\Vert \bar{\varrho}_{{\rm approx}}\right\Vert .
\end{array}\label{eq:T+ und T- TADM}
\end{equation}
 These two optimization methods are used in combination with two different
tensor network ansätze: 1)~MPO and 2)~double MPS. A double MPS is
a MPS of twice the size of a regular MPS and acts as an operator,
as is discussed in appendix~\ref{sec:Double-MPS tadm}. The double
MPS was chosen because only marginal adaptations are necessary to
run the MPO algorithm with a double MPS.

\begin{figure*}
\includegraphics[scale=0.2]{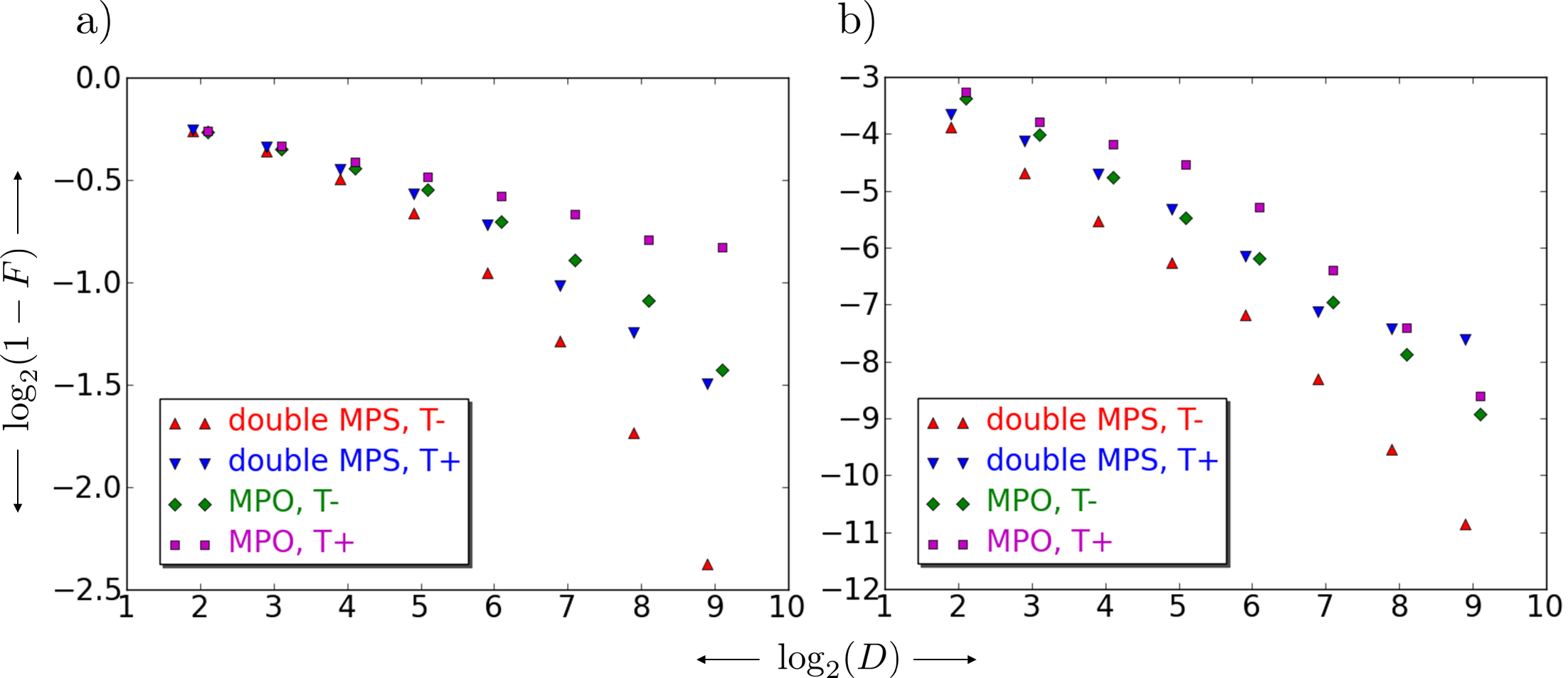}

\protect\caption{\label{fig:Fidelity TADM}Comparison of the fidelity $F$ (\ref{eq:Uhlmann Fidelity TADM})
for different methods. The logarithm of $1-F$ is plotted over the
logarithm of the bond dimension $D$, for a) $|\Psi_{+}\rangle$ (\ref{eq:RhoAllMixed TADM})
and b) $|\Psi_{\uparrow}\rangle$ (\ref{eq:RhoAllUp TADM}). In both
cases, the system size is $L=13$. The different methods are explained
in the main text (Sec.~\ref{sub:Performance-of-the different methods TADM}).
For better visibility, the data points are slightly shifted \---
$\log_{2}(D)$ is always an integer.}
\end{figure*}

Fig.~\ref{fig:AbleitungRhoAllTyps TADM} shows a log-log plot of
$q$ (\ref{eq:Guete again TADM}) in dependence of the bond dimension
$D$ for the initial states $|\Psi_{\uparrow}\rangle=|0\rangle^{\otimes25}$
and $|\Psi_{+}\rangle=|+\rangle^{\otimes25}$. As can be seen, the
double MPS allows to obtain better results than the MPO for these
states. Not surprisingly, we also find that the $T+$ method performs
better than the standard $T-$ method, since the $T+$ method was
designed to generate the highest possible $q$ values.

\subsubsection{Fidelity}

Nonetheless, theoretical considerations in appendix~\ref{sub:Alternative-optimization-ansatz tadm}
suggest that the $T-$ method should be better suited to compute physical
quantities than the $T+$ method. To verify this thesis, we studied
a small and hence exactly solvable spin chain of length $L=13$, which
allows us to calculate the fidelity $F$ 
\begin{equation}
F={\rm Tr}\sqrt{\sqrt{\bar{\varrho}}\bar{\varrho}_{{\rm approx}}\sqrt{\bar{\varrho}}}.\label{eq:Uhlmann Fidelity TADM}
\end{equation}
We remark that $F$ is normally used in the context of positive matrices
only, while the numerically approximated TADM $\bar{\varrho}_{{\rm approx}}$
might have a few small negative eigenvalues. Still, this has no essential
influence on our line of argumentation. 

In Fig.~\ref{fig:Fidelity TADM}, the results for $1-F$ are shown
in a log-log plot in dependence of the bond dimension $D$ for the
initial states $|\Psi_{\uparrow}\rangle=|0\rangle^{\otimes13}$ and
$|\Psi_{+}\rangle=|+\rangle^{\otimes13}$. We still find that the
double MPS performs better than the MPO and as expected, the standard
$T-$ method generates better results than the $T+$ method. 

As a consequence of these findings, we use the standard $T-$ method
to compute the physical properties of a TADM $\bar{\varrho}$, while
we employ the MPO based $T+$ method to compare the $q$ values of
different spin chains, as we do next.

\begin{figure*}
\includegraphics[scale=0.2]{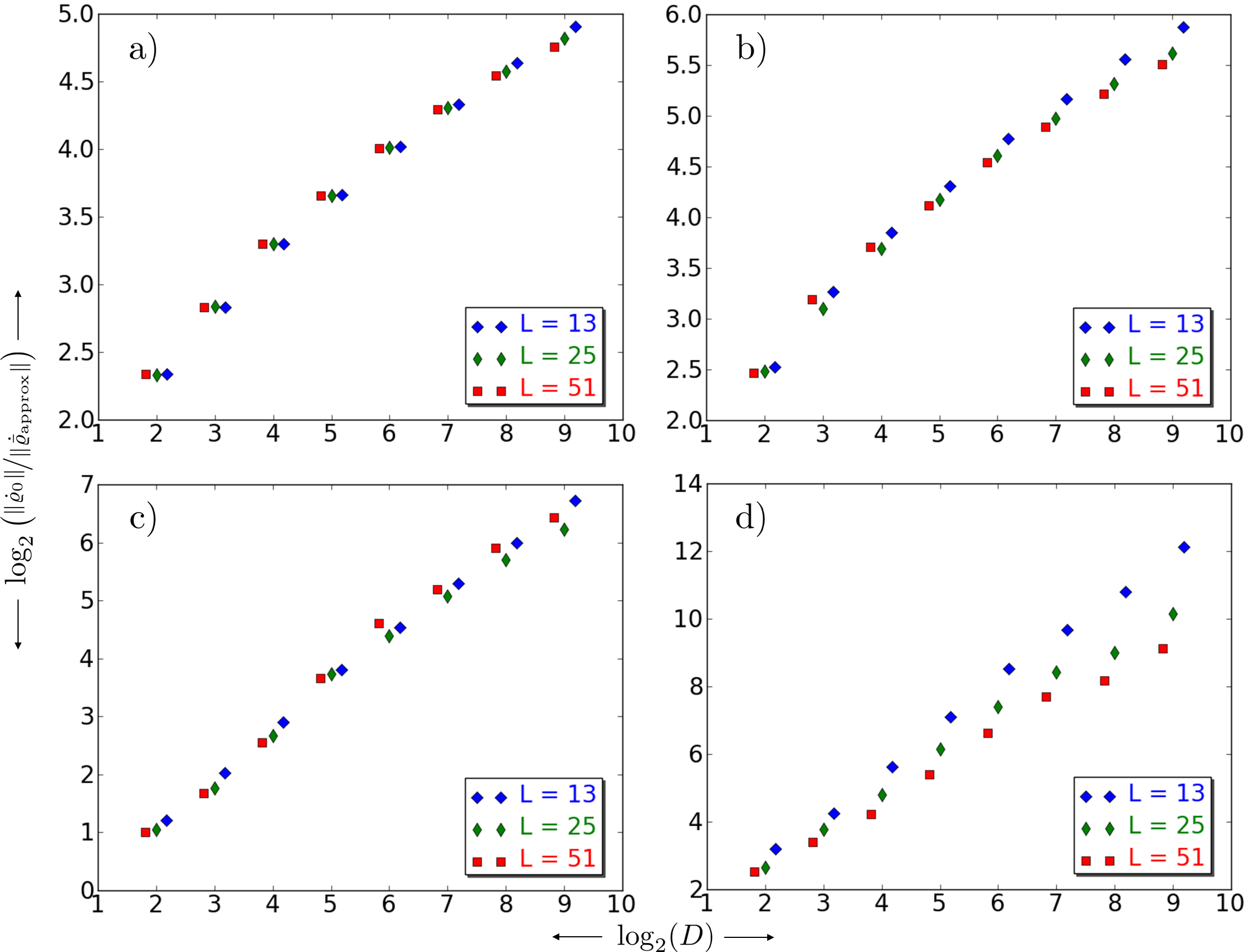}

\protect\caption{\label{fig:Ableitung VergleichLength TADM}Comparison of the $q$
values (\ref{eq:Guete again TADM}) for different system lengths $L=13,25,51$.
The logarithm of the $q$ value is plotted over the logarithm of the
bond dimension $D$, for a) $\sigma_{{\rm field}}^{{\rm central}}$
(\ref{eq:HeisenbergMitte TADM}), b) $S_{{\rm field}}$ (\ref{eq:HeisenbergAll TADM}),
c) $|\Psi_{+}\rangle$ (\ref{eq:RhoAllMixed TADM}) and d) $|\Psi_{\uparrow}\rangle$
(\ref{eq:RhoAllUp TADM}). In all four cases, the $T+$ method (\ref{eq:T+ und T- TADM})
was used combined with a MPO ansatz. For better visibility, the data
points are slightly shifted \--- $\log_{2}(D)$ is always an integer.}
\end{figure*}

\subsection{Spin chains of different length}

In Fig.~\ref{fig:Ableitung VergleichLength TADM}, the $q$ values
(\ref{eq:Guete again TADM}) of the initial operators $\sigma_{{\rm field}}^{{\rm central}}$
(\ref{eq:HeisenbergMitte TADM}), $S_{{\rm field}}$ (\ref{eq:HeisenbergAll TADM})
and the initial states $|\Psi_{+}\rangle$ (\ref{eq:RhoAllMixed TADM}),
$|\Psi_{\uparrow}\rangle$ (\ref{eq:RhoAllUp TADM}) are shown for
spin chains of length $L=13,25,51$. Especially for the time average
of the operator $\sigma_{{\rm field}}^{{\rm central}}$ (Fig.~\ref{fig:Ableitung VergleichLength TADM}~a),
the $q$ value is mostly independent of the length of the spin chain.
For $S_{{\rm field}}$ and $|\Psi_{+}\rangle$, this is roughly true,
as well, while for $|\Psi_{\uparrow}\rangle$, we see a pronounced
difference. At least for $\sigma_{{\rm field}}^{{\rm central}}$,
this weak dependence on the length of the spin chain can be understood
when we look at the operator space entanglement entropy of the time
averages, what we do next.

\subsection{Entanglement entropy\label{sub:Entanglement-entropy TADM}}

In the following five figures, we study the operator space entanglement
entropy (OSEE) \cite{Prosen2007_OSEE} in dependence of the position
where the spin chain is split into two parts. Each plot consists of
a family of curves, where each curve depicts the results obtained
for one specific bond dimension $D=2^{n}$ between $D=4$ and $D=512$.
To emphasize the symmetry of the plots, the center of the spin chain
is denoted as zero and the spin positions left from the center are
addressed with negative numbers. 

\begin{figure*}
\includegraphics[scale=0.2]{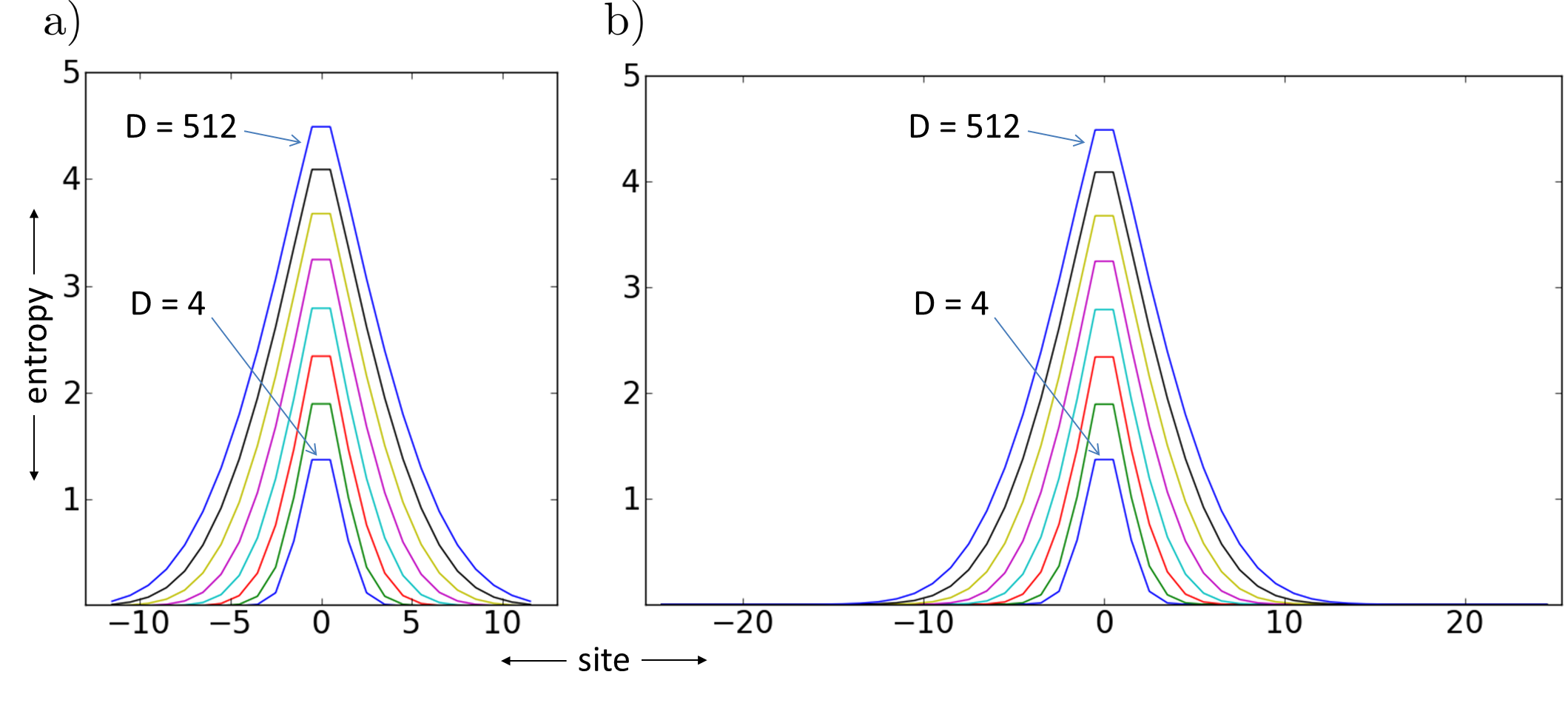}

\protect\caption{\label{fig:Entropie_HeisenbergMitte}For the time averaged operator
$\bar{\sigma}_{{\rm field}}^{{\rm central}}$ (\ref{eq:HeisenbergMitte TADM})
and system sizes a) $L=25$ and b) $L=51$, the operator space entanglement
entropy (OSEE) is plotted over the position of the bipartition. The
OSEE is plotted for all bond dimensions $D=2^{k}$ with $k=2,3,\dots,9$,
whereby the OSEE is monotone increasing with $D$. To emphasize the
symmetry, the center spin is denoted as position zero.}
\end{figure*}

\subsubsection{$\bar{\sigma}_{{\rm field}}^{{\rm central}}$}

Fig.~\ref{fig:Entropie_HeisenbergMitte} shows the OSEE of $\bar{\sigma}_{{\rm field}}^{{\rm central}}$
(the time average of the operator $\sigma_{{\rm field}}^{{\rm central}}$
(\ref{eq:HeisenbergMitte TADM})) for two spin chains of length $L=25$
and $L=51$. Next to the nearly equidistant scaling of the entropy
curves with the bond dimension, we notice a striking resemblance between
the plots for $L=25$ and $L=51$. The $L=51$ appears like the trivial
continuation of the $L=25$ plot, where the OSEE is zero for all bipartions
sufficiently far from the center. 

A closer inspection shows that the approximated $\bar{\sigma}_{{\rm field}}^{{\rm central}}$
acts as an identity operator on spins in the area of vanishing OSEE.
This explains the findings in Fig.~\ref{fig:Ableitung VergleichLength TADM}~a)
that the $q$ value~(\ref{eq:Guete again TADM}) is nearly independent
of the system's length. 

In the context of MPS approximations, it is quite common that the
limitation of the bond dimension induces exponentially decaying correlations.
Usually, this can be understood by the mere observation that the amount
of information a MPS can transmit is limited, while the amount of
transmittable information increases exponentially and hence, has to
be damped. In contrast, for the approximated $\bar{\sigma}_{{\rm field}}^{{\rm central}}$,
most of the MPO's capacity to transmit information appears widely
unused.

\begin{figure*}
\includegraphics[scale=0.2]{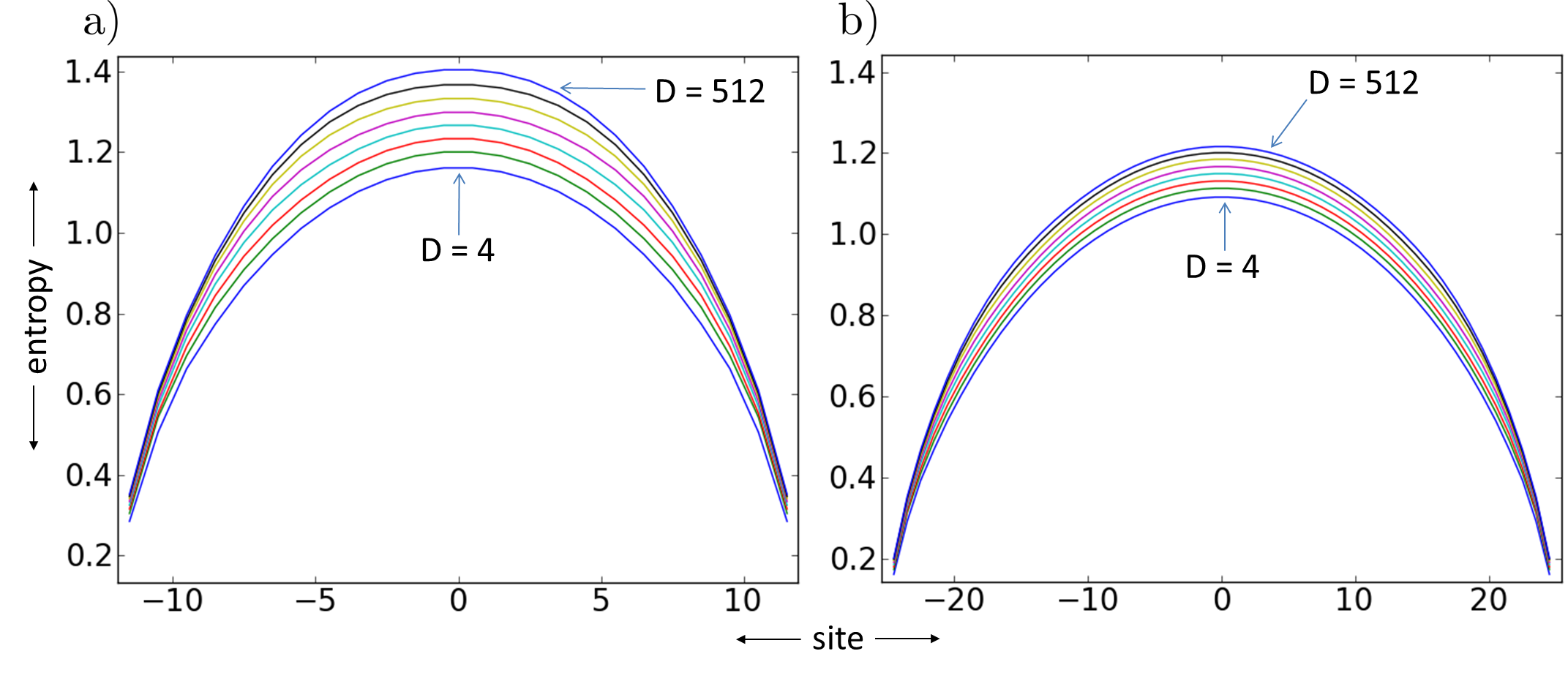}

\protect\caption{\label{fig:Entropie HeisenbergAll TADM}For the time averaged operator
$\bar{S}_{{\rm field}}$ (\ref{eq:HeisenbergMitte TADM}) and system
sizes a) $L=25$ and b) $L=51$, the operator space entanglement entropy
(OSEE) is plotted over the position of the bipartition. The different
curves belong to different bond dimensions $D=2^{k}$ with $k=2,3,\dots,9$,
whereby the OSEE is monotone increasing with $D$. To emphasize the
symmetry, the center spin is denoted as position zero.}
\end{figure*}

\subsubsection{$S_{{\rm field}}$}

Fig.~\ref{fig:Entropie HeisenbergAll TADM} shows the OSEE of the
approximated time averaged operator $\bar{S}_{{\rm field}}$ (\ref{eq:HeisenbergAll TADM})
for two spin chains of length $L=25$ and $L=51$. As for $\bar{\sigma}_{{\rm field}}^{{\rm central}}$
(Fig.~\ref{fig:Entropie_HeisenbergMitte} ), a nearly equidistant
scaling of the entropy curves with the bond dimension can be observed,
but with a much smaller spacing and a distinct offset. Further, the
entropy for $L=51$ is lower than for $L=25$. Interestingly, if one
uses the equidistant scaling for a bold extrapolation to the maximally
needed bond dimensions $D=2^{24}$ respectively $D=2^{50}$, one finds
that the maximal value of the OSEE $S$ is around $S\approx2$ for
both system lengths, $L=25$ as well as $L=51$.

\begin{figure*}
\includegraphics[scale=0.2]{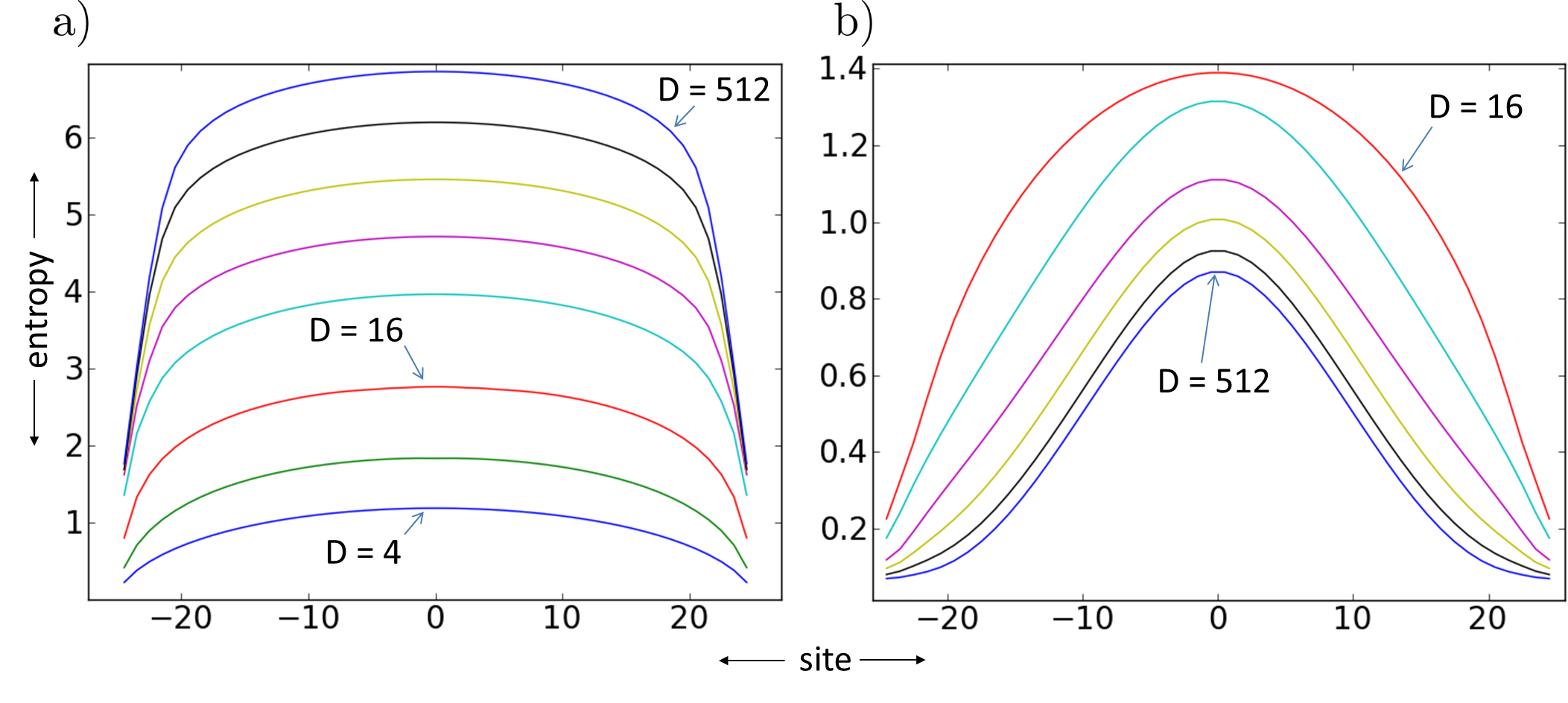}

\protect\caption{\label{fig:Entropie_Rho TADM}For the initial states a) $|\Psi_{+}\rangle$
(\ref{eq:RhoAllMixed TADM}) and b) $|\Psi_{\uparrow}\rangle$ (\ref{eq:RhoAllUp TADM})
and a system size $L=51$, the operator space entanglement entropy
(OSEE) of the TADM $\bar{\varrho}$ is plotted over the position of
the bipartition. The different curves belong to different bond dimensions
$D=2^{k}$, where for a) $k=2,3,\dots,9$ and b) $k=4,5,\dots,9$.
For $|\Psi_{+}\rangle$, the OSEE is monotone increasing with $D$,
while for $|\Psi_{\uparrow}\rangle$, the OSEE is monotone decreasing
with $D$. To have an unambiguously decreasing plot for $|\Psi_{\uparrow}\rangle$,
the bond dimensions $D=4,8$ were omitted, since for for them, the
OSEE is smaller than for $D=16$. To emphasize the symmetry, the center
spin is denoted as position zero.}
\end{figure*}

\subsubsection{\emph{$|\Psi_{+}\rangle$ and $|\Psi_{\uparrow}\rangle$}}

The OSEE of $\bar{\varrho}_{{\rm approx}}$ for the initial states
$|\Psi_{+}\rangle$ (\ref{eq:RhoAllMixed TADM}) and $|\Psi_{\uparrow}\rangle$
(\ref{eq:RhoAllUp TADM}) is shown in Fig.~\ref{fig:Entropie_Rho TADM}
($L=51$). The arguably more interesting plot is the one for $|\Psi_{\uparrow}\rangle$
(Fig.~\ref{fig:Entropie_Rho TADM}~b). Here, the entropy \emph{decreases
}with increasing bond dimension. This anomalous behavior might be
a consequence of the unorthodox optimization ansatz $\bar{\varrho}=\varrho_{0}-\mathfrak{C}M$
(\ref{eq:=0000DCbersichtsformel Time averaged d.m.}). 

Besides the anomalous\emph{ }decrease of the OSEE with increasing
bond dimension, we also notice a convergence of the OSEE. This convergence
is even more distinct for smaller spin chains (not shown here), which
is in accordance with the stronger dependence of the $q$ value on
the system size for $|\Psi_{\uparrow}\rangle$ (Fig.~\ref{fig:Ableitung VergleichLength TADM}~d).
Although the convergence of the OSEE does not necessarily imply the
convergence of the TADM $\bar{\varrho}_{{\rm approx}}\rightarrow\bar{\varrho}_{{\rm exact}}$,
it is still a good indicator.

\begin{figure*}
\includegraphics[scale=0.2]{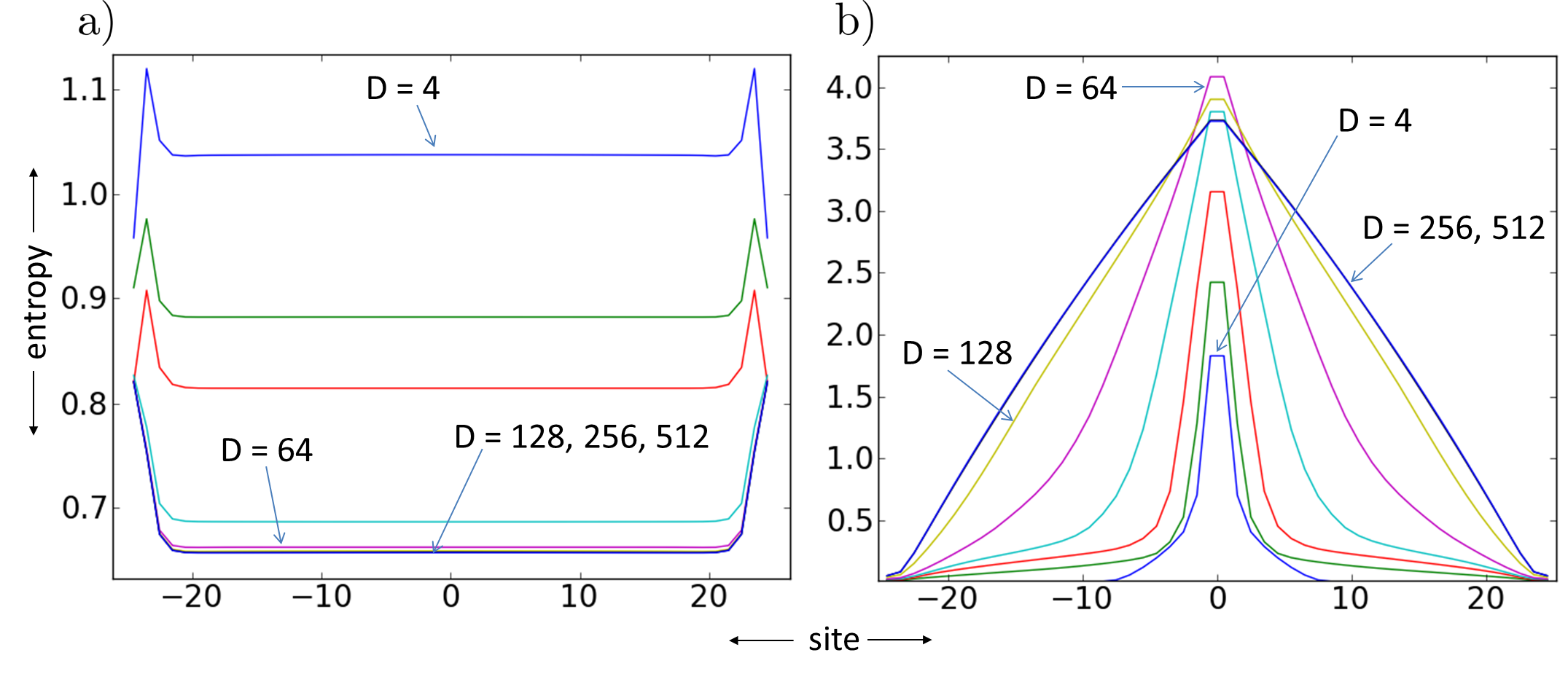}

\protect\caption{\label{fig:Entropie Eigen TADM}For the initial states a) $|\Psi_{\textrm{outer flip}}\rangle$
(\ref{eq:EigenRhoAussen}) and b) $|\Psi_{\textrm{central flip}}\rangle$
(\ref{eq:EigenRhoMitte}) and a system size $L=51$, the operator
space entanglement entropy (OSEE) of the TADM $\bar{\varrho}$ is
plotted over the position of the bipartition. The different curves
belong to different bond dimensions $D=2^{k}$, with $k=2,3,\dots,9$.
For $|\Psi_{\textrm{outer flip}}\rangle$, the OSEE is monotone decreasing
with $D$, while for $|\Psi_{\textrm{central flip}}\rangle$, the
curves broaden with increasing $D$. To emphasize the symmetry, the
center spin is denoted as position zero.}
\end{figure*}

\subsubsection{\textup{$|\Psi_{\textrm{outer flip}}\rangle$ and $|\Psi_{\textrm{central flip}}\rangle$}}

The convergence of the OSEE is even more pronounced for the initial
states $|\Psi_{\textrm{central flip}}\rangle$ (\ref{eq:EigenRhoMitte})
and $|\Psi_{\textrm{outer flip}}\rangle$ (\ref{eq:EigenRhoAussen}),
as shown in Fig.~\ref{fig:Entropie Eigen TADM}. For $|\Psi_{\textrm{outer flip}}\rangle$
(Fig.~\ref{fig:Entropie Eigen TADM}~a), the OSEE for the bond dimensions
$128$, $256$ and $512$ appear as one line and cannot be distinguished.

\begin{figure*}
\includegraphics[scale=0.2]{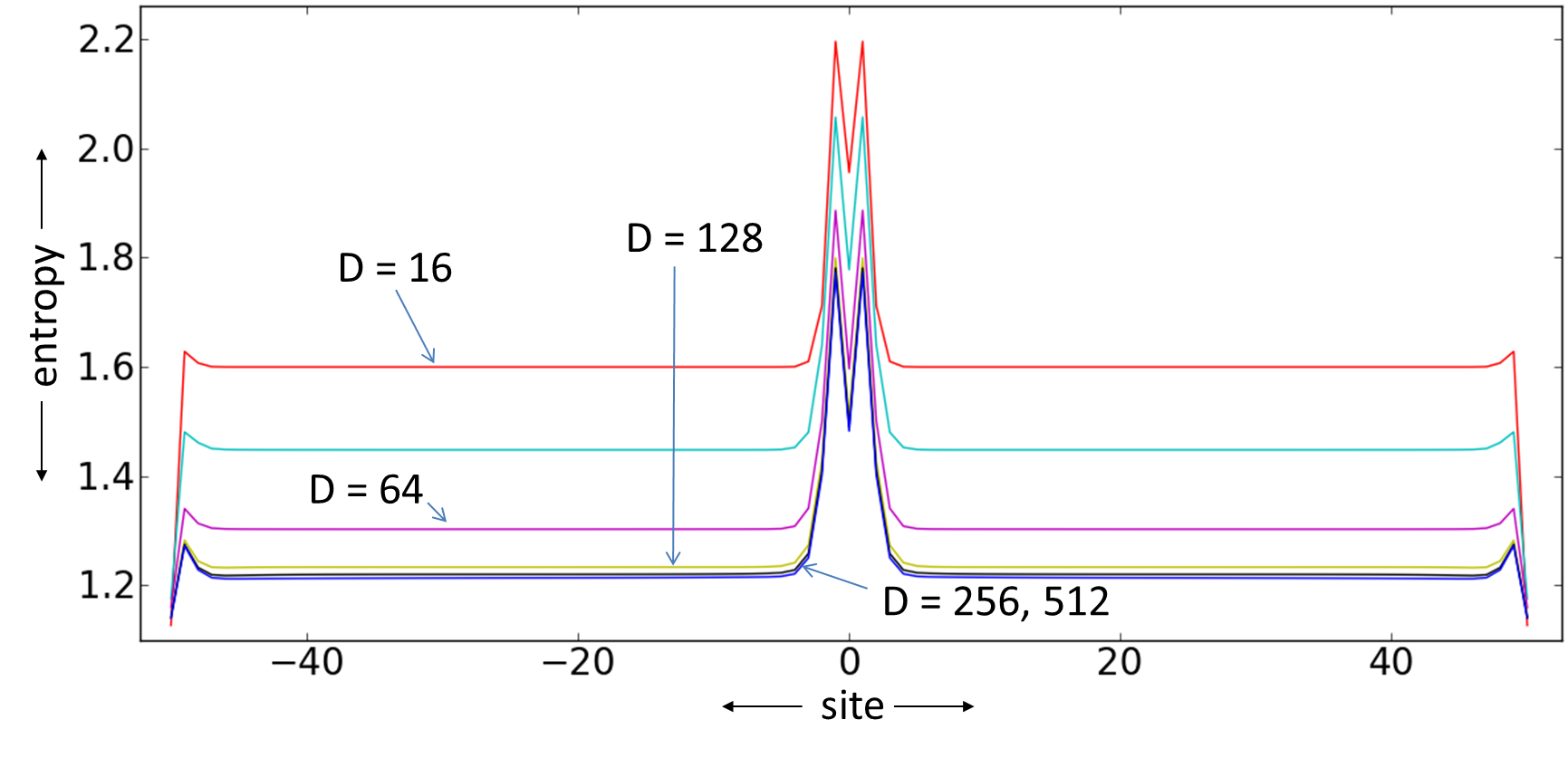}

\protect\caption{\label{fig:Entropie Eigen Doppelt TADM}For the initial state $\mathcal{P}$
(\ref{eq:Doppel System again TADM}), the operator space entanglement
entropy (OSEE) of the TADM $\bar{\varrho}$ is plotted over the position
of the bipartition. The different curves belong to different bond
dimensions $D=2^{k}$, with $k=4,5,\dots,9$. The OSEE is monotone
decreasing with $D$. To have an unambiguously decreasing plot, the
bond dimensions $D=4,8$ were omitted, since for for them, the OSEE
is smaller than for $D=16$. The state $\mathcal{P}$ is an artificial
double state, consisting of two system in a row, each of length $L=51$
in the $|\Psi_{\textrm{outer flip}}\rangle$ state. The zero position
denotes the bipartion which separates these two systems. For the OSEE
at this position and the consequences for the variances $\sigma^{2}$,
see also appendix~\ref{sub:Fully-equilibrated-systems TADM}.}
\end{figure*}

Due to the excellent convergence of $\bar{\varrho}_{{\rm approx}}$
for $|\Psi_{\textrm{outer flip}}\rangle$, we also calculated the
time average of the doubled system~$\mathcal{P}$ 
\begin{equation}
\mathcal{P}:=|\Psi_{\textrm{outer flip}}\rangle\langle\Psi_{\textrm{outer flip}}|\otimes|\Psi_{\textrm{outer flip}}\rangle\langle\Psi_{\textrm{outer flip}}|,\label{eq:Doppel System again TADM}
\end{equation}
whose TADM $\bar{\mathcal{P}}$ allows to compute variances $\sigma^{2}$
(\ref{eq:Doppeltes quanten system tadm}) for the time averaged expectation
values of $|\Psi_{\textrm{outer flip}}(t)\rangle$, as is explained
in appendix~\ref{sec:Variance-of-expectation values TADM} . The
OSEE of the time averaged $\bar{\mathcal{P}}$ is shown in Fig.~\ref{fig:Entropie Eigen Doppelt TADM}
and indicates a very good convergence, as well. 

\begin{figure*}
\includegraphics[scale=0.2]{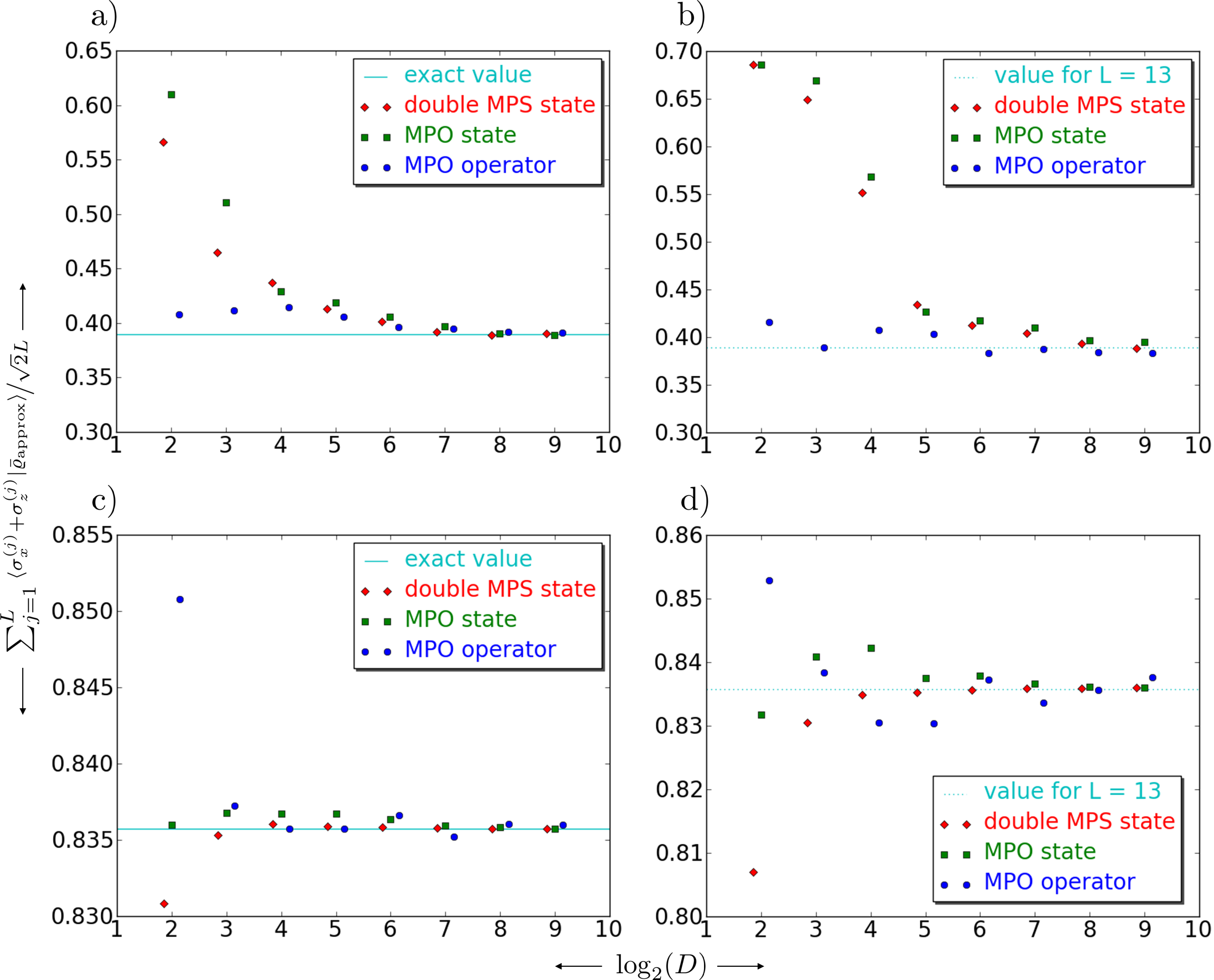}

\protect\caption{\label{fig:Erwartungswerte TADM}The approximation of the time averaged
expectation value $\overline{\langle S_{{\rm field}}\rangle}$ (\ref{eq:HeisenbergAll TADM})
is plotted over the logarithm of the bond dimension $D$. The initial
states are a) $|\Psi_{+}\rangle$ with a system length $L=13$, b)
$|\Psi_{+}\rangle$ with $L=51$, c) $|\Psi_{\uparrow}\rangle$ with
$L=13$ and d) $|\Psi_{\uparrow}\rangle$ with $L=51$. To determine
the time averaged expectation values, the time average $\bar{S}_{{\rm field}}$
of the operator was calculated, as well as the time averaged density
matrices $\bar{\varrho}$. For $\bar{\varrho}$, a MPO and a double
MPS (appendix~\ref{sec:Double-MPS tadm}) ansatz were used. For better
visibility, the data points are slightly shifted \--- $\log_{2}(D)$
is always an integer.}
\end{figure*}

\subsection{Expectation values}

While the OSEE of $\bar{\varrho}_{{\rm approx}}$ indicates an excellent
convergence for the initial state $|\Psi_{\textrm{outer flip}}\rangle$,
the convergence for $|\Psi_{\uparrow}\rangle$ is less clear and for
$|\Psi_{+}\rangle$, we see no convergence at all. Since this might
be the more common situation, we look at the time averaged expectation
values of $|\Psi_{+}\rangle$ and $|\Psi_{\uparrow}\rangle$ first.
As operator, we choose $S_{{\rm field}}$~(\ref{eq:HeisenbergAll TADM}),
for which we have calculated the time average, as well. 

Since we have chosen a non-integrable Hamiltonian~(\ref{eq:Hamilton Nichtintegrabel TADM}),
we do not know the correct results for large systems. The only indicators
we can provide are the common convergence of three different methods
(MPO and double MPS ansatz for $\bar{\varrho}$ and the MPO ansatz
for the time averaged operator $\bar{S}_{{\rm field}}$) and a comparison
with a small, exactly solvable system of 13 sites. The results are
shown in Fig.~\ref{fig:Erwartungswerte TADM}. Here, the worst result
is arguably the one for the 51 sites long $|\Psi_{+}\rangle$ state
(Fig.~\ref{fig:Erwartungswerte TADM}~b). But taking the difficulty
of the task into account, one might still find the results encouraging. 

\begin{figure*}
\includegraphics[scale=0.2]{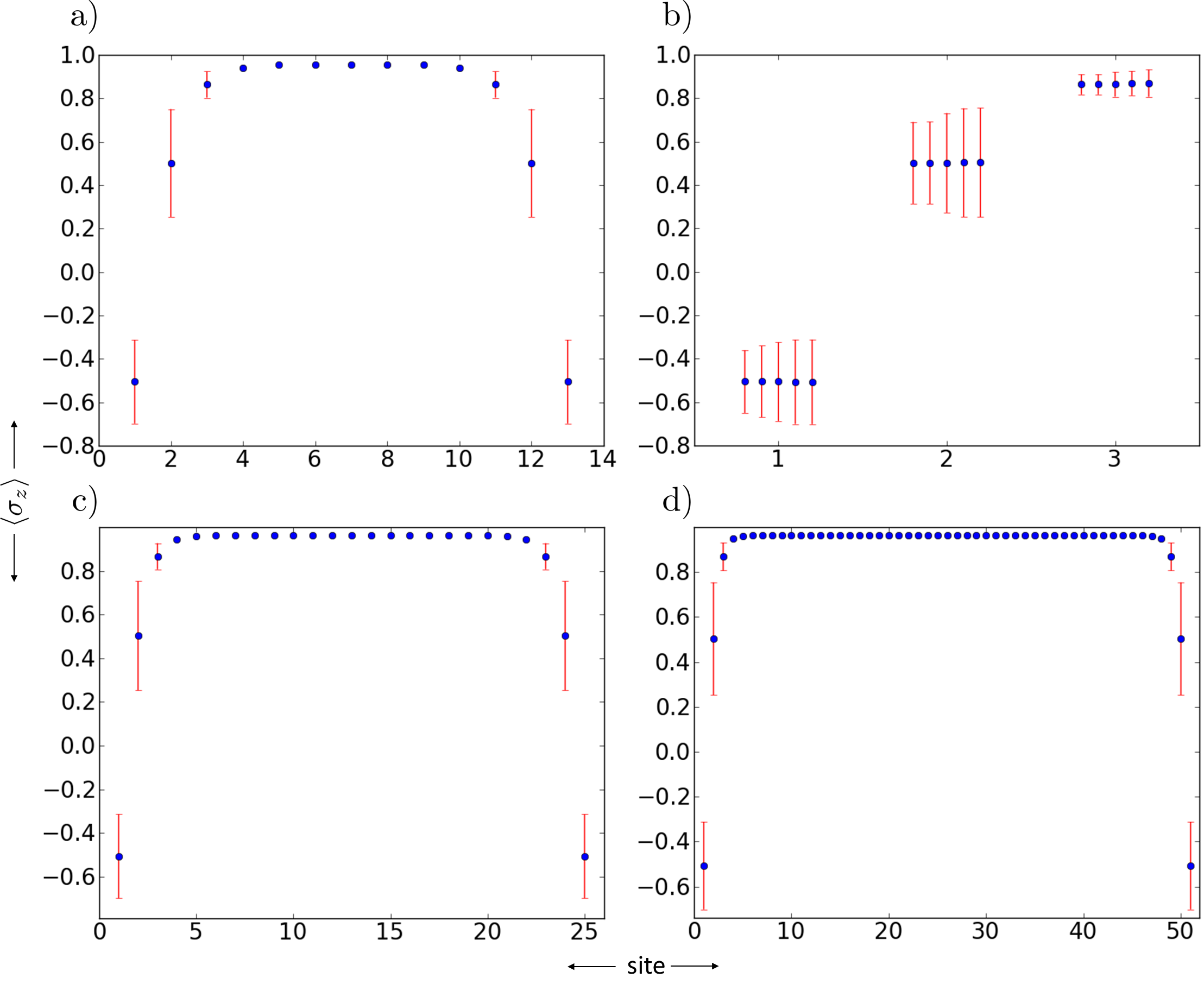}

\protect\caption{\label{fig:Varianz_EigenAussen TADM}For the initial state $|\Psi_{\textrm{outer flip}}\rangle$
(\ref{eq:EigenRhoAussen}), the time averaged expectation values $\overline{\langle\sigma_{z}^{(j)}\rangle}$
and their variances are plotted over the site index $j$, for system
lengths a) $L=13$, c)~$L=25$ and d)~$L=51$. Figure b) shows a
comparison of the variances of the three outer sites for $L=13$,
calculated with five different methods (the data points are slightly
shifted for better visibility). The methods used for the calculation
of theses three variances are (from left to right): 1/ exact result
according to Eq.~(\ref{eq:Varianz mit Eigenwerten}); 2/ sampling
of $|\Psi_{\textrm{outer flip}}(t)\rangle$ with $t=0\dots10^{10}$;
3/ sampling of $|\Psi_{\textrm{outer flip}}(t)\rangle$ with $t=0\dots10^{5}$;
4/ algorithm with double MPS ansatz (appendix~\ref{sec:Double-MPS tadm})
and 5/ algorithm with MPO ansatz. For other sites beyond the outer
three ones, the variances become to small to separate them reliably
from numerical imprecision. The bond dimension is always $D=512$.}
\end{figure*}

\begin{figure*}
\includegraphics[scale=0.2]{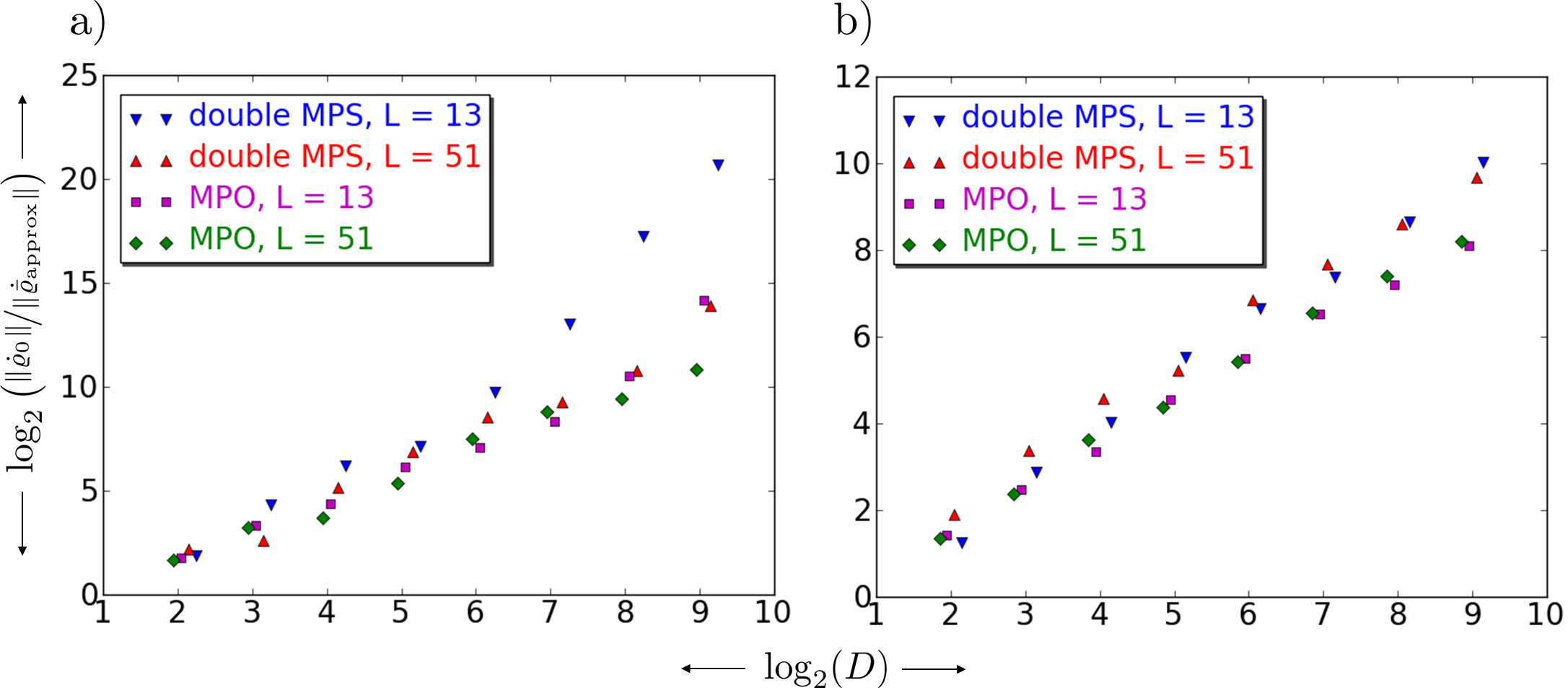}

\protect\caption{\label{fig:Ableitung_Eigen}Comparison of the $q$ value (\ref{eq:Guete again TADM})
for different system lengths $L=13,51$ combined with either a MPO
or a double MPS ansatz (appendix~\ref{sec:Double-MPS tadm}). The
logarithm of the $q$ value is plotted over the logarithm of the bond
dimension $D$, for a) $|\Psi_{\textrm{outer flip}}\rangle$ (\ref{eq:EigenRhoAussen})
and b) $\mathcal{P}$ (\ref{eq:Doppel System again TADM}). In both
cases, the $T-$ method (\ref{eq:T+ und T- TADM}) was used. For better
visibility, the data points are slightly shifted \--- $\log_{2}(D)$
is always an integer.}
\end{figure*}

\subsubsection{Variances\label{sub:Variances Results TADM}}

Finally, we come to the most precise results: The TADM for the initial
state $|\Psi_{\textrm{outer flip}}\rangle$~(\ref{eq:EigenRhoAussen}),
which we employ to determine the time averaged expectation values
of the local Pauli matrices $\sigma_{z}^{(j)}$. As already announced,
we can take advantage of the techniques described in appendix~\ref{sec:Variance-of-expectation values TADM}
and use the time average $\bar{\mathcal{P}}$ of the doubled system
to compute the variances 
\begin{eqnarray}
{\rm Var}\Bigl(\bigl\langle\sigma_{z}^{(j)}\bigr\rangle\Bigr) & = & \overline{\langle\Psi_{\textrm{outer flip}}|\sigma_{z}^{(j)}|\Psi_{\textrm{outer flip}}\rangle^{2}}\nonumber \\
 & - & \overline{\langle\Psi_{\textrm{outer flip}}|\sigma_{z}^{(j)}|\Psi_{\textrm{outer flip}}\rangle}^{2}.
\end{eqnarray}
We determined the time averaged expectation values $\overline{\langle\sigma_{z}^{(j)}\rangle}$
and the associated variances for spin systems of lengths $L=13,25,51$,
see Fig.~\ref{fig:Varianz_EigenAussen TADM}. Here, we find virtually
identical results for the MPO and the double MPS ansatz. 

For $L=13$, the time averaged expectation values $\overline{\langle\sigma_{z}^{(j)}\rangle}$
and their variances can also be calculated exactly (\ref{eq:Varianz mit Eigenwerten}).
The results of our algorithm display a slight overestimation of the
variances compared to the exact results. To put this finding into
the right perspective, we also determined the variances by sampling
over $1.5\cdot10^{5}$ different $|\Psi_{\textrm{outer flip}}(t)\rangle$
for times $t=0\dots10^{5}$ and $3\cdot10^{5}$ samples for $t=0\dots10^{10}$.
Hereby, we observed a greater difference between the variances belonging
to the time interval $t=0\dots10^{5}$ and the exact result than for
the result of the time interval $t=0\dots10^{5}$ and the outcome
of our algorithm. For smaller time intervals, this effect is even
more pronounced. But even for $t=0\dots10^{10}$, we still notice
a difference between the sampled variance and the exact result. This
indicates changes on timescales which are extremely long compared
to the timespans which are usually accessible for numerical simulations.
Independent of these differences, our findings clearly indicate that
also for larger systems, the three outer spins do not equilibrate
within a predictable timeframe.

Finally, to underpin the reliability (and also the limitations) of
our results, in Fig.~\ref{fig:Ableitung_Eigen}, we show the $q$
values (\ref{eq:Guete again TADM})	for $\bar{\varrho}_{{\rm approx}}=\overline{|\Psi_{\textrm{outer flip}}\rangle\langle\Psi_{\textrm{outer flip}}|}$
and $\bar{\mathcal{P}}_{{\rm approx}}$ (\ref{eq:Doppel System again TADM}).
We emphasize that these are not the values obtained from the $T+$
method as in Fig.~\ref{fig:Ableitung VergleichLength TADM}, but
the results of the $T-$ method~(\ref{eq:T+ und T- TADM}), which
were also used for the computation of the variances.

\section{Discussion and outlook\label{sec:Discussion-and-outlook TADM}}

We have presented a new method to compute time averages of density
matrices and operators based on a constraint overlap maximization.
A big advantage of this method is that it can be easily combined with
a tensor network ansatz, as we demonstrated for matrix product operators
(MPO) and double MPS (appendix~\ref{sec:Double-MPS tadm}). As a
new method, it should be compared with already existing ones. Of all
possible alternative methods, here, we consider exact diagonalization. 

Despite its name, the term \emph{exact} diagonalization is commonly
used for a numerical method. Often, the term exact diagonalization
is also used for solutions obtained by, e.g.,\ the Lanczos algorithm
\cite{Lanczos1951} or related iterative methods \cite{Saad2003}.
These methods allow to obtain faithful results for some eigenvectors
of the outer energy spectrum (especially the ground state) but the
results for other eigenvectors in the middle of the spectrum are usually
poor for systems of none-trivial sizes.

Here, we only consider system sizes which do not allow a \emph{complete}
diagonalization into all eigenvectors. If such a complete diagonalization
is possible, this should be the method of choice. Also the iterative
diagonalization algorithms can only handle systems up to a certain
size. Beyond this limit, one might still combine these algorithms
with a tensor network ansatz. This entails new complication, which
will not be listed here, but see e.g.\ Ref.~\cite{Dargel2012_MPS_Laczos}
for a detailed treatment. 

In case of an exact diagonalization of the outer spectrum only, the
decisive question is whether these outer eigenvectors suffice, e.g.,\ to
reconstruct the initial state $\Psi_{0}$. If this is possible, the
time averaged density matrix $\bar{\varrho}$ can be immediately constructed
from these eigenvectors (although for many tasks, the explicit construction
is not necessary). Therefore, various initial states $\Psi_{0}$ might
be categorized by terms like ``easy'', ''difficult'' or ``impossible'',
depending on their overlap with the eigenvectors of the outer energy
spectrum. For the Hamiltonian described by Eq.~(\ref{eq:Hamilton Nichtintegrabel TADM}),
initial states like $|\Psi_{+}\rangle$ should qualify for ``difficult''
up to ``impossible'' (we did not check this explicitly). Further,
determining the time average of an operator $\hat{O}(t)$ with exact
diagonalization should be impossible for nearly all commonly used
operators.

These tasks, which are difficult up to impossible for exact diagonalization,
are also difficult for our algorithm, in the sense that an exact solutions
requires huge bond dimensions which exceed our resources. Still, since
having a weak approximation is still better than having no solution
at all, this is probably the realm where our algorithm has its strongest
superiority compared to exact diagonalization. 

For the ``easy'' states, future investigations have to show which
approach is the most promising. Here, exact diagonalization algorithms
have a certain advantage, since they only deal with states and not
with density matrices, as our algorithm does. But one also has to
consider the task at hand. For expectation values, working with a
collection of eigenstates is numerically favorable, while for the
operator space entanglement entropy (OSEE) calculated in Sec.~\ref{sub:Entanglement-entropy TADM},
one needs the explicit form of $\bar{\varrho}$. Here, even for many
easy initial states, our algorithm should be favorable.

We also determined the variances (Sec.~\ref{sub:Variances Results TADM})
associated with the initial state $|\Psi_{\textrm{outer flip}}\rangle$~(\ref{eq:EigenRhoAussen})
and unveiled that the outer spins do not equilibrate. Although $|\Psi_{\textrm{outer flip}}\rangle$
should qualify as ``easy'' state, we are not aware of any previous
tensor network based approaches which did a similar calculation.

For future applications, it might be interesting to combine the TADM
algorithm with other types of tensor networks, which allow the handling
of greater amounts of entanglement and/or higher dimensions than the
one-dimensional case treated here. Several alternative tensor network
structures are known \cite{Verstraete2004PEPS,Vidal2007Mera,Huebener2009RAGE,CTS}
with different advantages and drawbacks. It remains to be seen which
of them can be integrated well in the algorithm presented here.

At the end, we like to add the speculation that there is a certain
chance that even flawed $\bar{\varrho}_{{\rm approx}}$ of difficult
states might give rise to suitable results for expectation values,
if the erroneous contributions average out. For a better understanding,
we start with the widely accepted assumption that for most closed
quantum states $\varrho(t)$, local expectation values reach an equilibrium
value, which they adopt most of the time. If such equilibrium value
exists, it has to be the same for $\varrho(t)$ as for the TADM $\bar{\varrho}$.
Looking at the off-diagonal elements of such $\varrho(t)=p_{jk}(t)|E_{j}\rangle\langle E_{k}|$,
we find that $|p_{jk}(t)|={\rm const.}$ That is, contrary to the
TADM $\bar{\varrho}$, the off-diagonal elements do not vanish. Still,
both density matrices $\varrho(t)$ and $\bar{\varrho}$ have the
same expectation values. Here, the general assumption is that the
initially aligned $p_{jk}(t=0)$ dephase and as a consequence, average
out.

Now, this dephasing is also an interesting aspect in the context of
flawed $\bar{\varrho}_{{\rm approx}}$. For difficult states, the
algorithm might fail to remove all off-diagonal elements in $\bar{\varrho}_{{\rm approx}}$,
but if the residual off-diagonal elements are sufficiently randomized,
their influence on expectation values might simply average out, as
well. At this point, further investigations are needed to decide,
whether or not the algorithm really randomizes the residual off-diagonal
elements. In any case, we remind the reader that our algorithm does
not introduce errors by altering the diagonal elements of the TADM
$\bar{\varrho}$, which is e.g.\ not true for a flawed diagonalization
algorithm.

\section*{Acknowledgments}

This research was funded by the Austrian Science Fund (FWF): P24273-N16
and by the Austrian Ministry of Science BMWF as part of the UniInfrastrukturprogramm
of the Focal Point Scientific Computing at the University of Innsbruck.
I like to thank Lars Bonnes and Andreas Läuchli for bringing the TADM
problem to my attention. Further, I like to thank them, Jens Eisert
and Tomotoshi Nishino for animating discussions, as well as Wolfgang
Dür and Jens Eisert for reading the manuscript and providing me with
valuable advise. 

I'm currently looking for a PostDoc position.

\bibliographystyle{apsrev}
\bibliography{Zitate_Sammel}

\begin{thebibliography}{37}
\expandafter\ifx\csname natexlab\endcsname\relax\def\natexlab#1{#1}\fi
\expandafter\ifx\csname bibnamefont\endcsname\relax
  \def\bibnamefont#1{#1}\fi
\expandafter\ifx\csname bibfnamefont\endcsname\relax
  \def\bibfnamefont#1{#1}\fi
\expandafter\ifx\csname citenamefont\endcsname\relax
  \def\citenamefont#1{#1}\fi
\expandafter\ifx\csname url\endcsname\relax
  \def\url#1{\texttt{#1}}\fi
\expandafter\ifx\csname urlprefix\endcsname\relax\def\urlprefix{URL }\fi
\providecommand{\bibinfo}[2]{#2}
\providecommand{\eprint}[2][]{\url{#2}}

\bibitem[{\citenamefont{Neumann}(1929)}]{Neumann1929}
\bibinfo{author}{\bibfnamefont{J.}~\bibnamefont{Neumann}},
  \bibinfo{journal}{Zeitschrift für Physik} \textbf{\bibinfo{volume}{57}},
  \bibinfo{pages}{30} (\bibinfo{year}{1929}).

\bibitem[{\citenamefont{Goldstein et~al.}(2010)\citenamefont{Goldstein,
  Lebowitz, Tumulka, and Zanghi}}]{Goldstein2010_Neumann}
\bibinfo{author}{\bibfnamefont{S.}~\bibnamefont{Goldstein}},
  \bibinfo{author}{\bibfnamefont{J.~L.} \bibnamefont{Lebowitz}},
  \bibinfo{author}{\bibfnamefont{R.}~\bibnamefont{Tumulka}}, \bibnamefont{and}
  \bibinfo{author}{\bibfnamefont{N.}~\bibnamefont{Zanghi}},
  \bibinfo{journal}{European Phys. J. H} \textbf{\bibinfo{volume}{35}},
  \bibinfo{pages}{173} (\bibinfo{year}{2010}).

\bibitem[{\citenamefont{Kinoshita et~al.}(2006)\citenamefont{Kinoshita, Wenger,
  and Weiss}}]{Kinoshita2006}
\bibinfo{author}{\bibfnamefont{T.}~\bibnamefont{Kinoshita}},
  \bibinfo{author}{\bibfnamefont{T.}~\bibnamefont{Wenger}}, \bibnamefont{and}
  \bibinfo{author}{\bibfnamefont{D.}~\bibnamefont{Weiss}},
  \bibinfo{journal}{Nature} \textbf{\bibinfo{volume}{440}},
  \bibinfo{pages}{900} (\bibinfo{year}{2006}).

\bibitem[{\citenamefont{Yukalov}(2011)}]{Yukalov2011_IonThermo}
\bibinfo{author}{\bibfnamefont{V.}~\bibnamefont{Yukalov}},
  \bibinfo{journal}{Laser Phys. Lett.} \textbf{\bibinfo{volume}{8}},
  \bibinfo{pages}{485} (\bibinfo{year}{2011}).

\bibitem[{\citenamefont{Deutsch}(1991)}]{Deutsch1991_ETH}
\bibinfo{author}{\bibfnamefont{J.}~\bibnamefont{Deutsch}},
  \bibinfo{journal}{Phys. Rev. A} \textbf{\bibinfo{volume}{43}},
  \bibinfo{pages}{4} (\bibinfo{year}{1991}).

\bibitem[{\citenamefont{Srednicki}(1994)}]{Srednicki1994}
\bibinfo{author}{\bibfnamefont{M.}~\bibnamefont{Srednicki}},
  \bibinfo{journal}{Phys. Rev. E} \textbf{\bibinfo{volume}{50}},
  \bibinfo{pages}{2} (\bibinfo{year}{1994}).

\bibitem[{\citenamefont{Rigol et~al.}(2007)\citenamefont{Rigol, Dunjko,
  Yurovsky, and Olshanii}}]{Rigol2007_GeneralizedGibbs}
\bibinfo{author}{\bibfnamefont{M.}~\bibnamefont{Rigol}},
  \bibinfo{author}{\bibfnamefont{V.}~\bibnamefont{Dunjko}},
  \bibinfo{author}{\bibfnamefont{V.}~\bibnamefont{Yurovsky}}, \bibnamefont{and}
  \bibinfo{author}{\bibfnamefont{M.}~\bibnamefont{Olshanii}},
  \bibinfo{journal}{Phys. Rev. Lett.} \textbf{\bibinfo{volume}{98}},
  \bibinfo{pages}{050405} (\bibinfo{year}{2007}).

\bibitem[{\citenamefont{Reimann}(2008)}]{Reimann2008}
\bibinfo{author}{\bibfnamefont{P.}~\bibnamefont{Reimann}},
  \bibinfo{journal}{Phys. Rev. Lett.} \textbf{\bibinfo{volume}{101}},
  \bibinfo{pages}{190403} (\bibinfo{year}{2008}).

\bibitem[{\citenamefont{Rigol et~al.}(2008)\citenamefont{Rigol, Dunjko, and
  Olshanii}}]{Rigol2008}
\bibinfo{author}{\bibfnamefont{M.}~\bibnamefont{Rigol}},
  \bibinfo{author}{\bibfnamefont{V.}~\bibnamefont{Dunjko}}, \bibnamefont{and}
  \bibinfo{author}{\bibfnamefont{M.}~\bibnamefont{Olshanii}},
  \bibinfo{journal}{Nature} \textbf{\bibinfo{volume}{452}},
  \bibinfo{pages}{854} (\bibinfo{year}{2008}).

\bibitem[{\citenamefont{Cramer et~al.}(2008)\citenamefont{Cramer, Dawson,
  Eisert, and Osborne}}]{EisertCramer2008_Therm0}
\bibinfo{author}{\bibfnamefont{M.}~\bibnamefont{Cramer}},
  \bibinfo{author}{\bibfnamefont{C.~M.} \bibnamefont{Dawson}},
  \bibinfo{author}{\bibfnamefont{J.}~\bibnamefont{Eisert}}, \bibnamefont{and}
  \bibinfo{author}{\bibfnamefont{T.~J.} \bibnamefont{Osborne}},
  \bibinfo{journal}{Phys. Rev. Lett.} \textbf{\bibinfo{volume}{100}},
  \bibinfo{pages}{030602} (\bibinfo{year}{2008}).

\bibitem[{\citenamefont{Linden et~al.}(2009)\citenamefont{Linden, Popescu,
  Short, and Winter}}]{Linden2009_Thermo}
\bibinfo{author}{\bibfnamefont{N.}~\bibnamefont{Linden}},
  \bibinfo{author}{\bibfnamefont{S.}~\bibnamefont{Popescu}},
  \bibinfo{author}{\bibfnamefont{A.~J.} \bibnamefont{Short}}, \bibnamefont{and}
  \bibinfo{author}{\bibfnamefont{A.}~\bibnamefont{Winter}},
  \bibinfo{journal}{Phys. Rev. E} \textbf{\bibinfo{volume}{79}},
  \bibinfo{pages}{061103} (\bibinfo{year}{2009}).

\bibitem[{\citenamefont{Bañuls et~al.}(2011)\citenamefont{Bañuls, Cirac, and
  Hastings}}]{Banuls2011_Thermo}
\bibinfo{author}{\bibfnamefont{M.~C.} \bibnamefont{Bañuls}},
  \bibinfo{author}{\bibfnamefont{J.~I.} \bibnamefont{Cirac}}, \bibnamefont{and}
  \bibinfo{author}{\bibfnamefont{M.~B.} \bibnamefont{Hastings}},
  \bibinfo{journal}{Phys. Rev. Lett.} \textbf{\bibinfo{volume}{106}},
  \bibinfo{pages}{050405} (\bibinfo{year}{2011}).

\bibitem[{\citenamefont{Rigol}(2009)}]{Rigol2009}
\bibinfo{author}{\bibfnamefont{M.}~\bibnamefont{Rigol}},
  \bibinfo{journal}{Phys. Rev. Lett.} \textbf{\bibinfo{volume}{103}},
  \bibinfo{pages}{100403} (\bibinfo{year}{2009}).

\bibitem[{\citenamefont{Gogolin et~al.}(2011)\citenamefont{Gogolin, Müller, and
  Eisert}}]{Gogolin2011_Thermo}
\bibinfo{author}{\bibfnamefont{C.}~\bibnamefont{Gogolin}},
  \bibinfo{author}{\bibfnamefont{M.~P.} \bibnamefont{Müller}},
  \bibnamefont{and} \bibinfo{author}{\bibfnamefont{J.}~\bibnamefont{Eisert}},
  \bibinfo{journal}{Phys. Rev. Lett.} \textbf{\bibinfo{volume}{106}},
  \bibinfo{pages}{040401} (\bibinfo{year}{2011}).

\bibitem[{\citenamefont{Cazalilla and
  Rigol}(2010)}]{Cazalilla2010_OverviewThermo}
\bibinfo{author}{\bibfnamefont{M.~A.} \bibnamefont{Cazalilla}}
  \bibnamefont{and} \bibinfo{author}{\bibfnamefont{M.}~\bibnamefont{Rigol}},
  \bibinfo{journal}{New J. of Phys.} \textbf{\bibinfo{volume}{12}},
  \bibinfo{pages}{055006} (\bibinfo{year}{2010}).

\bibitem[{\citenamefont{Polkovnikov et~al.}(2011)\citenamefont{Polkovnikov,
  Sengupta, and ans M.~Vengalattore}}]{Polkovnikov2011_ThermoOverview}
\bibinfo{author}{\bibfnamefont{A.}~\bibnamefont{Polkovnikov}},
  \bibinfo{author}{\bibfnamefont{K.}~\bibnamefont{Sengupta}}, \bibnamefont{and}
  \bibinfo{author}{\bibfnamefont{A.~S.} \bibnamefont{ans M.~Vengalattore}},
  \bibinfo{journal}{Rev. Mod. Phys.} \textbf{\bibinfo{volume}{83}},
  \bibinfo{pages}{863} (\bibinfo{year}{2011}).

\bibitem[{\citenamefont{Eisert et~al.}(2015)\citenamefont{Eisert, Friesdorf,
  and Gogolin}}]{Eisert2014_Thermo}
\bibinfo{author}{\bibfnamefont{J.}~\bibnamefont{Eisert}},
  \bibinfo{author}{\bibfnamefont{M.}~\bibnamefont{Friesdorf}},
  \bibnamefont{and} \bibinfo{author}{\bibfnamefont{C.}~\bibnamefont{Gogolin}},
  \bibinfo{journal}{Nature Physics} \textbf{\bibinfo{volume}{11}}
  (\bibinfo{year}{2015}).

\bibitem[{\citenamefont{Calabrese et~al.}(2011)\citenamefont{Calabrese, Essler,
  and Fagotti}}]{Calabrese2011_QuenschIsing}
\bibinfo{author}{\bibfnamefont{P.}~\bibnamefont{Calabrese}},
  \bibinfo{author}{\bibfnamefont{F.~H.~L.} \bibnamefont{Essler}},
  \bibnamefont{and} \bibinfo{author}{\bibfnamefont{M.}~\bibnamefont{Fagotti}},
  \bibinfo{journal}{Phys. Rev. Lett.} \textbf{\bibinfo{volume}{106}},
  \bibinfo{pages}{227203} (\bibinfo{year}{2011}).

\bibitem[{\citenamefont{Kormos et~al.}(2014)\citenamefont{Kormos, Collura, and
  Calabrese}}]{KormosCalabrese2014}
\bibinfo{author}{\bibfnamefont{M.}~\bibnamefont{Kormos}},
  \bibinfo{author}{\bibfnamefont{M.}~\bibnamefont{Collura}}, \bibnamefont{and}
  \bibinfo{author}{\bibfnamefont{P.}~\bibnamefont{Calabrese}},
  \bibinfo{journal}{Phys. Rev. A} \textbf{\bibinfo{volume}{89}},
  \bibinfo{pages}{013609} (\bibinfo{year}{2014}).

\bibitem[{\citenamefont{Schuch et~al.}(2008)\citenamefont{Schuch, Wolf,
  Vollbrecht, and Cirac}}]{EntropyTimeSchuch2008}
\bibinfo{author}{\bibfnamefont{N.}~\bibnamefont{Schuch}},
  \bibinfo{author}{\bibfnamefont{M.~M.} \bibnamefont{Wolf}},
  \bibinfo{author}{\bibfnamefont{K.~G.~H.} \bibnamefont{Vollbrecht}},
  \bibnamefont{and} \bibinfo{author}{\bibfnamefont{J.~I.} \bibnamefont{Cirac}},
  \bibinfo{journal}{New J. Phys.} \textbf{\bibinfo{volume}{10}},
  \bibinfo{pages}{033032} (\bibinfo{year}{2008}).

\bibitem[{\citenamefont{Kim et~al.}(2014)\citenamefont{Kim, Bañuls, Cirac,
  Hastings, and Huse}}]{Kim2014_Banuls}
\bibinfo{author}{\bibfnamefont{H.}~\bibnamefont{Kim}},
  \bibinfo{author}{\bibfnamefont{M.~C.} \bibnamefont{Bañuls}},
  \bibinfo{author}{\bibfnamefont{J.~I.} \bibnamefont{Cirac}},
  \bibinfo{author}{\bibfnamefont{M.~B.} \bibnamefont{Hastings}},
  \bibnamefont{and} \bibinfo{author}{\bibfnamefont{D.~A.} \bibnamefont{Huse}},
  \emph{\bibinfo{title}{Slowest local operators in quantum spin chains}},
  \bibinfo{howpublished}{E-print: arXiv:1410.4186v1} (\bibinfo{year}{2014}).

\bibitem[{\citenamefont{Verstraete et~al.}(2008)\citenamefont{Verstraete,
  Cirac, and Murg}}]{Verstraete2008Review}
\bibinfo{author}{\bibfnamefont{F.}~\bibnamefont{Verstraete}},
  \bibinfo{author}{\bibfnamefont{J.~I.} \bibnamefont{Cirac}}, \bibnamefont{and}
  \bibinfo{author}{\bibfnamefont{V.}~\bibnamefont{Murg}},
  \bibinfo{journal}{Adv. Phys.} \textbf{\bibinfo{volume}{57}},
  \bibinfo{pages}{143} (\bibinfo{year}{2008}).

\bibitem[{\citenamefont{Schollw\"ock}(2011)}]{Schollwoeck2011}
\bibinfo{author}{\bibfnamefont{U.}~\bibnamefont{Schollw\"ock}},
  \bibinfo{journal}{Annals of Physics} \textbf{\bibinfo{volume}{326}},
  \bibinfo{pages}{96} (\bibinfo{year}{2011}).

\bibitem[{\citenamefont{Eisert}(2013)}]{Eisert2013OverviewTNS}
\bibinfo{author}{\bibfnamefont{J.}~\bibnamefont{Eisert}},
  \bibinfo{journal}{Modeling and Simulation} \textbf{\bibinfo{volume}{3}},
  \bibinfo{pages}{520} (\bibinfo{year}{2013}).

\bibitem[{\citenamefont{Orus}(2014)}]{Orus2014_TensorNetworks}
\bibinfo{author}{\bibfnamefont{R.}~\bibnamefont{Orus}},
  \bibinfo{journal}{Annals of Physics} \textbf{\bibinfo{volume}{349}},
  \bibinfo{pages}{117} (\bibinfo{year}{2014}).

\bibitem[{\citenamefont{Prosen and Znidaric}(2007)}]{Prosen2007_Integrabel}
\bibinfo{author}{\bibfnamefont{T.}~\bibnamefont{Prosen}} \bibnamefont{and}
  \bibinfo{author}{\bibfnamefont{M.}~\bibnamefont{Znidaric}},
  \bibinfo{journal}{Phys.Rev.E} \textbf{\bibinfo{volume}{75}},
  \bibinfo{pages}{015202} (\bibinfo{year}{2007}).

\bibitem[{\citenamefont{Prosen and Pizorn}(2007)}]{Prosen2007_OSEE}
\bibinfo{author}{\bibfnamefont{T.}~\bibnamefont{Prosen}} \bibnamefont{and}
  \bibinfo{author}{\bibfnamefont{I.}~\bibnamefont{Pizorn}},
  \bibinfo{journal}{Phys. Rev. A} \textbf{\bibinfo{volume}{76}},
  \bibinfo{pages}{032316} (\bibinfo{year}{2007}).

\bibitem[{\citenamefont{Lanczos}(1951)}]{Lanczos1951}
\bibinfo{author}{\bibfnamefont{C.}~\bibnamefont{Lanczos}},
  \bibinfo{journal}{Journal of research of the National Bureau of Standards}
  \textbf{\bibinfo{volume}{45}}, \bibinfo{pages}{255} (\bibinfo{year}{1951}).

\bibitem[{\citenamefont{Saad}(2003)}]{Saad2003}
\bibinfo{author}{\bibfnamefont{Y.}~\bibnamefont{Saad}},
  \emph{\bibinfo{title}{Iterative methods for sparse linear systems (2nd ed.)}}
  (\bibinfo{publisher}{SIAM. ISBN 0898715342}, \bibinfo{year}{2003}).

\bibitem[{\citenamefont{Dargel et~al.}(2012)\citenamefont{Dargel, Wöllert,
  Honecker, McCulloch, Schollwöck, and Pruschke}}]{Dargel2012_MPS_Laczos}
\bibinfo{author}{\bibfnamefont{P.~E.} \bibnamefont{Dargel}},
  \bibinfo{author}{\bibfnamefont{A.}~\bibnamefont{Wöllert}},
  \bibinfo{author}{\bibfnamefont{A.}~\bibnamefont{Honecker}},
  \bibinfo{author}{\bibfnamefont{I.~P.} \bibnamefont{McCulloch}},
  \bibinfo{author}{\bibfnamefont{U.}~\bibnamefont{Schollwöck}},
  \bibnamefont{and} \bibinfo{author}{\bibfnamefont{T.}~\bibnamefont{Pruschke}},
  \bibinfo{journal}{Phys. Rev. B} \textbf{\bibinfo{volume}{85}}
  (\bibinfo{year}{2012}).

\bibitem[{\citenamefont{Verstraete and Cirac}(2004)}]{Verstraete2004PEPS}
\bibinfo{author}{\bibfnamefont{F.}~\bibnamefont{Verstraete}} \bibnamefont{and}
  \bibinfo{author}{\bibfnamefont{J.~I.} \bibnamefont{Cirac}},
  \emph{\bibinfo{title}{Renormalization algorithms for quantum-many body
  systems in two and higher dimensions}}, \bibinfo{howpublished}{E-print:
  arXiv: cond-mat/0407066} (\bibinfo{year}{2004}).

\bibitem[{\citenamefont{Vidal}(2007)}]{Vidal2007Mera}
\bibinfo{author}{\bibfnamefont{G.}~\bibnamefont{Vidal}},
  \bibinfo{journal}{Phys. Rev. Lett.} \textbf{\bibinfo{volume}{99}},
  \bibinfo{pages}{220405} (\bibinfo{year}{2007}).

\bibitem[{\citenamefont{Hübener et~al.}(2009)\citenamefont{Hübener, Kruszynska,
  Hartmann, Dür, Verstraete, Eisert, and Plenio}}]{Huebener2009RAGE}
\bibinfo{author}{\bibfnamefont{R.}~\bibnamefont{Hübener}},
  \bibinfo{author}{\bibfnamefont{C.}~\bibnamefont{Kruszynska}},
  \bibinfo{author}{\bibfnamefont{L.}~\bibnamefont{Hartmann}},
  \bibinfo{author}{\bibfnamefont{W.}~\bibnamefont{Dür}},
  \bibinfo{author}{\bibfnamefont{F.}~\bibnamefont{Verstraete}},
  \bibinfo{author}{\bibfnamefont{J.}~\bibnamefont{Eisert}}, \bibnamefont{and}
  \bibinfo{author}{\bibfnamefont{M.}~\bibnamefont{Plenio}},
  \bibinfo{journal}{Phys. Rev. A} \textbf{\bibinfo{volume}{79}},
  \bibinfo{pages}{022317} (\bibinfo{year}{2009}).

\bibitem[{\citenamefont{Hübener et~al.}(2010)\citenamefont{Hübener, Nebendahl,
  and Dür}}]{CTS}
\bibinfo{author}{\bibfnamefont{R.}~\bibnamefont{Hübener}},
  \bibinfo{author}{\bibfnamefont{V.}~\bibnamefont{Nebendahl}},
  \bibnamefont{and} \bibinfo{author}{\bibfnamefont{W.}~\bibnamefont{Dür}},
  \bibinfo{journal}{New J. Phys.} \textbf{\bibinfo{volume}{12}},
  \bibinfo{pages}{025004} (\bibinfo{year}{2010}).

\bibitem[{\citenamefont{Arnoldi}(1951)}]{Arnoldi1951}
\bibinfo{author}{\bibfnamefont{W.}~\bibnamefont{Arnoldi}},
  \bibinfo{journal}{Quarterly of Applied Mathematics}
  \textbf{\bibinfo{volume}{9}}, \bibinfo{pages}{17} (\bibinfo{year}{1951}).

\bibitem[{\citenamefont{Crosswhite and Bacon}(2008)}]{Crosswhite2008_Automata}
\bibinfo{author}{\bibfnamefont{G.~M.} \bibnamefont{Crosswhite}}
  \bibnamefont{and} \bibinfo{author}{\bibfnamefont{D.}~\bibnamefont{Bacon}},
  \bibinfo{journal}{Phys. Rev. A} \textbf{\bibinfo{volume}{78}},
  \bibinfo{pages}{012356} (\bibinfo{year}{2008}).

\bibitem[{\citenamefont{Fröwis et~al.}(2010)\citenamefont{Fröwis, Nebendahl,
  and Dür}}]{Froewis2010}
\bibinfo{author}{\bibfnamefont{F.}~\bibnamefont{Fröwis}},
  \bibinfo{author}{\bibfnamefont{V.}~\bibnamefont{Nebendahl}},
  \bibnamefont{and} \bibinfo{author}{\bibfnamefont{W.}~\bibnamefont{Dür}},
  \bibinfo{journal}{Phys. Rev. A} \textbf{\bibinfo{volume}{81}},
  \bibinfo{pages}{062337} (\bibinfo{year}{2010}).

\end{thebibliography}

\appendix

\section{Inverse problem\label{sub: Inverse Problem tadm}}

Here, we study the possibility to solve the time averaged density
matrix (TADM) $\bar{\varrho}$ as inverse problem. We start with Eq.~(\ref{eq:Rho off as HM tadm}),
which expresses the off-diagonal elements $\varrho_{\textrm{off-diag}}$
of $\varrho_{0}$ as commutator

\begin{equation}
\varrho_{\textrm{off-diag}}=[H,M]\overset{\eqref{eq:Kommutatorzeiche C def t.a.d.m.}}{=}\mathfrak{C}M.\label{eq:Rho Off gleich HM_Anhang}
\end{equation}
Our task is to find a suitable matrix $M$. Formally, this is solved
by
\begin{equation}
M=\mathfrak{C}^{-1}\varrho_{\textrm{off-diag}}.\label{eq:M als invers tadm}
\end{equation}
We do not know the inverse operator $\mathfrak{C}^{-1}$ and generally,
it is much more demanding to construct $\mathfrak{C}^{-1}$ than to
find a suitable $M$. Still, it seems beneficial to have a short look
at the formal structure of the inverse problem, since with Eq.~(\ref{eq:M als invers tadm}),
we have establish a link to a well known class of problems. As a bonus,
the algorithms derived for the calculation of the matrix $M$ might
also be used for other inverse problems, arising from completely different
tasks. 

In our particular problem, we have to keep in mind that $\mathfrak{C}$
has a non-vanishing kernel $\mathfrak{C}M_{\textrm{diag}}=0$ (\ref{eq:C vernichtet rho_diag}).
Hence, $\mathfrak{C}^{-1}$ should be a well defined Pseudo-Inverse.
That is, while $\mathfrak{C\cdot}\mathfrak{C}^{-1}\varrho_{\textrm{off-diag}}=\varrho_{\textrm{off-diag}}$,
we also have to demand that 
\begin{equation}
\mathfrak{C\cdot}\mathfrak{C}^{-1}\varrho_{\textrm{diag}}=0.
\end{equation}
 As a consequence, we find 
\begin{eqnarray}
\mathfrak{C\cdot}\mathfrak{C}^{-1}\varrho_{0} & = & \underbrace{\mathfrak{C\cdot}\mathfrak{C}^{-1}\varrho_{\textrm{diag}}}_{0}+\underbrace{\mathfrak{C\cdot}\mathfrak{C}^{-1}\varrho_{\textrm{off-diag}}}_{\varrho_{\textrm{off-diag}}}\nonumber \\
 & = & \varrho_{\textrm{off-diag}}.
\end{eqnarray}
With that, the ansatz described by Eq.~(\ref{eq:Ansatz Struktur tadm})
reads
\begin{equation}
\bar{\varrho}=\varrho_{0}-\mathfrak{C\cdot}\mathfrak{C}^{-1}\varrho_{0},
\end{equation}
which evidently would not work if $\mathfrak{C}^{-1}$ were a regular
inverse.

\subsection{Quasi-degenerate eigenstates}

A key aspect of our algorithm is the distinction between diagonal
and off-diagonal elements, respectively the difference between the
two cases $E_{j}=E_{k}$ and $E_{j}\neq E_{k}$. Numerically, this
distinction becomes blurred for quasi-degenerate energy eigenvalues
$E_{q}$, $E_{r}$ with $E_{q}-E_{r}=\varepsilon$, where $0<\varepsilon\ll E_{r}\lesssim E_{q}$.
As a consequence, if the energy difference $\varepsilon$ becomes
to small, the calculated time averaged density matrix $\bar{\varrho}$
might contain non-zero matrix elements $\bar{p}_{qr}\neq0$, which
should actually be zero. 

To be fair, one has to mention that quasi-degenerate energy eigenvalues
pose a general problem, which is not restricted to the method presented
here. To illustrate this, let us replace the idealized limit $T\rightarrow\infty$
in the definition (\ref{eq:Def time averaged desity matrix}) of the
time averaged density matrix $\bar{\varrho}=\lim_{T\rightarrow\infty}\left(\frac{1}{T}\cdot\int_{0}^{T}\varrho(t)\cdot dt\right)$
by a more realistic finite value of $T$. As long as for this $T$
the condition
\begin{equation}
(E_{q}-E_{r})T\gg\hbar\label{eq:EnegieZeit TADM}
\end{equation}
holds, Eq.~(\ref{eq:Delta-Integral time averaged density matrix})
is still a good approximation

\begin{equation}
\int_{0}^{T}\exp\left(-\frac{i}{\hbar}(E_{q}-E_{r})t\right)dt\approx0.
\end{equation}
But for quasi-degenerate energy eigenvalues, the condition $(E_{q}-E_{r})T\gg\hbar$
might no longer hold and the density matrix $\bar{\varrho}_{T}$ averaged
over the finite timespan $T$ might contain off-diagonal matrix elements
$\bar{p}_{qr}$ which deviate substantially from zero.

\section{Alternative optimization ansatz\label{sub:Alternative-optimization-ansatz tadm}}

Throughout this paper, we follow the strategy to express the TADM
as $\bar{\varrho}=\varrho_{0}-\mathfrak{C}M$ (\ref{eq:=0000DCbersichtsformel Time averaged d.m.}).
In a realistic setting, the optimal $M$ will be a highly structured
object, which is often far to complex to be represented faithfully
on a classical computer. Therefore, we have to settle for good approximations
of $M$. This raises the question, how the optimal approximation of
$M$ respectively $\bar{\varrho}$ should look like. The answer to
this question is not unique and closely related to the question how
we actually measure the quality of a given $\bar{\varrho}_{{\rm approx}}$. 

In general, we need a quality measure which is numerically calculable
with the limited computational resources at hand and which does not
require knowledge of the exact TADM $\bar{\varrho}$. Under these
conditions, the arguably best choice is the residual time dependence
of $\bar{\varrho}_{{\rm approx}}$, respectively 
\begin{equation}
\left\Vert \dot{\bar{\varrho}}\mathfrak{_{{\rm approx}}}\right\Vert =\left\Vert \mathfrak{C\bar{\varrho}_{{\rm approx}}}\right\Vert \geqslant0,\label{eq:Quality Kommutator tadm}
\end{equation}
which vanishes for a perfect approximation. Normally, in case of $\left\Vert \mathfrak{C\bar{\varrho}_{{\rm approx}}}\right\Vert =0$,
the diagonal elements of $\bar{\varrho}_{{\rm approx}}$ (expressed
in energy eigenstates) could still be erroneous. But for our method,
which is based on the ansatz $\bar{\varrho}_{{\rm approx}}=\varrho_{0}-\mathfrak{C}M$
(\ref{eq:=0000DCbersichtsformel Time averaged d.m.}), the diagonal
elements are guaranteed to be flawless, since $\mathfrak{C}M$ is
a purely off-diagonal matrix. Hence, $\left\Vert \mathfrak{C\bar{\varrho}_{{\rm approx}}}\right\Vert =0$
if and only if the approximation is perfect.

Now, let us try to find the matrices $M$ which minimize~$\left\Vert \mathfrak{C\bar{\varrho}_{{\rm approx}}}\right\Vert =\left\Vert \mathfrak{C}\left(\varrho_{0}-\mathfrak{C}M\right)\right\Vert $
\begin{eqnarray}
M & = & \textrm{arg min}\Bigl(\left\Vert \mathfrak{C}\left(\varrho_{0}-\mathfrak{C}M'\right)\right\Vert ^{2}\Bigr)\nonumber \\
 & = & \textrm{arg min}\Bigl(\underbrace{\xcancel{\langle\varrho_{0}|\mathfrak{C}^{2}|\varrho_{0}\rangle}}_{={\rm const.}}-2{\rm Re}\bigl(\langle\varrho_{0}|\mathfrak{C}^{3}|M'\rangle\bigr)\nonumber \\
 &  & \qquad\qquad+\langle M'|\mathfrak{C}^{4}|M'\rangle\Bigr).\label{eq:Min for C^3M}
\end{eqnarray}
This minimization can be done in two steps. First, we determine the
optimal $M_{{\rm normed}}$ under the condition that $\langle M|\mathfrak{C}^{4}|M\rangle=1$,
i.e.
\begin{equation}
M_{{\rm normed}}=\underset{\langle M'|\mathfrak{C}^{4}|M'\rangle=1}{\textrm{arg max}}\Bigl(\langle\varrho_{0}|\mathfrak{C}^{3}|M'\rangle\Bigr).\label{eq:Best Kommu M tadm}
\end{equation}
Once this $M_{{\rm normed}}$ is found, we can rescale it similar
to Eq.~(\ref{eq:=0000DCbersichtsformel Time averaged d.m.}). Hence,
Eq.~(\ref{eq:Min for C^3M}) becomes minimal for 
\begin{equation}
M=\left\langle \mathfrak{C}^{3}M_{{\rm normed}}|\varrho_{0}\right\rangle \cdot M_{{\rm normed}}.
\end{equation}

\subsection{Which method is better\label{sub:Which-method-is better T+ T-}?}

The minimization of Eq.~(\ref{eq:Min for C^3M}) results in a $\mathfrak{\bar{\varrho}_{{\rm approx}}}$
with the smallest possible time dependence. But how about other physical
properties as expectation values? Here, the standard method which
minimizes $\left\Vert \mathfrak{\bar{\varrho}_{{\rm approx}}}\right\Vert $
(\ref{eq: Minimierung Rho_Approx TADM}) should still be the better
choice, as we discuss in the following.

The error in any expectation value made by using $\bar{\varrho}_{{\rm approx}}$
instead of the exact $\bar{\varrho}$ stems entirely from residual
off-diagonal elements $\bar{p}_{jk}|E_{j}\rangle\langle E_{k}|$ in
$\bar{\varrho}_{{\rm approx}}$ with $\bar{p}_{jk}\neq0$. Both methods
try to remove these off-diagonal elements, but for the minimization
of $\left\Vert \mathfrak{C\bar{\varrho}_{{\rm approx}}}\right\Vert $,
elements with small energy differences $|E_{j}-E_{k}|$ have lower
priority than elements with high energy differences. Generally, there
is no reason why elements with high energy differences should have
a stronger impact on the error than elements with low energy differences.
It is even more likely to assume that for a local operator $O_{{\rm local}}$,
the unwanted contribution $\langle E_{k}|O_{{\rm local}}|E_{j}\rangle$
is close to zero if the energy difference $|E_{j}-E_{k}|$ is high. 

For the remainder of the paper, only the minimization of $\left\Vert \mathfrak{\bar{\varrho}_{{\rm approx}}}\right\Vert $
respectively Eq.~(\ref{eq:Optimization Objective tadm}) are explained.
However, the adaptations to be made to handle Eq.~(\ref{eq:Best Kommu M tadm})
and with that the minimization of $\left\Vert \mathfrak{C\bar{\varrho}_{{\rm approx}}}\right\Vert $
should be quite obvious, once the principles of the algorithm are
understood.

\section{Time averaged operator and error reduction \label{sec:Time-averaged-operator variance and error reduction tadm}}

The TADM $\bar{\varrho}$ is an object which is associated with the
Schrödinger picture, where states are time dependent and operators
are time independent. In the Heisenberg picture, where states are
time independent and the operators are time dependent, the analog
to the TADM is a time averaged operator $\bar{O}$
\begin{equation}
\bar{O}=\lim_{T\rightarrow\infty}\left(\frac{1}{T}\int_{0}^{T}O(t)dt\right).\label{eq:Def t.a. Operator}
\end{equation}
As done for $\varrho_{0}$, the initial operator $O_{0}$ at time
$t=0$ can be written as 
\begin{equation}
O_{0}=\sum_{j,k}o_{jk}|E_{j}\rangle\langle E_{k}|.
\end{equation}
Analog to Eq.~(\ref{eq:Ansatz Struktur tadm}), the time averaged
operator $\bar{O}$ consists of (block-)diagonal elements only and
can be expressed as
\begin{equation}
\bar{O}=O_{\textrm{diag}}=O_{0}-O_{\textrm{off-diag}}.
\end{equation}
Following the same line of argumentation that led to Eq.~(\ref{eq:=0000DCbersichtsformel Time averaged d.m.})
for the TADM $\bar{\varrho}$, we find that $\bar{O}$ can be obtained
as
\begin{equation}
\bar{O}=O_{0}-\langle\mathfrak{C}\tilde{M}|O_{0}\rangle\cdot\mathfrak{C}\tilde{M},\label{eq:=0000DCbersichtsformel operator average}
\end{equation}
where $\tilde{M}$ has to be chosen such that the inner product $\langle O_{0}|\mathfrak{C}\tilde{M}\rangle$
is maximized and $\Vert\mathfrak{C}\tilde{M}\Vert^{2}=1$.

Using the decomposition into diagonal and off-diagonal elements $\bar{\varrho}=\varrho_{\textrm{diag}}$
and $O_{0}=O_{\textrm{diag}}+O_{\textrm{off-diag}}$, we find for
time averaged expectations values $\bigl\langle\bar{O}\bigr\rangle$
\begin{eqnarray}
\bigl\langle\bar{O}\bigr\rangle & := & \bigl\langle\bar{\varrho}\big|O_{0}\bigr\rangle\nonumber \\
 & = & \bigl\langle\varrho_{\textrm{diag}}\big|O_{\textrm{diag}}+O_{\textrm{off-diag}}\bigr\rangle\nonumber \\
 & = & \bigl\langle\varrho_{\textrm{diag}}\big|O_{\textrm{diag}}\bigr\rangle+\underbrace{\xcancel{\bigl\langle\varrho_{\textrm{diag}}\big|O_{\textrm{off-diag}}\bigr\rangle}}_{=0}\nonumber \\
 & = & \bigl\langle\bar{\varrho}\big|\bar{O}\bigr\rangle,\label{eq:Doppelte Zeitmittelung}
\end{eqnarray}
where we used that $\left\langle A_{\textrm{diag}}|B_{\textrm{off-diag}}\right\rangle \equiv0$
for any matrices $A$, $B$ due to structure of the inner product
$\left\langle A|B\right\rangle =\sum_{j,k}a_{jk}^{*}b_{jk}$. Similarly,
we find
\begin{equation}
\bigl\langle\bar{\varrho}\big|O_{0}\bigr\rangle=\bigl\langle\bar{\varrho}\big|\bar{O}\bigr\rangle=\bigl\langle\varrho_{0}\big|\bar{O}\bigr\rangle.\label{eq:Drei Wege zum erwartungswert TADM}
\end{equation}
The possibility to use both time averaged matrices $\bar{\varrho}$
and $\bar{O}$ to calculate the time averaged expectation value $\left\langle \bar{O}\right\rangle =\left\langle \bar{\varrho}|\bar{O}\right\rangle $
offers a potential way to reduce numerical errors, as we show next.

\subsection{Error reduction\label{sub:Error-reduction tadm}}

For most non-trivial cases, any numerical optimization routine will
only be able to find approximated matrices $M$ (\ref{eq:=0000DCbersichtsformel Time averaged d.m.})
and $\tilde{M}$ (\ref{eq:=0000DCbersichtsformel operator average}),
which deviate from the optimal ones. In this case, the off-diagonal
elements of the obtained $\bar{\varrho}_{\textrm{approx}}$ and $\bar{O}_{\textrm{approx}}$
do not vanish completely, as they should. That is
\begin{eqnarray}
\bar{\varrho}_{\textrm{approx}} & = & \varrho_{\textrm{diag}}+\mathcal{E}_{\textrm{off-diag}}^{[\bar{\varrho}]}\nonumber \\
\bar{O}_{\textrm{approx}} & = & O_{\textrm{diag}}+\mathcal{E}_{\textrm{off-diag}}^{[\bar{O}]},\label{eq:Diag plus Error t.a.d.m}
\end{eqnarray}
where, $\mathcal{E}_{\textrm{off-diag}}^{[\bar{\varrho}]}$ and $\mathcal{E}_{\textrm{off-diag}}^{[\bar{O}]}$
represent the erroneous off-diagonal elements. In any case, the diagonal
elements are always correct for the approach presented here. With
this, we find for the time averaged expectation value $\left\langle \bar{O}\right\rangle _{\textrm{approx}}$
\begin{eqnarray}
\bigl\langle\bar{O}_{\textrm{approx}}\bigr\rangle & \overset{\eqref{eq:Doppel System again TADM}}{=} & \bigl\langle\bar{\varrho}_{\textrm{approx}}\big|\bar{O}_{\textrm{approx}}\bigr\rangle\nonumber \\
 & \overset{\eqref{eq:Diag plus Error t.a.d.m}}{=} & \bigl\langle\varrho_{\textrm{diag}}+\mathcal{E}_{\textrm{off-diag}}^{[\bar{\varrho}]}\big|O_{\textrm{diag}}+\mathcal{E}_{\textrm{off-diag}}^{[\bar{O}]}\bigr\rangle\nonumber \\
 & = & \underbrace{\bigl\langle\varrho_{\textrm{diag}}\big|O_{\textrm{diag}}\bigr\rangle}_{=\left\langle \bar{O}\right\rangle }+\bigl\langle\mathcal{E}_{\textrm{off-diag}}^{[\bar{\varrho}]}\big|\mathcal{E}_{\textrm{off-diag}}^{[\bar{O}]}\bigr\rangle\nonumber \\
 &  & \,
\end{eqnarray}
where we used again that $\left\langle A_{\textrm{diag}}|B_{\textrm{off-diag}}\right\rangle \equiv0$.
Evidently, in case we use both approximated time averages $\bar{\varrho}_{\textrm{approx}}$
and $\bar{O}_{\textrm{approx}}$, the error $E^{[\bar{\varrho},\bar{O}]}$
in the time averaged expectation value $\bigl\langle\bar{O}_{\textrm{approx}}\bigr\rangle$
is given as
\begin{equation}
E^{[\bar{\varrho},\bar{O}]}=\bigl\langle\bar{\varrho}_{\textrm{approx}}\big|\bar{O}_{\textrm{approx}}\bigr\rangle-\bigl\langle\bar{O}\bigr\rangle=\bigl\langle\mathcal{E}_{\textrm{off-diag}}^{[\bar{\varrho}]}\big|\mathcal{E}_{\textrm{off-diag}}^{[\bar{O}]}\bigr\rangle.\label{eq:Quadratic error t.a.d.m.}
\end{equation}
If one resorts to $\left\langle \bar{O}\right\rangle =\bigl\langle\bar{\varrho}|O_{0}\bigr\rangle$
or $\left\langle \bar{O}\right\rangle =\left\langle \varrho_{0}|\bar{O}\right\rangle $
instead, the expressions for the errors is obtained by the same line
of reasoning are
\begin{align}
E^{[\bar{\varrho}]}=\bigl\langle\bar{\varrho}_{\textrm{approx}}\big|O_{0}\bigr\rangle-\bigl\langle\bar{O}\bigr\rangle & =\bigl\langle\mathcal{E}_{\textrm{off-diag}}^{[\bar{\varrho}]}\big|O_{\textrm{off-diag}}\bigr\rangle\nonumber \\
E^{[\bar{O}]}=\bigl\langle\varrho_{0}\big|\bar{O}_{\textrm{approx}}\bigr\rangle-\bigl\langle\bar{O}\bigr\rangle & =\bigl\langle\varrho_{\textrm{off-diag}}\big|\mathcal{E}_{\textrm{off-diag}}^{[\bar{O}]}\bigr\rangle.\label{eq:Standard time averaged expect value}
\end{align}
To compare the errors $E^{[\bar{\varrho},\bar{O}]}$, $E^{[\bar{\varrho}]}$
and $E^{[\bar{O}]}$, we start with a simple error model, which assumes
that all off-diagonal elements are roughly damped by the same factor
$\varepsilon<1$, giving us
\begin{eqnarray}
\mathcal{E}_{\textrm{off-diag}}^{[\bar{\varrho}]} & = & \varepsilon\varrho_{\textrm{off-diag}}\nonumber \\
\mathcal{E}_{\textrm{off-diag}}^{[\bar{O}]} & = & \varepsilon O_{\textrm{off-diag}}.
\end{eqnarray}
In this case, $E^{[\bar{\varrho},\bar{O}]}=\varepsilon^{2}\left\langle \varrho_{\textrm{off-diag}}|O_{\textrm{off-diag}}\right\rangle $
scales with the square of $\varepsilon$, while $E^{[\bar{\varrho}]}=E^{[\bar{O}]}=\varepsilon\left\langle \varrho_{\textrm{off-diag}}|O_{\textrm{off-diag}}\right\rangle $
are only linear in $\varepsilon$.

Unfortunately, for our MPO based calculations, this error model proved
to be insufficient. Instead of using one damping factor $\varepsilon$
for all matrix elements, it seems more adequate to use individual
damping factors $\varepsilon_{jk}$ for each matrix element
\begin{eqnarray}
\mathcal{E}_{\textrm{off-diag}}^{[\bar{\varrho}]} & = & \sum_{E_{j}\neq E_{k}}\varepsilon_{jk}^{[\bar{\varrho}]}p_{jk}|E_{j}\rangle\langle E_{k}|\nonumber \\
\mathcal{E}_{\textrm{off-diag}}^{[\bar{O}]} & = & \sum_{E_{j}\neq E_{k}}\varepsilon_{jk}^{[\bar{O}]}o_{jk}|E_{j}\rangle\langle E_{k}|,
\end{eqnarray}
where $p_{jk}$ and $o_{jk}$ are the matrix elements of $\varrho_{0}$
respectively $O_{0}$. With that, we obtain
\begin{eqnarray}
E^{[\bar{\varrho},\bar{O}]} & \overset{\eqref{eq:Quadratic error t.a.d.m.}}{=} & \sum_{E_{j}\neq E_{k}}\varepsilon_{jk}^{[\bar{\varrho}]*}\varepsilon_{jk}^{[\bar{O}]}p_{jk}^{*}\cdot o_{jk}\nonumber \\
E^{[\bar{\varrho}]} & \overset{\eqref{eq:Standard time averaged expect value}}{=} & \sum_{E_{j}\neq E_{k}}\varepsilon_{jk}^{[\bar{\varrho}]*}p_{jk}^{*}o_{jk}\nonumber \\
E^{[\bar{O}]} & \overset{\eqref{eq:Standard time averaged expect value}}{=} & \sum_{E_{j}\neq E_{k}}\varepsilon_{jk}^{[\bar{O}]}p_{jk}^{*}o_{jk}.
\end{eqnarray}
For statistically distributed $\bigl|\varepsilon_{jk}^{[\bar{\varrho}]}\bigr|\leqslant1\geqslant\bigl|\varepsilon_{jk}^{[\bar{O}]}\bigr|$,
the error $E^{[\bar{\varrho},\bar{O}]}$ will generally be smaller
than the errors $E^{[\bar{\varrho}]}$ and $E^{[\bar{O}]}$. But we
can also find an error model, where this is not the case. Let us suppose
that most matrix elements belonging to the bases $|E_{j}\rangle\langle E_{k}|$
are either extremely easy to approximate or extremely difficult. In
the first case, we expect $\varepsilon_{jk}^{[\bar{\varrho}]}\approx\varepsilon_{jk}^{[\bar{O}]}\approx0$,
while in the second case, $\varepsilon_{jk}^{[\bar{\varrho}]}\approx\varepsilon_{jk}^{[\bar{O}]}\approx1$
seems an adequate assumption. Evidently, for this error model $\varepsilon_{jk}^{[\bar{\varrho}]*}\cdot\varepsilon_{jk}^{[\bar{O}]}\approx\varepsilon_{jk}^{[\bar{\varrho}]*}\approx\varepsilon_{jk}^{[\bar{O}]}$
and with that, all three errors $E^{[\bar{\varrho},\bar{O}]}$, $E^{[\bar{\varrho}]}$
and $E^{[\bar{O}]}$ are roughly the same.

Even for the model, where $\varepsilon_{jk}^{[\bar{\varrho}]}$ and
$\varepsilon_{jk}^{[\bar{O}]}$ only adopt the two values zero and
one, using both time averages $\bar{\varrho}_{\textrm{approx}}$ and
$\bar{O}_{\textrm{approx}}$ to calculate $\left\langle \bar{O}\right\rangle $
offers still an advantage, if the values of $\varepsilon_{jk}^{[\bar{\varrho}]}$
and $\varepsilon_{jk}^{[\bar{O}]}$ are not correlated, i.e.,  if
the combinations $\varepsilon_{jk}^{[\bar{\varrho}]}=1$, $\varepsilon_{jk}^{[\bar{O}]}=0$
and $\varepsilon_{jk}^{[\bar{\varrho}]}=0$, $\varepsilon_{jk}^{[\bar{O}]}=1$
are realistic. The likelihood of having uncorrelated $\varepsilon_{jk}^{[\bar{\varrho}]}$
and $\varepsilon_{jk}^{[\bar{O}]}$ might strongly increase, if $\bar{\varrho}_{\textrm{approx}}$
and $\bar{O}_{\textrm{approx}}$ are calculated by two different methods.
Unfortunately, the error reduction relies on the property of the commutator
based method (\ref{eq:=0000DCbersichtsformel Time averaged d.m.})
that the error is entirely restricted to the off-diagonal matrix elements
of $\bar{\varrho}_{\textrm{approx}}$ and $\bar{O}_{\textrm{approx}}$,
while the diagonal elements are $100\%$ accurate. For other methods,
this is not necessarily the case.

Further, we have to ponder the computational effort to contract $\bigl\langle\bar{\varrho}_{\textrm{approx}}\big|\bar{O}_{\textrm{approx}}\bigr\rangle$.
While for two MPO, this effort is acceptable, this is e.g.\ no longer
necessarily true if $\bar{\varrho}_{\textrm{approx}}$ is given as
double MPS (appendix~\ref{sec:Double-MPS tadm}) and $\bar{O}_{\textrm{approx}}$
as MPO. For two double MPS, the effort would be acceptable again,
but it seems unlikely that a good operator approximation $\bar{O}_{\textrm{approx}}$
can be obtained with an double MPS ansatz. In our MPO based applications,
the error correction only provided a slight improvement.

\section{Variance of expectation values\label{sec:Variance-of-expectation values TADM}}

Here, we study the possibility to compute the variance $\sigma^{2}=\textrm{Var}\Bigl(\bigl\langle\varrho(t)|O\bigr\rangle\Bigr)$
of the expectation value $\bigl\langle\varrho(t)|O\bigr\rangle$ with
respect to time for an arbitrary operator $O$
\begin{equation}
\sigma^{2}=\overline{\bigl\langle\varrho(t)|O\bigr\rangle^{2}}-\overline{\bigl\langle\varrho(t)|O\bigr\rangle}^{2},\label{eq:Varianz Def TADM}
\end{equation}
where the overbar $\overline{\overset{}{\ldots}}$ indicates time
average. 

The second term in Eq.~(\ref{eq:Varianz Def TADM}) is simply $\overline{\bigl\langle\varrho(t)|O\bigr\rangle}^{2}=\bigl\langle\bar{\varrho}|O\bigr\rangle^{2}$
(\ref{eq:Drei Wege zum erwartungswert TADM}), while the first term
$\overline{\bigl\langle\varrho(t)|O\bigr\rangle^{2}}$ needs a more
thorough treatment. In order to find the time average with the means
presented in this paper, we write 
\begin{eqnarray}
\left\langle \varrho(t)|O\right\rangle ^{2} & = & \left\langle \varrho(t)|O\right\rangle \left\langle \varrho(t)|O\right\rangle \nonumber \\
 & = & \bigl\langle\underbrace{\varrho(t)\otimes\varrho(t)}_{\mathcal{P}(t)}|\underbrace{O\otimes O}_{\mathcal{O}}\bigr\rangle\nonumber \\
 & = & \left\langle \mathcal{P}(t)|\mathcal{O}\right\rangle .\label{eq:Doppeltes quanten system tadm}
\end{eqnarray}
That is, by squaring the Hilbert space respectively doubling the quantum
system, we formally transformed the quadratic expression $\left\langle \varrho(t)|O\right\rangle ^{2}$
into the linear expression $\left\langle \mathcal{P}(t)|\mathcal{O}\right\rangle $.
This allows us to proceed as follows 
\begin{eqnarray}
\overline{\bigl\langle\varrho(t)|O\bigr\rangle^{2}} & = & \lim_{T\rightarrow\infty}\Bigl(\frac{1}{T}\int_{0}^{T}\left\langle \varrho(t)|O\right\rangle ^{2}dt\Bigr)\nonumber \\
 & \overset{\eqref{eq:Doppeltes quanten system tadm}}{=} & \lim_{T\rightarrow\infty}\Bigl(\frac{1}{T}\int_{0}^{T}\left\langle \mathcal{P}(t)|\mathcal{O}\right\rangle dt\Bigr)\nonumber \\
 & = & \Bigl\langle\lim_{T\rightarrow\infty}\Bigl(\frac{1}{T}\int_{0}^{T}\mathcal{P}(t)dt\Bigr)\Big|\mathcal{O}\Bigr\rangle\nonumber \\
 & = & \left\langle \bar{\mathcal{P}}|\mathcal{O}\right\rangle .
\end{eqnarray}
Alternatively, we could also compute the time averaged $\bar{\mathcal{O}}$
instead of $\bar{\mathcal{P}}$, since $\left\langle \bar{\mathcal{P}}|\mathcal{O}\right\rangle =\left\langle \mathcal{P}_{0}|\bar{\mathcal{O}}\right\rangle $~(\ref{eq:Drei Wege zum erwartungswert TADM}).
To calculate $\bar{\mathcal{P}}=\mathcal{P}_{\textrm{diag}}$, we
can use Eq.~(\ref{eq:=0000DCbersichtsformel Time averaged d.m.})
$\bar{\mathcal{P}}=\mathcal{P}_{0}-[\mathcal{H},\mathcal{M}]$, where
$\mathcal{M}$ is a suitable matrix we have to find and $\mathcal{H}=H\otimes{\mathbbm1}+{\mathbbm1}\otimes H$
is the Hamiltonian of the doubled quantum system. The chosen structure
of the Hamiltonian $\mathcal{H}$ is easily understood when we look
at the time evolution of $\mathcal{P}(t)=\varrho(t)\otimes\varrho(t)$
\begin{eqnarray}
\varrho(t)\otimes\varrho(t) & = & \left(e^{-iHt}\varrho_{0}e^{iHt}\right)\otimes\left(e^{-iHt}\varrho_{0}e^{iHt}\right)\nonumber \\
 & = & e^{-iH\otimes{\mathbbm1}t}e^{-i{\mathbbm1}\otimes Ht}\left(\varrho_{0}\otimes\varrho_{0}\right)e^{iH\otimes{\mathbbm1}t}e^{i{\mathbbm1}\otimes Ht}\nonumber \\
 & = & e^{-i\left(H\otimes{\mathbbm1}+{\mathbbm1}\otimes H\right)t}\left(\varrho_{0}\otimes\varrho_{0}\right)e^{i\left(H\otimes{\mathbbm1}+{\mathbbm1}\otimes H\right)t}\nonumber \\
 & = & e^{-i\mathcal{H}t}\left(\varrho_{0}\otimes\varrho_{0}\right)e^{i\mathcal{H}t},
\end{eqnarray}
with $\hbar\coloneqq1$.

\subsection{Fully equilibrated systems\label{sub:Fully-equilibrated-systems TADM}}

Analog to Eq.~(\ref{eq:Doppeltes quanten system tadm}), we can write
$\overline{\bigl\langle\varrho(t)|O\bigr\rangle}^{2}=\bigl\langle\bar{\varrho}|O\bigr\rangle^{2}$
as $\bigl\langle\bar{\varrho}|O\bigr\rangle^{2}=\bigl\langle\bar{\varrho}\otimes\bar{\varrho}|O\otimes O\bigr\rangle$
and with that
\begin{equation}
\sigma^{2}\overset{\eqref{eq:Varianz Def TADM},\eqref{eq:Doppeltes quanten system tadm}}{=}\left\langle \bar{\mathcal{P}}|O\otimes O\right\rangle -\bigl\langle\bar{\varrho}\otimes\bar{\varrho}|O\otimes O\bigr\rangle.\label{eq:Zwischenschritt unentangled doppelsystem TADM}
\end{equation}
If a quantum system can fully equilibrate in the sense that the variances
$\sigma^{2}$ vanish for \emph{all} operators $O$, we must have 
\begin{equation}
\forall O:\sigma^{2}=0\;\overset{\eqref{eq:Zwischenschritt unentangled doppelsystem TADM}}{\Longleftrightarrow}\;\bar{\mathcal{P}}=\bar{\varrho}\otimes\bar{\varrho}.
\end{equation}
That is, $\bar{\mathcal{P}}$ is a product state consisting of two
$\bar{\varrho}$. Conversely, the amount of entanglement between the
two subsystems in $\bar{\mathcal{P}}$ can be regarded as indirect
measure for incomplete equilibration. For most systems, demanding
that the variances $\sigma^{2}$ vanish for $all$ operators (not
just the local ones) should be too strong. On the other hand, we can
always restrict the operators to a certain subsystem $A$ by tracing
out all parts of $\bar{\mathcal{P}}$ which do not belong to $A$
and look for the entanglement in the remaining system.

\subsection{Energy eigenstates}

The eigenstates $|\mathcal{E}\rangle$ of the Hamiltonian $\mathcal{H}=H\otimes{\mathbbm1}+{\mathbbm1}\otimes H$
are simply given by $|E_{j}\rangle\otimes|E_{k}\rangle$, where $|E_{j}\rangle$
are the eigenstates of $H$. Both eigenstates $|E_{j}\rangle\otimes|E_{k}\rangle$
and $|E_{k}\rangle\otimes|E_{j}\rangle$ have the same eigenvalue
$E_{j}+E_{k}$. Hence, the spectrum of $\mathcal{H}$ is always degenerate
and $\bar{\mathcal{P}}$ is a block-diagonal matrix. 

On page~\pageref{Diag gleich Blocj Diag TADM}, we introduced the
convention to refer to the block-diagonal elements as ``diagonal'',
as well (see Eq.~(\ref{eq:Del Rho diag tadm})). Here, for once,
we have to discriminate between diagonal and block-diagonal. Actually,
$\bar{\varrho}\otimes\bar{\varrho}$ and $\bar{\mathcal{P}}$ have
the same diagonal elements, but in addition, $\bar{\mathcal{P}}$
has some extra block-diagonal elements due to the afore-mentioned
degeneration. Assuming that the spectrum of $H$ itself is non-degenerate
and that the spectrum of $\mathcal{H}$ exhibits no further degeneration
beyond the one identified above, we find that
\begin{align}
\bar{\varrho}\otimes\bar{\varrho} & =\sum_{j,k}p_{jj}p_{kk}|E_{j}\rangle\langle E_{j}|\otimes|E_{k}\rangle\langle E_{k}|\nonumber \\
\bar{\mathcal{P}} & =\bar{\varrho}\otimes\bar{\varrho}+\sum_{j\neq k}p_{jk}p_{kj}|E_{j}\rangle\langle E_{k}|\otimes|E_{k}\rangle\langle E_{j}|,
\end{align}
with $p_{jk}=\langle E_{j}|\varrho_{0}|E_{k}\rangle$. Inserting this
result into Eq.~(\ref{eq:Zwischenschritt unentangled doppelsystem TADM}),
we find for the variance $\sigma^{2}$ of the time averaged expectation
value $\left\langle \bar{O}\right\rangle $
\begin{eqnarray}
\sigma^{2} & = & \sum_{j\neq k}\|p_{jk}\langle E_{k}|O|E_{j}\rangle\|^{2}\nonumber \\
 & = & \sum_{j\neq k}\|p_{jk}o_{kj}\|^{2},\label{eq:Varianz mit Eigenwerten}
\end{eqnarray}
with $o_{jk}=\langle E_{j}|O|E_{k}\rangle$.

\section{Solving the optimization problem \--- general approach\label{sec:Solving-the-optimization problem general approach tadm}}

In Sec.~\ref{sec:Time-averaged-density as optimizatio problem},
we have seen how the calculation of the time averaged density matrix
(TADM) $\bar{\varrho}$ can be formulated as a common linear optimization
problem with quadratic normalization condition. Here, we study a general
approach to solve this problem, whereby, we assume that the needed
matrix operations can all be executed. Due to the exponential scaling
of the Hilbert space, this assumption is usually only justified for
quantum systems consisting of very few particles. Still, other more
powerful methods might adopt the ideas of this section, as we will
demonstrate for the MPO based approach in appendix~\ref{sec:Time-averaged-density as MPO}. 

We like to construct a matrix $M$ which fulfills Eq.~(\ref{eq:Rho Off gleich HM})
\begin{equation}
\varrho_{\textrm{off-diag}}\overset{\eqref{eq:Rho Off gleich HM}}{=}\mathfrak{C}M\overset{\eqref{eq:Kommutatorzeiche C def t.a.d.m.}}{:=}[H,M]\label{eq:Wiederholung Rho Off gleich HM}
\end{equation}
and hence allows to calculate $\bar{\varrho}=\varrho_{0}-\varrho_{\textrm{off-diag}}$
(\ref{eq:Ansatz Struktur tadm}). The strategy we are going to pursue
is to approximate $M$ as a sum of iteratively generated matrices~$\mathcal{M}_{j}$
\begin{eqnarray}
M & = & \sum_{j}\alpha_{j}\mathcal{M}_{j}\nonumber \\
\varrho_{\textrm{off-diag}} & \overset{\eqref{eq:Wiederholung Rho Off gleich HM}}{=} & \sum_{j}\alpha_{j}\mathfrak{C}\mathcal{M}_{j},\label{eq:Summen Ansatz rho off}
\end{eqnarray}
with $\alpha_{j}\in\mathbb{C}$. The matrices $\mathcal{M}_{j}$ are
modified in such a way that the commutators $\mathfrak{C}\mathcal{M}_{j}$
build an orthonormal system 
\begin{equation}
\langle\mathfrak{C}\mathcal{M}_{j}|\mathfrak{C}\mathcal{M}_{k}\rangle=\delta_{jk}.\label{eq:Weighted norm for HM}
\end{equation}
To do so, we can use an iterative method similar to the Gram-Schmidt
orthonormalization. Any time we generate a new matrix $\tilde{\mathcal{M}}_{n}$,
it is orthonormalized $\tilde{\mathcal{M}}_{n}\rightarrow\mathcal{M}_{n}$
against its precursors $\mathcal{M}_{j<n}$ by the following steps
\begin{eqnarray}
\tilde{\mathcal{M}}'{}_{n} & = & \tilde{\mathcal{M}}{}_{n}-\sum_{j=1}^{n-1}\langle\mathfrak{C}\mathcal{M}_{j}|\mathfrak{C}\tilde{\mathcal{M}}{}_{n}\rangle\mathcal{M}_{j}\nonumber \\
\mathcal{M}_{n} & = & \bigl\Vert\mathfrak{C}\tilde{\mathcal{M}}'{}_{n}\bigr\Vert^{-1}\tilde{\mathcal{M}}'{}_{n}.\label{eq:Gram Schmidt angepasst t.a.d.m.}
\end{eqnarray}
The correctness of this procedure is easily seen when we multiply
the two equations from the left with $\mathfrak{C}$, which turns
the procedure into the standard Gram-Schmidt orthonormalization for
matrices $A_{n}=\mathfrak{C}\tilde{\mathcal{M}}_{n}$.

Using Eq.~(\ref{eq:Summen Ansatz rho off}) for $\varrho_{\textrm{off-diag}}$
and the orthonormalization Eq.~(\ref{eq:Weighted norm for HM}),
we find 
\begin{eqnarray}
\langle\mathfrak{C}\mathcal{M}_{k}|\varrho_{\textrm{off-diag}}\rangle & \overset{\eqref{eq:Summen Ansatz rho off}}{=} & \sum_{j}\alpha_{j}\langle\mathfrak{C}\mathcal{M}_{k}|\mathfrak{C}\mathcal{M}_{j}\rangle,\nonumber \\
 & \overset{\eqref{eq:Weighted norm for HM}}{=} & \alpha_{k}.\label{eq:Pre alpha t.a.d.m.}
\end{eqnarray}
According to Eq.~(\ref{eq:Gleicher Overlar rho null Rho off}), the
overlap $\langle\mathfrak{C}\mathcal{M}_{k}|\varrho_{\textrm{off-diag}}\rangle$
with the unknown $\varrho_{\textrm{off-diag}}$ is the same as the
overlap $\langle\mathfrak{C}\mathcal{M}_{k}|\varrho_{0}\rangle$ with
the known $\varrho_{0}$, such that
\begin{equation}
\alpha_{k}=\langle\mathfrak{C}\mathcal{M}_{k}|\varrho_{0}\rangle\label{eq:Alpha t.a.d.m}
\end{equation}
can be calculated for any given $\mathcal{M}_{k}$.

In other words, we are able to project the unknown matrix $\varrho_{\textrm{off-diag}}$
onto a set of orthonormalized commutators $\mathfrak{C}\mathcal{M}_{j}$.
After $n$ iteration steps, the absolute difference between $\varrho_{\textrm{off-diag}}$
and its approximation $\sum_{j=1}^{n}\alpha_{j}\mathfrak{C}\mathcal{M}_{j}$
is 
\begin{equation}
\Bigl\Vert\varrho_{\textrm{off-diag}}-\sum_{j=1}^{n}\alpha_{j}\mathfrak{C}\mathcal{M}_{j}\Bigr\Vert=\Bigl(\left\Vert \varrho_{\textrm{off-diag}}\right\Vert ^{2}-\sum_{j=1}^{n}\left|\alpha_{j}\right|^{2}\Bigr){}^{\frac{1}{2}}.
\end{equation}
Since we do not know the exact value of $\left\Vert \varrho_{\textrm{off-diag}}\right\Vert \leqslant\left\Vert \varrho_{0}\right\Vert $,
we have no good estimator for the quality of the approximation. If
we had used the numerically more demanding Eq.~(\ref{eq:Best Kommu M tadm})
instead of Eq.~(\ref{eq:Optimization Objective tadm}) as optimization
objective, we would have obtained $\sum_{j=1}^{n}\left|\alpha_{j}\right|^{2}\underset{n\rightarrow\infty}{\longrightarrow}1$.
But here, we can only state the obvious that the approximation gets
better with each new matrix $\mathcal{M}_{n+1}$ with $\alpha_{n+1}\neq0$.
For fast convergence, we like to find new matrices $\mathcal{M}_{n+1}$
with $|\alpha_{n+1}|$ which are preferably as big as possible. This
is what we are going to study next.

\subsection{Generating the matrices $\mathcal{M}_{j}$\label{sub:Generating-the-matrices M_j} }

In this subsection, we present a method to generate suitable matrices
$\mathcal{M}_{j}$ for Eq.~(\ref{eq:Summen Ansatz rho off}). To
start with the first matrix $\mathcal{M}_{1}$, we like to find a
$\mathcal{M}_{1}$ with a big absolute value $|\alpha_{1}|$~ (\ref{eq:Summen Ansatz rho off}),
which is according to Eq.~(\ref{eq:Alpha t.a.d.m})
\begin{equation}
\alpha_{1}\overset{\eqref{eq:Alpha t.a.d.m}}{=}\left\langle \mathcal{\mathfrak{C}M}_{1}|\varrho_{0}\right\rangle \overset{\eqref{eq:Selbstadjungierter kommutator t.a.d.m.}}{=}\left\langle \mathcal{M}_{1}|\mathfrak{C}\varrho_{0}\right\rangle .\label{eq:erster Overlap im algorithmus tadm}
\end{equation}
Therefore, as educated guess, we choose $\tilde{\mathcal{M}}_{1}=\mathfrak{C}\varrho_{0}$.
Here, a tilde is used to distinguish the matrices $\tilde{\mathcal{M}}_{j}$
from the matrices $\mathcal{M}_{j}$ which are already correctly orthonormalized
according to Eq.~(\ref{eq:Weighted norm for HM}). For $\mathcal{M}_{1}$,
this simply means $\mathcal{M}_{1}=\bigl\Vert\mathfrak{C}\tilde{\mathcal{M}}_{1}\bigr\Vert^{-1}\tilde{\mathcal{M}}_{1}$. 

Now, what do we choose as second matrix $\tilde{\mathcal{M}}_{2}$?
If we use the same ansatz as for $\tilde{\mathcal{M}}_{1}$, we have
$\alpha_{2}=\left\langle \mathcal{M}_{2}|\mathfrak{C}\varrho_{0}\right\rangle $
and hence $\tilde{\mathcal{M}}_{2}=\mathfrak{C}\varrho_{0}$ \---
but that is the result we already had for $\tilde{\mathcal{M}}_{1}$,
so we cannot use it again. This problem actually occurs for all $\tilde{\mathcal{M}}_{j>1}$.
To solve it, we have to go back to the first line of Eq.~(\ref{eq:Pre alpha t.a.d.m.})
\begin{eqnarray}
\left\langle \mathcal{\mathfrak{C}M}_{n}|\varrho_{\textrm{off-diag}}\right\rangle  & \overset{\eqref{eq:Pre alpha t.a.d.m.}}{=} & \sum_{j}\alpha_{j}\bigl\langle\mathcal{\mathfrak{C}M}_{n}|\mathfrak{C}\mathcal{M}_{j}\bigr\rangle\nonumber \\
\overset{\eqref{eq:Gleicher Overlar rho null Rho off}}{\Leftrightarrow}\left\langle \mathcal{\mathfrak{C}M}_{n}|\varrho_{0}\right\rangle  & = & \sum_{j}\alpha_{j}\bigl\langle\mathcal{\mathfrak{C}M}_{n}|\mathfrak{C}\mathcal{M}_{j}\bigr\rangle\nonumber \\
\bigl\langle\mathcal{\mathfrak{C}M}_{n}\big|\varrho_{0}-\sum_{j\neq n}\alpha_{j}\mathfrak{C}\mathcal{M}_{j}\bigr\rangle & = & \alpha_{n}\bigl\langle\mathcal{\mathfrak{C}M}_{n}|\mathfrak{C}\mathcal{M}_{n}\bigr\rangle\nonumber \\
\overset{\eqref{eq:Selbstadjungierter kommutator t.a.d.m.}}{\Leftrightarrow}\bigl\langle\mathcal{M}_{n}\big|\mathfrak{C}\varrho_{0}-\sum_{j\neq n}\alpha_{j}\mathfrak{C}^{2}\mathcal{M}_{j}\bigr\rangle & = & \alpha_{n}\bigl\langle\mathcal{\mathfrak{C}M}_{n}|\mathfrak{C}\mathcal{M}_{n}\bigr\rangle.\nonumber \\
 &  & \,\label{eq:eq:Vorschlag M t.a.d.m}
\end{eqnarray}
Now, we introduce two approximations: First, we ignore the quadratic
term $\bigl\langle\mathcal{\mathfrak{C}M}_{n}|\mathfrak{C}\mathcal{M}_{n}\bigr\rangle$
(i.e.,\ treat it as $=1$). Second, since we do not know $\mathcal{M}_{j>n}$,
we simply ignore them (which is the same as demanding that $\mathcal{M}_{n}$
will be perfect and no more $\mathcal{M}_{j>n}$ are needed). With
that, our educated guess becomes 
\begin{equation}
\tilde{\mathcal{M}}_{n}=\mathfrak{C}\varrho_{0}-\sum_{j=1}^{n-1}\alpha_{j}\mathfrak{C}^{2}\mathcal{M}_{j}.\label{eq:Optimal M tilde ohne Ortho tadm}
\end{equation}
This can also be written as $\tilde{\mathcal{M}}_{n}=\tilde{\mathcal{M}}_{n-1}-\alpha_{n-1}\mathfrak{C}^{2}\mathcal{M}_{n-1}$.
We still have to orthonormalize $\tilde{\mathcal{M}}_{n}$~(\ref{eq:Gram Schmidt angepasst t.a.d.m.}).
Therefore, we can drop the term $\tilde{\mathcal{M}}_{n-1}$ in $\tilde{\mathcal{M}}_{n}$,
since it can be expressed as linear combination of the previous matrices
$\mathcal{M}_{j<n}$ 
\begin{equation}
\tilde{\mathcal{M}}_{n}\rightarrow\mathcal{\hat{M}}_{n}=\mathfrak{C}^{2}\mathcal{M}_{n-1}.\label{eq:M dach t.a.d.m.}
\end{equation}
Now, $\mathcal{M}_{n}$ is obtained orthonormalizing $\mathcal{\hat{M}}_{n}$.
From Eq.~(\ref{eq:M dach t.a.d.m.}), we find by induction that the
matrices $\mathcal{M}_{n}$ can be expressed as
\begin{equation}
\mathcal{M}_{n}=\sum_{j=1}^{n}\gamma_{nj}\mathfrak{C}^{2j-1}\varrho_{0},\label{eq:Kummutator Polynom M t.a.d.m}
\end{equation}
with some appropriate coefficients $\gamma_{nj}$. Hence, the subspace
spanned by the matrices $\mathcal{M}_{j}$ is 
\begin{align}
\textrm{span}\left\{ \mathcal{M}_{1},\mathcal{M}_{2},\ldots,\mathcal{M}_{n}\right\} =\textrm{span}\left\{ \right. & \mathfrak{C}\varrho_{0},\mathfrak{C}^{3}\varrho_{0},\dots\nonumber \\
 & \ldots,\mathfrak{C}^{2n-1}\varrho_{0}\left.\right\} ,\label{eq:Krylov t.a.d.m}
\end{align}
which has the structure of a Krylov subspace.

\subsubsection{Orthogonality\label{sub:Orthogonality of the M matricen tadm}}

At the end, the matrix $\mathcal{M}_{n}$ can be obtained orthonormalizing
(\ref{eq:Gram Schmidt angepasst t.a.d.m.}) either $\tilde{\mathcal{M}}_{n}$
(\ref{eq:Optimal M tilde ohne Ortho tadm}) or $\mathcal{\hat{M}}_{n}$~(\ref{eq:M dach t.a.d.m.})
against the previous matrices $\mathcal{M}_{j<n}$. Actually, for
$\mathcal{\hat{M}}_{n}$ it already suffices to orthonormalize it
against the last two matrices $\mathcal{M}_{n-2}$ and $\mathcal{M}_{n-1}$
to ensure that it is orthonormal to all matrices $\mathcal{M}_{j<n}$.
For $\tilde{\mathcal{M}}_{n}$, it is even sufficient to orthonormalize
it against the last matrix $\mathcal{M}_{n-1}$ only. The proofs are
given in appendix~\ref{sec:Orthogonality-proofs TADM}. Still, in
practical calculations, such orthogonality results for iteratively
generated Krylov subspace basis are often undermined by numerical
imprecision and should be handled with care. Therefore, one might
still consider to orthonormalize each new matrix against all previous
$\mathcal{M}_{j}$.

\subsection{Discussion of the general method\label{sub:Discussion-of-the method tadm}}

To generate suitable $\mathcal{M}_{j}$ (\ref{eq:Summen Ansatz rho off}),
we have studied a method building up a Krylov subspace, which is a
common approach, as, e.g.,\ used by the Lanczos algorithm \cite{Lanczos1951}
to find eigenvectors. Therefore, we have a closer look at the inner
structure of the solution, which might help to put the method into
the right perspective compared to alternative methods. This is a special
topic, which might be skipped. 

Combing Eq.~(\ref{eq:Summen Ansatz rho off}) and Eq.~(\ref{eq:Kummutator Polynom M t.a.d.m}),
we find that after $n$ iterations, $\varrho_{\textrm{off-diag}}$
is approximated as 
\begin{eqnarray}
\varrho_{\textrm{off-diag}} & \overset{\eqref{eq:Summen Ansatz rho off}}{\approx} & \sum_{j=1}^{n}\alpha_{j}\mathcal{\mathfrak{C}M}_{j},\nonumber \\
 & \overset{\eqref{eq:Kummutator Polynom M t.a.d.m}}{=} & \sum_{j=1}^{n}\alpha_{j}\mathcal{\mathfrak{C}}\sum_{k=1}^{j}\gamma_{jk}\mathfrak{C}^{2k-1}\varrho_{0}\nonumber \\
 & = & \sum_{j=1}^{n}\beta_{j}\mathfrak{C}\varrho_{0},\;\textrm{with}\;\beta_{j}=\sum_{p=j}^{n}\alpha_{p}\gamma_{pj}.\nonumber \\
 &  & \,\label{eq:Summe gerader potenzen C tadm}
\end{eqnarray}
Evidently, this ansatz uses only even powers of $\mathfrak{C}$. This
corresponds with the general eigenvector problem in Eq.~(\ref{eq:General Eigenvector problem TADM}).
Here, we find the two operators $\mathfrak{C}^{2}$ and $\mathfrak{C}|\varrho_{0}\rangle\langle\varrho_{0}|\mathfrak{C}$,
where the latter can be absorbed into the normalization~(\ref{eq:Weighted norm for HM}).
Solving the general eigenvector problem by a Lanczos algorithm using
the same initial vector would result into the same Krylov subspace.
Still, one might wonder whether odd powers of $\mathfrak{C}$ could
possibly be useful, as well. For the commutator operator~ (\ref{eq:Kommutatorzeiche C def t.a.d.m.}),
the answer is definitely no. For odd powers of the commutator operator,
the matrix $\mathfrak{C}^{2j+1}\varrho_{0}$ has zero overlap with
the matrices $\varrho_{0}$ and $\varrho_{\textrm{off-diag}}$, as
we prove below~(\ref{eq:Null Binominal ungerade t.a.d.m}).
\begin{equation}
\left\langle \varrho_{\textrm{off-diag}}|\mathfrak{C}^{2j+1}\varrho_{0}\right\rangle \overset{\eqref{eq:Gleicher Overlar rho null Rho off}}{=}\left\langle \varrho_{0}|\mathfrak{C}^{2j+1}\varrho_{0}\right\rangle \overset{\eqref{eq:Null Binominal ungerade t.a.d.m}}{=}0.\label{eq:Ungerage Kommutator Null overap t.a.d.m}
\end{equation}
Even with this vanishing overlap, the odd power terms $\mathfrak{C}^{2j+1}\varrho_{0}$
could still be useful in an indirect way, if they had an overlap with
any of the even power terms $\mathfrak{C}^{2j}\varrho_{0}$ which
have non-vanishing overlaps with $\varrho_{0}$. But this is never
the case

\begin{equation}
\left\langle \mathfrak{C}^{2k}\varrho_{0}|\mathfrak{C}^{2j+1}\varrho_{0}\right\rangle \overset{\eqref{eq:Selbstadjungierter kommutator t.a.d.m.}}{=}\left\langle \varrho_{0}|\mathfrak{C}^{2(j+k)+1}\varrho_{0}\right\rangle \overset{\eqref{eq:Null Binominal ungerade t.a.d.m}}{=}0.
\end{equation}
Hence, the terms with odd powers of $\mathfrak{C}$ are of no help
to approximate $\varrho_{\textrm{off-diag}}.$

We still have to prove Eq.~(\ref{eq:Ungerage Kommutator Null overap t.a.d.m})
$\left\langle \varrho_{0}|\mathfrak{C}^{2j+1}\varrho_{0}\right\rangle =0$.
To see this, it is useful to expand the commutators~$\mathfrak{C}^{n}$
\begin{equation}
\left\langle \varrho_{0}|\mathfrak{C}^{n}\varrho_{0}\right\rangle \overset{\eqref{eq:Kommutatorzeiche C def t.a.d.m.}}{=}\sum_{p=0}^{n}\left(-1\right)^{p}\binom{n}{p}\left\langle \varrho_{0}|H^{n-p}\varrho_{0}H^{p}\right\rangle ,\label{eq:Binominal Commutator power t.a.d.m.}
\end{equation}
with the binomial coefficients $\tbinom{n}{p}$ as in $(a-b)^{n}=\sum_{p=0}^{n}\left(-1\right)^{p}\binom{n}{p}a^{n-p}b^{p}$.
Since $\varrho_{0}$ and $H$ are a Hermitian matrices, we can use
the cyclic property of the trace to derive
\begin{equation}
\left\langle \varrho_{0}|H^{n-p}\varrho_{0}H^{p}\right\rangle =\left\langle \varrho_{0}|H^{p}\varrho_{0}H^{n-p}\right\rangle .
\end{equation}
Using this equation together with the identity $\tbinom{n}{p}=\tbinom{n}{n-p}$,
we can add the terms for $p$ and $n-p$ in Eq.~(\ref{eq:Binominal Commutator power t.a.d.m.})
and use their average to obtain 
\begin{equation}
\left\langle \varrho_{0}|\mathfrak{C}^{n}\varrho_{0}\right\rangle =\sum_{p=0}^{n}\frac{\left(-1\right)^{p}+\left(-1\right)^{n-p}}{2}\binom{n}{p}\left\langle \varrho_{0}|H^{n-p}\varrho_{0}H^{p}\right\rangle .\label{eq:Null Binominal ungerade t.a.d.m}
\end{equation}
Since $\left(-1\right)^{p}+\left(-1\right)^{n-p}$ vanishes for all
odd numbers $n$ independent of $p\in\mathbb{N}$, the inner product
$\left\langle \varrho_{0}|\mathfrak{C}^{n}\varrho_{0}\right\rangle $
is zero for odd powers $n=2j+1$ of $\mathfrak{C}$, as claimed in
Eq.~(\ref{eq:Ungerage Kommutator Null overap t.a.d.m}).

The ansatz described by Eq.~(\ref{eq:Summe gerader potenzen C tadm})
does not make use of any terms $H^{n-p}\varrho_{0}H^{p}$ with odd
$n$. It is not that these terms are not useful per se. They are just
not generated in a useful combination applying powers of the commutator
$\mathfrak{C}$ on $\varrho_{0}$. If we produce these terms by other
means, they can be part of the approximation (\ref{eq:Summen Ansatz rho off})
of $\varrho_{\textrm{off-diag}}$, as well. 

The reason for using the commutator $\mathfrak{C}$ was that it generates
matrices which have no overlap with the time averaged density matrix
$\bar{\varrho}$. In other words, the components $\left|E_{j}\right\rangle \left\langle E_{k}\right|$
vanish in the matrices generated by the commutator $\mathfrak{C}$
for identical energy eigenvalues $E_{j}=E_{k}$. Another way to guaranty
the vanishing of these components is resorting to matrices
\begin{equation}
\mathscr{M}=\sum_{p=0}^{n}a_{p}H^{n-p}\varrho_{0}H^{p},\quad\textrm{with}\quad\sum_{p=0}^{n}a_{p}=0,\quad a_{p}\in\mathbb{C}.\label{eq:Alternatives M tadm}
\end{equation}
This can be seen inserting $\varrho_{0}=\sum p_{jk}\left|E_{j}\right\rangle \left\langle E_{k}\right|$
in Eq.~(\ref{eq:Alternatives M tadm}). Actually, the matrices $\mathfrak{C}^{n}\varrho_{0}$
build a subset of the matrices $\mathscr{M}$ (\ref{eq:Alternatives M tadm}).
Another example for a subset of the matrices $\mathscr{M}$ are the
matrices $\mathfrak{M=\mathfrak{C}}\left(H^{p}\varrho_{0}H^{q}\right)$,
with arbitrary $p,q\in\mathbb{N}$.

A special situation arises, when the initial state is pure $\varrho_{0}=\left|\Psi_{0}\right\rangle \left\langle \Psi_{0}\right|$.
Then, inner products as
\begin{equation}
\left\langle \varrho_{0}|H^{p}\varrho_{0}H^{q}\right\rangle =\left\langle \Psi_{0}|H^{p}|\Psi_{0}\right\rangle \left\langle \Psi_{0}|H^{q}|\Psi_{0}\right\rangle 
\end{equation}
are easily calculated if we know all $H^{r}|\Psi_{0}\rangle$. In
this case, we could also use a diagonalization method based on the
Krylov subspace $\mathcal{K}_{n}$
\begin{equation}
\mathcal{K}_{n}=\textrm{span}\left\{ \left|\Psi_{0}\right\rangle ,H\left|\Psi_{0}\right\rangle ,\ldots,H^{n-1}\left|\Psi_{0}\right\rangle \right\} ,
\end{equation}
as the Lanczos or Arnoldi algorithm \cite{Lanczos1951,Arnoldi1951}
to obtain approximated energy eigenstates $|\tilde{E}_{j}\rangle$.
With these, the TADM $\bar{\varrho}$ can be approximated as
\begin{equation}
\bar{\varrho}_{\textrm{approx}}=\sum_{j}|\tilde{E}_{j}\rangle\langle\tilde{E_{j}}|.\label{eq:lanczos Approx rho tadm}
\end{equation}
We emphasize that the $\bar{\varrho}_{\textrm{approx}}$  obtained
by this Eq.~(\ref{eq:lanczos Approx rho tadm}) is not suitable for
the error reduction method introduced in appendix~\ref{sub:Error-reduction tadm}.
For this method to work, it is essential that the diagonal elements
of the approximated TADM $\bar{\varrho}_{\textrm{approx}}$ are error-free,
i.e.
\begin{equation}
\left\langle E_{j}|\bar{\varrho}_{\textrm{approx}}|E_{k}\right\rangle \overset{\textrm{for }E_{j}=E_{k}}{=}\left\langle E_{j}|\bar{\varrho}|E_{k}\right\rangle .
\end{equation}
This is generally not true for Eq.~(\ref{eq:lanczos Approx rho tadm}),
while it is guaranteed for the method introduced here (\ref{eq:=0000DCbersichtsformel Time averaged d.m.}).

\section{Time averaged density matrix as matrix product operator\label{sec:Time-averaged-density as MPO}}

In appendix~\ref{sec:Solving-the-optimization problem general approach tadm},
we explained an algorithm for finding the time averaged density matrix
(TADM) $\bar{\varrho}$. But so far, this algorithm does not solve
the main numerical problem which usually hinders us to calculate $\bar{\varrho}$:
The exponential scaling of the Hilbert space with the number of the
constituents and the associated demand for computational resources.
In this section, we address this problem and present a monotone converging
optimization algorithm based on a matrix product operator (MPO) approximation,
which allows the handling of the exponential scaling (see overview
articles \cite{Verstraete2008Review,Schollwoeck2011,Eisert2013OverviewTNS,Orus2014_TensorNetworks}).

\subsection{Short introduction to MPO\label{sub:Short-introduction-to MPO in TADM}}

Any matrix operator $M$ acting on a system consisting of $n$ sites
$s_{j}$ can be written as

\begin{equation}
M=\sum_{s_{1}\cdots s_{n};s'_{1}\cdots s'_{n}}\mathfrak{M}_{s_{1}\cdots s_{n}}^{s'_{1}\cdots s'_{n}}|s_{1}\rangle\langle s_{1}'|\otimes\cdots\otimes|s_{n}\rangle\langle s_{n}'|,\label{eq:MPO Intro1 TADM}
\end{equation}
with a high-dimensional tensor $\mathfrak{M}_{s_{1}\cdots s_{n}}^{s'_{1}\cdots s'_{n}}$.
The idea of a MPO is to express this high-dimensional tensor $\mathfrak{M}_{s_{1}\cdots s_{n}}^{s'_{1}\cdots s'_{n}}$
as a product of $n$ low-dimensional tensors $\mathsf{M}_{[j]}$.
\begin{eqnarray}
\mathfrak{M}_{s_{1}\cdots s_{n}}^{s'_{1}\cdots s'_{n}} & = & \sum_{\alpha_{1}\cdots\alpha_{n}}\mathsf{M}_{[1]s_{1}s'_{1}}^{\alpha_{n}\alpha_{1}}\cdot\mathsf{M}_{[2]s_{2}s'_{2}}^{\alpha_{1}\alpha_{2}}\cdot\ldots\nonumber \\
 &  & \quad\ldots\cdot\mathsf{M}_{[n-1]s_{n-1}s'_{n-1}}^{\alpha_{n-2}\alpha_{n-1}}\cdot\mathsf{M}_{[n]s_{n}s'_{n}}^{\alpha_{n-1}\alpha_{n}},\label{eq:MPO Intro2 TADM}
\end{eqnarray}
where $s_{j}$ and $s'_{j}$ represent the physical indices, while
the $\alpha_{j}$ are auxiliary indices which are summed over. In
case of closed boundary conditions (which we use here), the dimension
of the outer index $\alpha_{n}$ is set to one (i.e., we can ignore
this index).

\begin{figure}
\includegraphics[width=1\columnwidth]{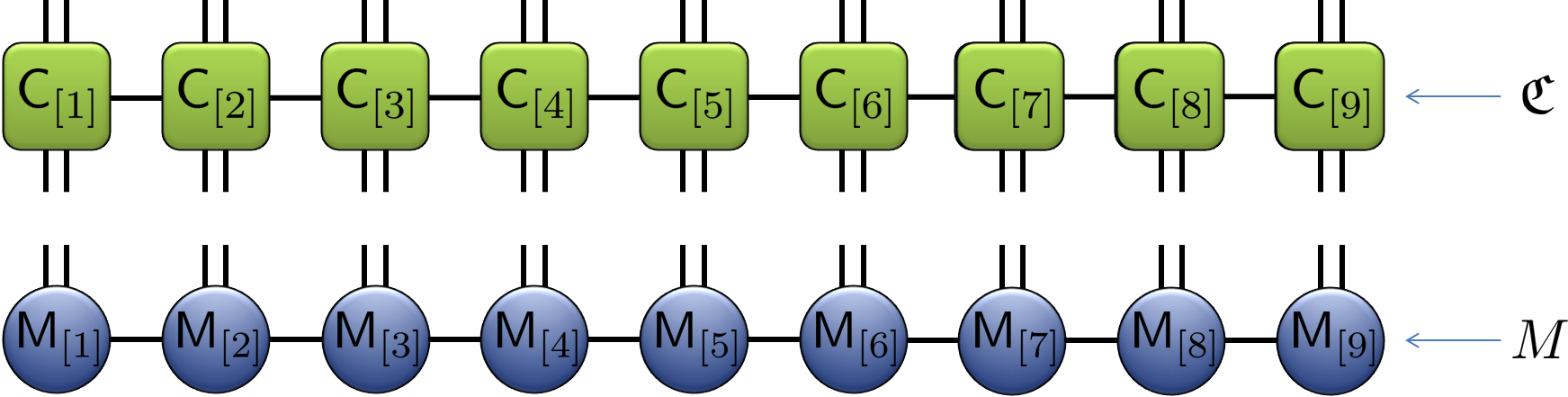}

\protect\caption{\label{fig:MPO Beispiel TADM}Graphical representation of the MPO
for the commutator operator $\mathfrak{C}$ (\ref{eq:Kommutatorzeiche C def t.a.d.m.})
and the matrix $M$. Each MPO tensor is symbolized by a shape. Auxiliary
indices are depicted by horizontal lines, while open vertical indices
represent physical indices. }
\end{figure}

Since Eq.~(\ref{eq:MPO Intro2 TADM}) is not really a well-readable
expression (and it becomes even worse if we consider e.g.\ products
of operators), one often resorts to graphical representations, as
in Fig.~\ref{fig:MPO Beispiel TADM}. Here, we also use the (non-standard)
symbolic shorthand notation

\begin{equation}
M=\prod_{j}\mathsf{M}_{[j]},\label{eq:Symbolisch MPO Eq tadm}
\end{equation}
which represents the joint information of Eq.~(\ref{eq:MPO Intro1 TADM})
and Eq.~(\ref{eq:MPO Intro2 TADM}). In the same way, we write $\mathfrak{C}=\prod_{j}\mathsf{C}_{[j]}$
for the commutator operator, where each tensor $\mathsf{C}_{[j]s_{j}s'_{j}\bar{s}_{j}\bar{s}'_{j}}^{\alpha_{j-1}\alpha_{j}}$
carries the four physical indices $s_{j},s'_{j},\bar{s}_{j},\bar{s}'_{j}$.
For the construction of the MPO tensors of $\mathfrak{C}$, see also
appendix~\ref{sec:Constructing-a-MPO for C TADM}. To calculate an
expression like $\mathfrak{C}M=\prod_{j}\mathsf{C}_{[j]}\cdot\mathsf{M}_{[j]}$,
we need to sum over the common physical indices
\begin{equation}
\mathsf{C}_{[j]}\mathsf{M}_{[j]}=\sum_{\bar{s}_{j},\bar{s}'_{j}}\mathsf{C}_{[j]s_{j}s'_{j}\bar{s}_{j}\bar{s}'_{j}}^{\alpha_{j-1}\alpha_{j}}\mathsf{M}_{[j]\bar{s}_{j}\bar{s}'_{j}}^{\beta_{j-1}\beta_{j}}
\end{equation}
 and the auxiliary indices. To shorten the notation, in the following
we will often use the multi-index $\sigma_{j}=(s_{j},s'_{j})$ for
the two physical indices $s_{j},s'_{j}$ of a MPO tensor. 

Many operators which are relevant for practical calculations allow
an exact MPO representation based on tensors $\mathsf{M}_{[j]}$ (\ref{eq:Symbolisch MPO Eq tadm})
which are of relatively small and constant size, independent of the
total number of the constituents. Here, we assume that this is also
true for the initial density matrix $\varrho_{0}=\prod_{j}\mathsf{P}_{[j]}$
and the Hamiltonian $H=\prod_{j}\mathsf{H}_{[j]}$. If it is true
for the Hamiltonian, it is also true for the commutator operator $\mathfrak{C}=\prod_{j}\mathsf{C}_{[j]}$
(\ref{eq:Kommutatorzeiche C def t.a.d.m.}). 

Unfortunately, this is not true for all operators. For a general operator
$\hat{O}$, an exact MPO representation $\hat{O}=\prod_{j}\mathsf{O}_{[j]}$
might demand tensors $\mathsf{O}_{[j]}$ whose size scales exponentially
with the number of constituents. In this case, one can still use tensors
$\tilde{\mathsf{O}}_{[j]}$ of limited size to obtain an approximation
$\hat{O}\approx\prod_{j}\tilde{\mathsf{O}}_{[j]}$. 

Once we put restrictions to the size of the MPO tensors, MPO no longer
build a vector space. That is, adding two MPO might create a sum MPO
whose tensors exceed the preset size limit. The same is even more
likely for the product of two MPO. Therefore, we cannot use a simple
one-to-one mapping to cast the algorithm presented in appendix~\ref{sec:Solving-the-optimization problem general approach tadm}
or any other algorithm into MPO form.

\subsection{MPO tensor optimization\label{sub:MPO-tensor-optimization tadm}}

Assume we have two MPO $M'=\prod_{j}\mathsf{M}'_{[j]}$ and $M''=\prod_{j}\mathsf{M}''_{[j]}$
differing only by one tensor $\mathsf{M}'_{[k]}\neq\mathsf{M}''_{[k]}$
(which still have the same dimensions), while all other tensors are
identical $\mathsf{M}'_{[j\neq k]}=\mathsf{M}''_{[j\neq k]}=\mathsf{M}{}_{[j\neq k]}$.
Symbolically, we write these MPO as
\begin{eqnarray}
M' & = & \mathsf{M}'_{[k]}\prod_{j\neq k}\mathsf{M}_{[j]}\nonumber \\
M'' & = & \mathsf{M}''_{[k]}\prod_{j\neq k}\mathsf{M}_{[j]}.
\end{eqnarray}
For $\alpha,\beta\in\mathbb{C}$, we find
\begin{equation}
\alpha M'+\beta M''=\left(\alpha\mathsf{M}'_{[k]}+\beta\mathsf{M}''_{[k]}\right)\prod_{j\neq k}\mathsf{M}_{[j]}.\label{eq:linearer MPO mit einem tensor tadm}
\end{equation}
The important observation is that the tensor $\alpha\mathsf{M}'_{[k]}+\beta\mathsf{M}''_{[k]}$
has the same dimensions as $\mathsf{M}'_{[k]}$ and $\mathsf{M}''_{[k]}$.
That is, adding the MPO $M'$ and $M''$ does not conflict with any
preset limits for the size of the MPO tensors. Hence, MPO of the type
$M'$ and $M''$ which differ only by a single tensor still build
a vector space. For this reason, theses MPO are much more suitable
to realize an adaptation of the algorithm explained in appendix~\ref{sec:Solving-the-optimization problem general approach tadm}. 

We express the TADM as $\bar{\varrho}=\varrho_{0}-c\mathfrak{C}M$
(\ref{eq:=0000DCbersichtsformel Time averaged d.m.}), where $c\in\mathbb{C}$
and 
\begin{equation}
M=\prod_{j}\mathsf{M}_{[j]}\label{eq:M als MPO Symbolisch tadm}
\end{equation}
is the MPO we have to find. Our task is to design an algorithm which
optimizes a single tensor $\mathsf{M}{}_{[k]}$ while all other tensors
$\mathsf{M}{}_{[j\neq k]}$ are kept constant. Once we have this algorithm,
we can apply it in repeated sweeps of the index $k$ to all tensors
$\mathsf{M}{}_{[k]}$ in the MPO. 

In appendix~\ref{sec:Solving-the-optimization problem general approach tadm},
we introduced a general algorithm, which expresses $M$ as a sum $M=\sum_{l}\alpha_{l}\mathcal{M}_{l}$~(\ref{eq:Summen Ansatz rho off}).
Here, we adopt this idea but extend it by the demand that the $\mathcal{M}_{l}$
are all MPO with the same tensors $\mathsf{M}{}_{[j\neq k]}$ 
\begin{equation}
\mathcal{M}_{l}=\mathsf{M}{}_{[k]}^{(l)}\prod_{j\neq k}\mathsf{M}_{[j]}.\label{eq:MPO-Summen M_l tadm}
\end{equation}
The MPO $\mathcal{M}_{l}$ differ only by the tensor $\mathsf{M}{}_{[k]}^{(l)}$.
Due to the linearity described by Eq.~(\ref{eq:linearer MPO mit einem tensor tadm}),
the summation of the MPO $M=\sum_{l}\alpha_{l}\mathcal{M}_{l}$ (\ref{eq:Summen Ansatz rho off})
is equivalent to the summation of the tensors $\mathsf{M}{}_{[k]}^{(l)}$
\begin{equation}
\mathsf{M}{}_{[k]}^{\textrm{optimized}}=\sum_{l}\alpha_{l}\mathsf{M}{}_{[k]}^{(l)},\label{eq:MPO tensors locally bad optimized tadm}
\end{equation}
resulting in the locally optimized tensor $\mathsf{M}{}_{[k]}^{\textrm{optimized}}$.

\begin{figure}
\includegraphics[width=1\columnwidth]{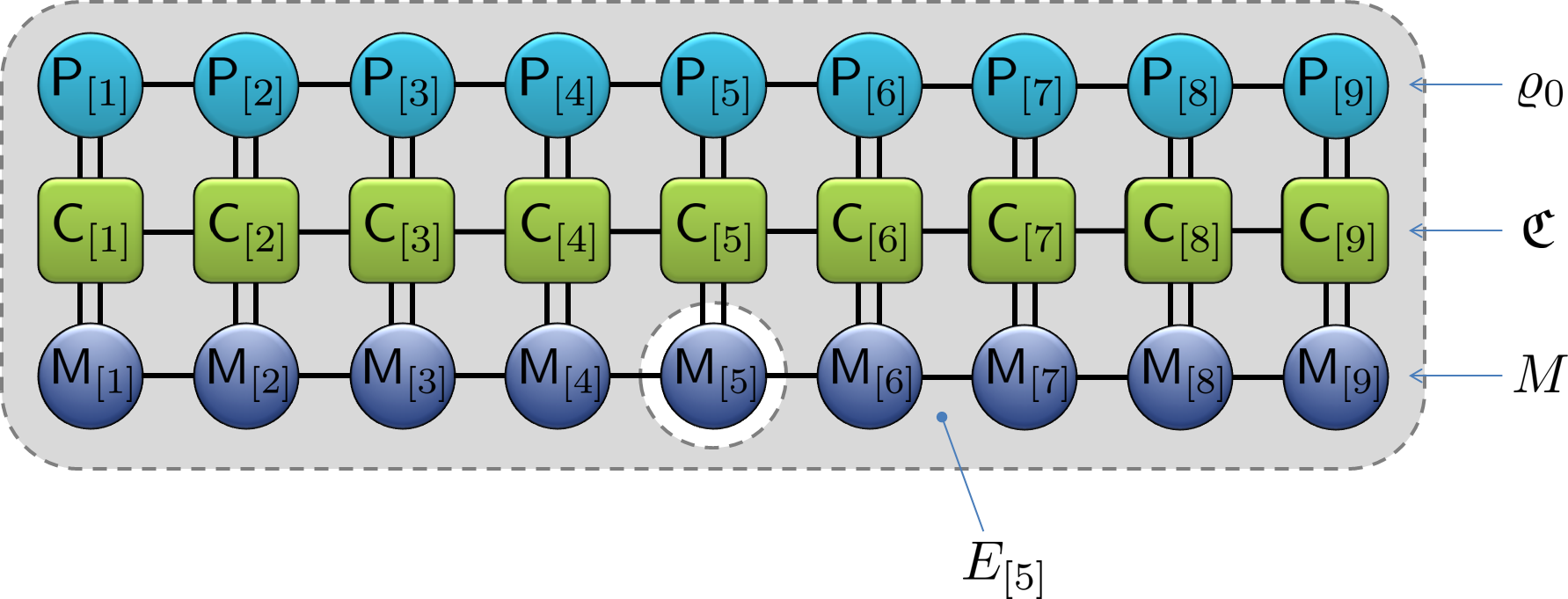}

\protect\caption{\label{fig:MPO-Umgebung TADM}Graphical representation of the MPO
overlap $\left\langle \varrho_{0}|\mathfrak{C}|M\right\rangle $.
The area $E_{[5]}=\varrho_{0}\mathfrak{C}\prod_{j\protect\neq5}\mathsf{M}_{[j]}$
defines the tensor environment of the tensor $\mathsf{M}_{[5]}$.
This corresponds to the situation in Eq.~(\ref{eq:Schlechter erster MPO tensor tadm}),
with $\tilde{\mathsf{M}}{}_{[5]}^{(1)}=E_{[5]}^{\dagger}$.}
\end{figure}

If we follow the ideas outlined in appendix~\ref{sub:Generating-the-matrices M_j}
to obtain suitable MPO $\mathcal{M}_{l}$, we need to generate linear
combinations of $\mathfrak{C}^{p}\varrho_{0}$ (\ref{eq:Krylov t.a.d.m}),
where $p$ denotes an exponent. How can we calculate $\mathfrak{C}^{p}\varrho_{0}$
while we are at the same time forced to keep all MPO tensors $\mathsf{M}{}_{[j\neq k]}$
unchanged? Here, we have to remind ourselves that the original intention
in appendix~\ref{sub:Generating-the-matrices M_j} was to maximize
the overlap $\left\langle \varrho_{0}|\mathfrak{C}M\right\rangle =\left\langle \varrho_{0}|\mathfrak{C}|M\right\rangle $
(\ref{eq:erster Overlap im algorithmus tadm}). Inserting Eq.~(\ref{eq:MPO-Summen M_l tadm})
in $\left\langle \varrho_{0}|\mathfrak{C}|\mathcal{M}_{1}\right\rangle $,
we obtain
\begin{eqnarray}
\bigl\langle\varrho_{0}\big|\mathfrak{C}\big|\mathcal{M}_{1}\bigr\rangle & \overset{\eqref{eq:MPO-Summen M_l tadm}}{=} & \bigl\langle\varrho_{0}\big|\mathfrak{C}\big|\mathsf{M}{}_{[k]}^{(1)}\prod_{j\neq k}\mathsf{M}_{[j]}\bigr\rangle\nonumber \\
 & = & \textrm{tr}\Bigl(\bigl(\varrho_{0}\mathfrak{C}\prod_{j\text{\ensuremath{\neq}}k}\mathsf{M}_{[j]}\bigr)\mathsf{M}{}_{[k]}^{(1)}\Bigr).\label{eq:tensor overlap optimization tadm}
\end{eqnarray}
This is maximized for
\begin{equation}
\tilde{\mathsf{M}}{}_{[k]}^{(1)}=\bigl(\varrho_{0}\mathfrak{C}\prod_{j\neq k}\mathsf{M}_{[j]}\bigr){}^{\dagger}=\bigl(\prod_{j\neq k}\mathsf{M}_{[j]}\bigr)^{\dagger}\mathfrak{C}\varrho_{0},\label{eq:Schlechter erster MPO tensor tadm}
\end{equation}
see also Fig.~\ref{fig:MPO-Umgebung TADM}. In the same fashion,
we can generate the tensors $\tilde{\mathsf{M}}{}_{[k]}^{(l)}$ analog
to Eq.~(\ref{eq:Optimal M tilde ohne Ortho tadm})
\begin{equation}
\tilde{\mathsf{M}}{}_{[k]}^{(l)}=\bigl(\prod_{j\neq k}\mathsf{M}_{[j]}\bigr)^{\dagger}\bigl(\mathfrak{C}\varrho_{0}-\sum_{p=1}^{l-1}\alpha_{p}\mathfrak{C}^{2}\mathcal{M}_{p}\bigr),\label{eq:MPO summen tensor iteration Formel tadm}
\end{equation}
with $\mathcal{M}_{p}=\mathsf{M}{}_{[k]}^{(p)}\prod_{j\neq k}\mathsf{M}_{[j]}$
(\ref{eq:MPO-Summen M_l tadm}). As in Eq.~(\ref{eq:Optimal M tilde ohne Ortho tadm}),
we used a tilde to mark that $\tilde{\mathsf{M}}{}_{[k]}^{(l)}$ has
still to be orthonormalized (\ref{eq:Weighted norm for HM}) against
the other tensors. Due to the linearity expressed by Eq.~(\ref{eq:linearer MPO mit einem tensor tadm}),
orthonormalizing the tensor $\tilde{\mathsf{M}}{}_{[k]}^{(l)}$ or
the associated MPO $\mathcal{\tilde{M}}_{l}=\tilde{\mathsf{M}}{}_{[k]}^{(l)}\prod_{j\neq k}\mathsf{M}_{[j]}$
is basically the same procedure. In the orthonormalization, as well
as in Eq.~(\ref{eq:MPO summen tensor iteration Formel tadm}), we
encounter the term $\bigl(\prod_{j\neq k}\mathsf{M}_{[j]}\bigr)^{\dagger}\mathfrak{C}^{2}\prod_{j\neq k}\mathsf{M}_{[j]}$,
which is also depictured in Fig.~\ref{fig:Effective MPO Norm Operator TADM}
(appendix~\ref{sub:Gauging-the-MPO tadm}).

\subsection{Advisable modification of the algorithm\label{sub:Advisable-modification-of the algorithm tadm} }

So far, we have presented a direct adaptation of the ideas presented
in appendix~\ref{sub:Generating-the-matrices M_j} to obtain the
locally optimized MPO tensors $\mathsf{M}{}_{[k]}^{\textrm{optimized}}$
(\ref{eq:MPO tensors locally bad optimized tadm}). But this adaptation
is far from being optimal and does not result in a monotone converging
algorithm. This will be mended in this subsection.

To see the problem, imagine that for some reason, we have already
found a perfect MPO $M$, where all MPO tensors $\mathsf{M}{}_{[j]}$
are optimal. Now, let us denote 
\begin{equation}
\mathsf{M}_{[k]}^{\textrm{old}}=\mathsf{M}{}_{[k]}
\end{equation}
and apply the optimization procedure presented in the last subsection~\ref{sub:MPO-tensor-optimization tadm}
to find a new tensor $\mathsf{M}{}_{[k]}^{\textrm{optimized}}$. Since
$\mathsf{M}^{\textrm{old}}{}_{[k]}$ is already optimal according
to our assumption, the optimization $\mathsf{M}{}_{[k]}^{\textrm{optimized}}=\sum_{l}\alpha_{l}\mathsf{M}{}_{[k]}^{(l)}$
(\ref{eq:MPO tensors locally bad optimized tadm}) should reproduce
the tensor $\mathsf{M}^{\textrm{old}}{}_{[k]}$. This is for sure
true, if we sum up sufficiently many tensors $\mathsf{M}{}_{[k]}^{(l)}$.
But in practical applications, one would usually just calculate a
handful of these tensors $\mathsf{M}{}_{[k]}^{(l)}$, resulting in
an $\mathsf{M}{}_{[k]}^{\textrm{optimized}}$ which is supposedly
worse than the tensor $\mathsf{M}^{\textrm{old}}{}_{[k]}$ we started
with.

For readers who are familiar with the Krylov subspace based MPO optimization
for ground states $|E_{0}\rangle$ 
\begin{equation}
|E_{0}\rangle=\underset{\left\langle \Psi|\Psi\right\rangle =1}{\textrm{arg min}}\left\langle \Psi|H|\Psi\right\rangle ,\label{eq:Beispiel Grund state tadm}
\end{equation}
we mention that here, the main difference is that the Krylov subspace
for the ground state search is based on \cite{Saad2003} 
\begin{equation}
\mathcal{K}_{n}=\textrm{span}\left\{ \Psi,H\Psi,H^{2}\Psi,\ldots,H^{n-1}\Psi\right\} ,\label{eq:Krylov f=0000FCr Grundzustand tadm}
\end{equation}
 while the algorithm for the TADM is based on 
\begin{equation}
\mathcal{K}_{n}=\textrm{span}\left\{ \mathfrak{C}\varrho_{0},\mathfrak{C}^{3}\varrho_{0},\mathfrak{C}^{5}\varrho_{0},\ldots,\mathfrak{C}^{2n-1}\varrho_{0}\right\} ,
\end{equation}
according to Eq.~(\ref{eq:Krylov t.a.d.m}). If we start the ground
state optimization with an optimal MPO $|\Psi\rangle=|E_{0}\rangle$,
this optimal $|\Psi\rangle$ is part of the Krylov subspace used for
the optimization. Therefore, the optimal solution is already obtained
in the first step. Contrary to the ground state optimization, the
algorithm for the TADM starts with $\mathfrak{C}\varrho_{0}$, which
does not convey any previously gained information about the optimal
solution.

Of course, starting the optimization with an already perfect solution
is just an extreme example to illustrate the problem: The algorithm,
as it is so far, does not learn from previous optimization steps.
This is actually easy to fix. We just include the old MPO tensor $\mathsf{M}_{[k]}^{\textrm{old}}$
into the subspace basis we use for finding the optimal tensor $\mathsf{M}{}_{[k]}^{\textrm{optimized}}$.
The subspace we obtain in this fashion is no longer a pure Krylov
subspace, but this is actually of no real importance.

So, we still use Eq.~(\ref{eq:MPO tensors locally bad optimized tadm})
\begin{equation}
\mathsf{M}{}_{[k]}^{\textrm{optimized}}=\sum_{l}\alpha_{l}\mathsf{M}{}_{[k]}^{(l)},\label{eq:Wiederholung guter Summen Tensor tadm}
\end{equation}
but for the first basis tensor $\mathsf{M}{}_{[k]}^{(1)}$, we now
take the normalized (\ref{eq:Weighted norm for HM}) old tensor 
\begin{equation}
\mathsf{M}{}_{[k]}^{(1)}=\frac{1}{\left\Vert \mathfrak{C}M\right\Vert }\mathsf{M}^{\textrm{old}}{}_{[k]}.\label{eq:Alter tensor recyle}
\end{equation}
For the second basis tensor $\mathsf{M}{}_{[k]}^{(2)}$, we could
use the formerly first tensor $\tilde{\mathsf{M}}{}_{[k]}^{(2)}\overset{?}{=}\bigl(\prod_{j\neq k}\mathsf{M}_{[j]}\bigr)^{\dagger}\mathfrak{C}\varrho_{0}$
(\ref{eq:Schlechter erster MPO tensor tadm}). But following the same
line of argumentation which led to Eq.~(\ref{eq:Optimal M tilde ohne Ortho tadm}),
we find it actually more advisable to use directly Eq.~(\ref{eq:MPO summen tensor iteration Formel tadm})
(the analog tensor equation of Eq.~(\ref{eq:Optimal M tilde ohne Ortho tadm}))
\begin{equation}
\tilde{\mathsf{M}}{}_{[k]}^{(l>1)}=\bigl(\prod_{j\neq k}\mathsf{M}_{[j]}\bigr)^{\dagger}\bigl(\mathfrak{C}\varrho_{0}-\sum_{p=1}^{l-1}\alpha_{p}\mathfrak{C}^{2}\mathcal{M}_{p}\bigr).\label{eq:Tensor M_l>1  tadm}
\end{equation}

\subsubsection{Further modifications\label{sub:Further-modifications TADM}}

The algorithm has still room for further improvements, which are explained
in detail in the following appendices. Here, we just outline what
can still be done:
\begin{itemize}
\item Tensor networks exhibit a versatile gauge freedom. Choosing an optimal
gauge is an essential ingredient for a successful tensor optimization.
It is advisable to use a non-standard gauge, which is tailored for
the weighted norm (\ref{eq:Weighted norm for HM}) used in the algorithm.
This is explained in appendix~\ref{sub:Gauging-the-MPO tadm}.
\item The tensor optimization is done in many successive sweeps. With a
small alteration, we can take advantage of previous optimization sweeps
to speed up the convergence. This is a special feature of the TADM
algorithm, which we dubbed  overarching orthonormalization. For more
details, see appendix~\ref{sub:Speeding-up-convergence tadm}
\item Density matrices are always Hermitian matrices. This implies a symmetry
which can be exploited and allows to map complex valued MPO onto real
valued MPO with the same bond dimension. While most symmetries are
connected to some special properties of the physical system in question,
the Hermitian symmetry is common to all physical systems. For more,
see appendix~\ref{sec:Mapping-hermitian-matrices onto real tADM}.
\end{itemize}

\section{Orthogonality p\label{sec:Orthogonality-proofs TADM}roofs}

Here, we provide some orthogonality proofs. First, we start with the
matrices $\tilde{\mathcal{M}}_{n}$ (\ref{eq:Optimal M tilde ohne Ortho tadm})
respectively $\mathcal{\hat{M}}_{n}$ (\ref{eq:M dach t.a.d.m.})
introduced in appendix~\ref{sec:Solving-the-optimization problem general approach tadm}
and after that, we look at their tensor network version $\tilde{\mathsf{M}}{}_{[k]}^{(l)}$
(\ref{eq:Alter tensor recyle}) and (\ref{eq:Tensor M_l>1  tadm}).
That is, we first show for the general method of appendix~(\ref{sec:Solving-the-optimization problem general approach tadm}) 

\begin{align}
\bigl\langle\mathfrak{C}\mathcal{M}_{j<n-2}\big|\mathfrak{C}\mathcal{\hat{M}}_{n}\bigr\rangle & =0\label{eq:Ortho von M dach tadm}\\
\bigl\langle\mathfrak{C}\mathcal{M}_{j<n-1}\big|\mathfrak{C}\mathcal{\tilde{M}}_{n}\bigr\rangle & =0,\label{eq:Ortho von M tilde tadm}
\end{align}
which expresses the demanded orthogonality according to Eq.~(\ref{eq:Weighted norm for HM}). 

We start with the proof of Eq.~(\ref{eq:Ortho von M dach tadm}),
which is a standard proof for Krylov subspaces

\begin{eqnarray}
\bigl\langle\mathcal{\mathfrak{C}M}_{j}|\mathfrak{C}\mathcal{\hat{M}}_{n}\bigr\rangle & \overset{\eqref{eq:M dach t.a.d.m.}}{=} & \bigl\langle\mathcal{\mathfrak{C}M}_{j}|\mathfrak{C}\left(\mathfrak{C}^{2}\mathcal{M}_{n-1}\right)\bigr\rangle\nonumber \\
 & \overset{\eqref{eq:Selbstadjungierter kommutator t.a.d.m.}}{=} & \bigl\langle\mathfrak{C}\left(\mathfrak{C}^{2}\mathcal{M}_{j}\right)|\mathfrak{C}\mathcal{M}_{n-1}\bigr\rangle\nonumber \\
 & \overset{\eqref{eq:M dach t.a.d.m.}}{=} & \bigl\langle\mathcal{\mathfrak{C}\hat{M}}_{j+1}|\mathfrak{C}\mathcal{M}_{n-1}\bigr\rangle\label{eq:Norm zero fuer j+1<n+1}
\end{eqnarray}
The matrix $\mathcal{\hat{M}}_{j+1}$ lies in the subspace $\textrm{span}\left\{ \mathcal{M}_{1},\mathcal{M}_{2},\ldots,\mathcal{M}_{j+1}\right\} $,
which is orthogonal to $\mathcal{M}_{n-1}$ (according to the definition~(\ref{eq:Weighted norm for HM}))
for $j+1<n-1$. Hence, for $j<n-2$, the overlap $\bigl\langle\mathcal{\mathfrak{C}M}_{j}|\mathfrak{C}\mathcal{\hat{M}}_{n}\bigr\rangle$
is zero such that $\mathcal{\hat{M}}_{n}$ needs only to be orthonormalized
against $\mathcal{M}_{n-2}$ and $\mathcal{M}_{n-1}$, as claimed
before.

While the poof of Eq.~(\ref{eq:Ortho von M dach tadm}) only uses
typical features of Krylov subspaces, the proof of Eq.~(\ref{eq:Ortho von M tilde tadm})
takes in addition advantage of Eq.~(\ref{eq:Optimal M tilde ohne Ortho tadm}),
which is specific to the problem at hand. We start by rewriting $\tilde{\mathcal{M}}_{n}$
as
\begin{eqnarray}
\tilde{\mathcal{M}}_{n} & \overset{\eqref{eq:Optimal M tilde ohne Ortho tadm}}{=} & \mathfrak{C}\varrho_{0}-\sum_{j=1}^{n-1}\alpha_{j}\mathfrak{C}^{2}\mathcal{M}_{j}\nonumber \\
 & = & \mathfrak{C}\Bigl(\varrho_{0}-\sum_{j=1}^{n-1}\alpha_{j}\mathfrak{C}\mathcal{M}_{j}\Bigr)\nonumber \\
 & = & \mathfrak{C}\mathcal{M}_{\bot},\label{eq:M tilde als C mal M orth tadm}
\end{eqnarray}
with
\begin{equation}
\mathcal{M}_{\bot}\coloneqq\varrho_{0}-\sum_{j=1}^{n-1}\alpha_{j}\mathfrak{C}\mathcal{M}_{j}.
\end{equation}
In this equation, the $\alpha_{j}=\left\langle \mathfrak{C}\mathcal{M}_{j}|\varrho_{0}\right\rangle $
(\ref{eq:Alpha t.a.d.m}) are chosen such that the $\alpha_{j}\mathfrak{C}\mathcal{M}_{j}$
annihilate the $\mathfrak{C}\mathcal{M}_{j}$ components in $\varrho_{0}$.
Hence, we find
\begin{equation}
\bigl\langle\mathfrak{C}\mathcal{M}_{j<n}\big|\mathcal{M}_{\bot}\bigr\rangle=0.\label{eq:M ortho ist Ortho CM_j tadk}
\end{equation}
With this result, we can prove Eq.~(\ref{eq:Ortho von M tilde tadm})

\begin{eqnarray}
\bigl\langle\mathfrak{C}\mathcal{M}_{j}\big|\mathfrak{C}\tilde{\mathcal{M}}_{n}\bigr\rangle & \overset{\eqref{eq:M tilde als C mal M orth tadm}}{=} & \bigl\langle\mathfrak{C}\mathcal{M}_{j}\big|\mathfrak{C}^{2}\mathcal{M}_{\bot}\bigr\rangle\nonumber \\
 & \overset{\eqref{eq:Selbstadjungierter kommutator t.a.d.m.}}{=} & \bigl\langle\mathfrak{C}^{3}\mathcal{M}_{j}\big|\mathcal{M}_{\bot}\bigr\rangle\nonumber \\
 & \overset{\eqref{eq:M dach t.a.d.m.}}{=} & \bigl\langle\mathfrak{C}\mathcal{\hat{M}}_{j+1}\big|\mathcal{M}_{\bot}\bigr\rangle.\label{eq:Ortho beweis fuer M tilde tadm}
\end{eqnarray}
The matrix $\mathcal{\hat{M}}_{j+1}$ lies in the subspace $\textrm{span}\left\{ \mathcal{M}_{1},\mathcal{M}_{2},\ldots,\mathcal{M}_{j+1}\right\} $.
Hence, according to Eq.~(\ref{eq:M ortho ist Ortho CM_j tadk}),
we find $\langle\mathfrak{C}\mathcal{\hat{M}}_{j+1}|\mathcal{M}_{\bot}\rangle=0$
for $j+1<n$ and with that $\langle\mathfrak{C}\mathcal{M}_{j<n-1}|\mathfrak{C}\tilde{\mathcal{M}}_{n}\rangle=0$
(\ref{eq:Ortho beweis fuer M tilde tadm}), as claimed in Eq.~(\ref{eq:Ortho von M tilde tadm}).

\subsection{Tensor network method\label{sub:Orthogonality f=0000FCr tensoren bei altered method}}

Finally, we have a look at the tensor network based method, as it
was explained in appendix~\ref{sub:Advisable-modification-of the algorithm tadm}
for MPO. Here, we face two main differences compared to the general
case: As a first difference, we only alter one MPO tensor $\mathsf{M}_{[k]}$
at a time, which implies that we keep all other MPO tensors $\mathsf{M}_{[j\neq k]}$
constant. This is e.g.\ the reason for the appearance of the term
$\bigl(\prod_{j\neq k}\mathsf{M}_{[j]}\bigr)^{\dagger}$ in Eq.~(\ref{eq:Tensor M_l>1  tadm}).
But with due diligence, one finds that this does not alter the line
of argumentation used above. As a second difference, we no longer
deal with a pure Krylov subspace, because of the extra role of the
MPO tensor $\mathsf{M}{}_{[k]}^{(1)}=\mathsf{M}^{\textrm{old}}{}_{[k]}$~(\ref{eq:Alter tensor recyle}).
That is, the MPO $\mathcal{M}_{1}=\mathsf{M}{}_{[k]}^{(1)}\prod_{j\neq k}\mathsf{M}_{[j]}$~(\ref{eq:MPO-Summen M_l tadm})
is generally not orthogonal to $\mathcal{\tilde{M}}_{n}=\tilde{\mathsf{M}}{}_{[k]}^{(n)}\prod_{j\neq k}\mathsf{M}_{[j]}$
\begin{equation}
\bigl\langle\mathfrak{C}\mathcal{M}_{1}\big|\mathfrak{C}\mathcal{\tilde{M}}_{n}\bigr\rangle\neq0,
\end{equation}
while for all other MPO $\mathcal{M}_{l}=\mathsf{M}{}_{[k]}^{(l)}\prod_{j\neq k}\mathsf{M}_{[j]}$
(\ref{eq:MPO-Summen M_l tadm}), the line of argumentation used above
still holds, i.e.
\begin{equation}
\bigl\langle\mathfrak{C}\mathcal{M}_{1<l<n-1}\big|\mathfrak{C}\mathcal{\tilde{M}}_{n}\bigr\rangle=0.
\end{equation}
In other words, each new tensor $\tilde{\mathsf{M}}{}_{[k]}^{(n)}$
needs to be orthonormalized against $\mathsf{M}{}_{[k]}^{(1)}$ and
$\mathsf{M}{}_{[k]}^{(n-1)}$. This result will be of great importance
for the speed up explained in appendix~\ref{sub:Speeding-up-convergence tadm}.

\begin{figure}
\includegraphics[width=1\columnwidth]{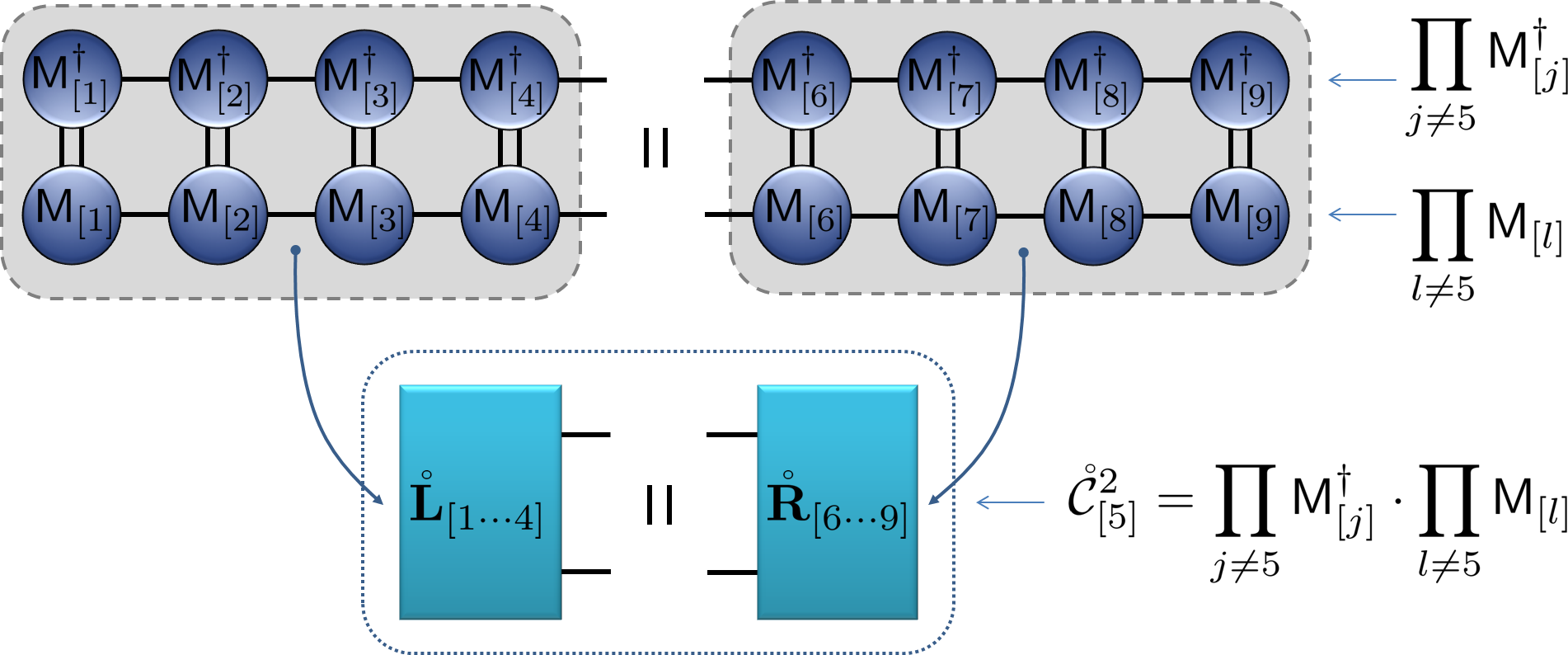}

\protect\caption{\label{fig:EffectiveSimpleNorm tadm}The MPO norm $\bigl\langle\prod_{j}\mathsf{M}_{[j]}\bigl|\prod_{l}\mathsf{M}_{[l]}\bigr\rangle$
(\ref{eq:Effective MPO Norm TADM}) can be expressed as weighted tensor
norm $\bigl\langle\mathsf{M}{}_{[k]}\bigl|\mathring{\mathcal{C}}{}_{[k]}^{2}\bigr|\mathsf{M}{}_{[k]}\bigr\rangle$.
The MPO tensors can be gauged such that $\mathring{\mathbf{L}}_{[1\cdots k-1]}={\mathbbm1}$
and $\mathring{\mathbf{R}}_{[k+1\cdots n]}={\mathbbm1}$. This situation
should be compared with the situation in Fig.~\ref{fig:Effective MPO Norm Operator TADM},
where such a gauge is generally not possible.}
\end{figure}

\begin{figure}
\includegraphics[width=1\columnwidth]{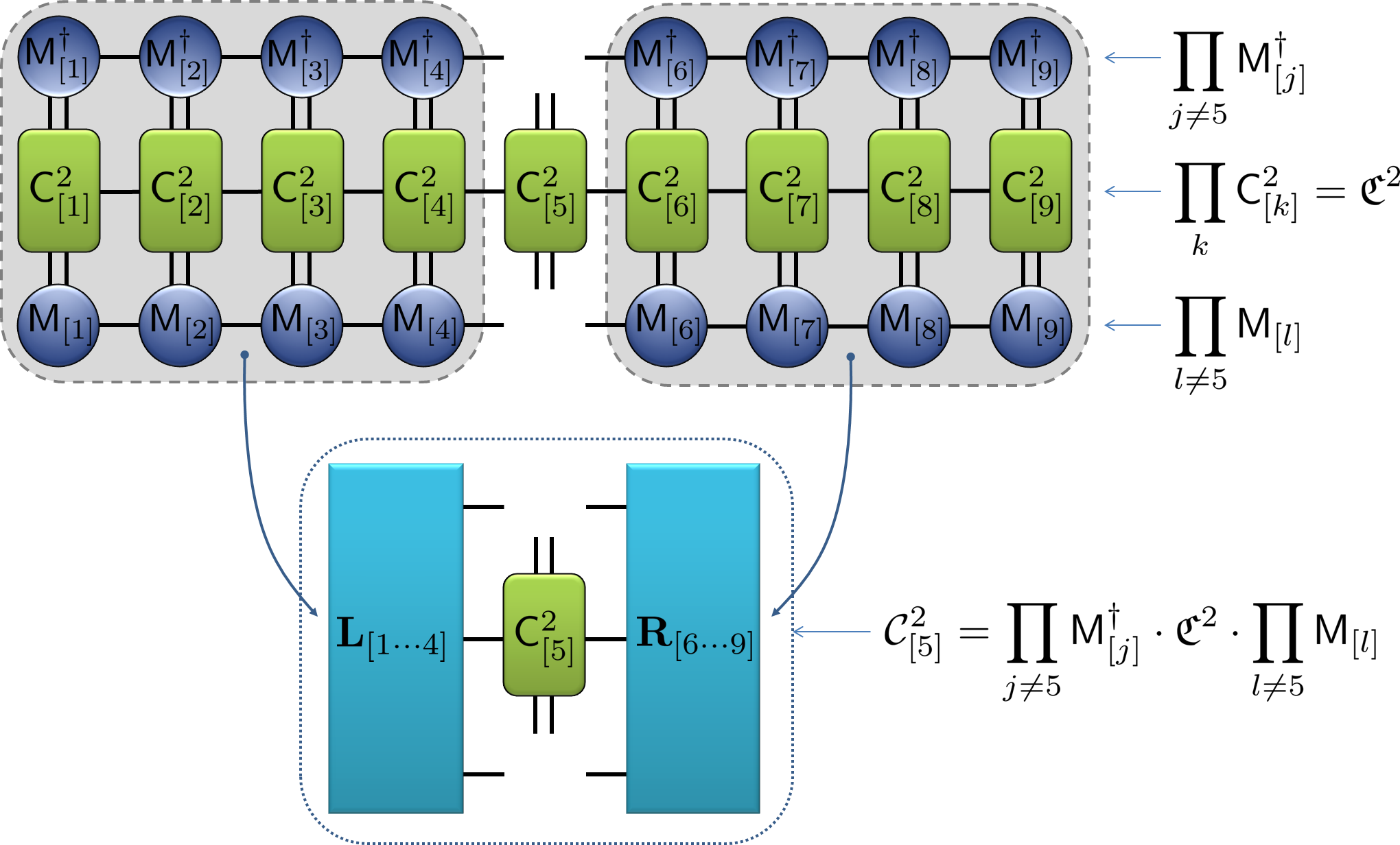}

\protect\caption{\label{fig:Effective MPO Norm Operator TADM}With the help of the
effective tensor operator $\mathcal{C}\text{\texttwosuperior}_{[k]},$
the weighted MPO norm $\bigl\langle\prod_{j}\mathsf{M}_{[j]}\bigl|\mathfrak{C}^{2}\bigr|\prod_{l}\mathsf{M}_{[l]}\bigr\rangle$
(\ref{eq:Effective MPO Norm TADM}) can be expressed as weighted tensor
norm $\bigl\langle\mathsf{M}{}_{[k]}\bigl|\mathcal{C}\text{\texttwosuperior}_{[k]}\bigr|\mathsf{M}{}_{[k]}\bigr\rangle$.}
\end{figure}

\section{Gauging the MPO\label{sub:Gauging-the-MPO tadm}}

MPS and MPO contain a gauge freedom\index{gauge freedom}, which can
be exploited to improve the performance of the algorithm. In our case,
the optimal gauge is not given by the standard canonical forms, which
are used in most other algorithms \cite{Schollwoeck2011}. 

The gauge freedom of the MPS/MPO stems from the simple fact that one
can always insert a matrix $a_{[j]}$ and its inverse $a_{[j]}^{-1}$
between two MPS tensors $\mathsf{M}_{[j]}$ and $\mathsf{M}_{[j+1]}$
\begin{eqnarray}
\mathsf{M}_{[j]\sigma_{j}}^{\alpha\beta}\mathsf{M}_{[j+1]\sigma_{j+1}}^{\beta\gamma} & = & \underbrace{\mathsf{M}_{[j]\sigma_{j}}^{\alpha\beta}a_{[j]}^{\beta\mu}}_{\mathsf{\tilde{M}}_{[j]}}\underbrace{a_{[j]}^{-1\,\mu\nu\,}\mathsf{M}_{[j+1]\sigma_{j+1}}^{\nu\gamma}}_{\mathsf{\tilde{M}}_{[j+1]}}\label{eq:Eich matrix TADM}
\end{eqnarray}
and replace them by $\mathsf{M}_{[j]}\rightarrow\mathsf{M}_{[j]}a_{[j]}$
and $\mathsf{M}_{[j+1]}\rightarrow a_{[j]}^{-1}\mathsf{M}_{[j+1]}$.
The aim of this section is to describe a method for finding beneficial
matrices $a_{[j]}$. 

For many application, it is advisable gauging a MPS~$M$ 
\begin{equation}
M=\mathsf{M}{}_{[k]}\prod_{j\neq k}\mathsf{M}_{[j]}\label{eq:Eich-MPO als Tesnorprodukt TADM}
\end{equation}
 in such a fashion that the norm of the entire MPS $M$ reduces to
the norm of the single MPS tensor $\mathsf{M}{}_{[k]}$
\begin{equation}
\forall\mathsf{M}{}_{[k]}:\quad\|M\|\overset{\eqref{eq:Eich-MPO als Tesnorprodukt TADM}}{=}\|\mathsf{M}{}_{[k]}\prod_{j\neq k}\mathsf{M}_{[j]}\|\overset{\textrm{special gauge}}{=}\|\mathsf{M}{}_{[k]}\|.\label{eq:Simple Eichung TADM}
\end{equation}
This can be achieved by the canonical form\index{canonical form}
described e.g.\ in Sec.~4.4 of Ref.~\cite{Schollwoeck2011}. Since
the norm of the single tensor $\mathsf{M}{}_{[k]}$ is easily controlled,
this canonical form is extremely helpful for MPS based algorithm which
have to fulfill the common side condition $\|M\|=1$. 

In our case, we have to deal with the weighted norm $\left\langle M|\mathfrak{C}^{2}|M\right\rangle =1$
(\ref{eq:Norm Dichte Matrix}) as side condition, which favors a different
kind of gauge. If this side condition could be simplified in the same
way as Eq.~(\ref{eq:Simple Eichung TADM}), i.e.,\ $\left\langle M|\mathfrak{C}^{2}|M\right\rangle \overset{?}{=}\|\mathsf{M}{}_{[k]}\|$
for arbitrary $\mathsf{M}{}_{[k]}$, the remaining overlap optimization
(\ref{eq:tensor overlap optimization tadm}) would be trivially solved
by Eq.~(\ref{eq:Schlechter erster MPO tensor tadm}) without the
necessity to approximate the optimal MPO tensor $\mathsf{M}_{[k]}^{{\rm optimal}}$
as a sum of many tensors $\mathsf{M}_{[k]}^{(l)}$ as in Eq.~(\ref{eq:MPO tensors locally bad optimized tadm}).
Unfortunately, a general $\left\langle M|\mathfrak{C}^{2}|M\right\rangle \overset{?}{=}\|\mathsf{M}{}_{[k]}\|$
cannot be achieved by simply gauging the MPO. Still, we should attempt
to get as close as possible to this relation to improve the performance
of the entire algorithm. So, let us have a closer look at the weighted
norm
\begin{eqnarray}
\bigl\langle M\bigl|\mathfrak{C}^{2}\bigr|M\bigr\rangle & \overset{\eqref{eq:Eich-MPO als Tesnorprodukt TADM}}{=} & \bigl\langle\mathsf{M}{}_{[k]}\prod_{j\neq k}\mathsf{M}_{[j]}\bigl|\mathfrak{C}^{2}\bigr|\mathsf{M}{}_{[k]}\prod_{l\neq k}\mathsf{M}_{[l]}\bigr\rangle\nonumber \\
 & = & \bigl\langle\mathsf{M}{}_{[k]}\bigl|\underbrace{\prod_{j\neq k}\mathsf{M}_{[j]}^{\dagger}\mathfrak{C}^{2}\prod_{l\neq k}\mathsf{M}_{[l]}}_{\mathcal{C}{}_{[k]}^{2}}\bigr|\mathsf{M}{}_{[k]}\bigr\rangle\nonumber \\
 & = & \bigl\langle\mathsf{M}{}_{[k]}\bigl|\mathcal{C}{}_{[k]}^{2}\bigr|\mathsf{M}{}_{[k]}\bigr\rangle,\label{eq:Effective MPO Norm TADM}
\end{eqnarray}
where we have introduced the effective tensor operator $\mathcal{C}{}_{[k]}^{2}$,
see Fig~\ref{fig:Effective MPO Norm Operator TADM}. To have $\left\langle M|\mathfrak{C}^{2}|M\right\rangle \approx\|\mathsf{M}{}_{[k]}\|$
for all $\mathsf{M}{}_{[k]}$, we need $\mathcal{C}{}_{[k]}^{2}\approx{\mathbbm1}$. 

We remark that we actually never calculate $\mathcal{C}{}_{[k]}^{2}$
explicitly, since the effort to do so scales with the fourth power
of the bond dimensions of the MPO $M$, while the scaling of all operations
presented so far does not exceed the third power. Therefore, we only
calculate with the components of $\mathcal{C}{}_{[k]}^{2}$, which
are $\mathbf{L}_{[1\cdots k-1]}$, $\mathsf{C}_{[k]}^{2}$ and $\mathbf{R}_{[k+1\cdots n]}$,
\begin{equation}
\mathcal{C}{}_{[k]\bar{\sigma}_{k}\sigma_{k}}^{2\ \bar{\mu}_{k-1}\mu_{k-1}\bar{\mu}_{k}\mu_{k}}=\mathbf{L}_{[1\cdots k-1]}^{\bar{\mu}_{k-1}\gamma_{k-1}\mu_{k-1}}\mathsf{C}_{[k]\bar{\sigma}_{k}\sigma_{k}}^{2\ \gamma_{k-1}\gamma_{k}}\mathbf{R}_{[k+1\cdots n]}^{\bar{\mu}_{k}\gamma_{k}\mu_{k}},\label{eq:Effective C as LCR tadm}
\end{equation}
as shown in Fig.~\ref{fig:Effective MPO Norm Operator TADM}. Hereby,
$\mathsf{C}_{[j]}^{2}$ are the tensors of the MPO representing the
squared commutator operator $\mathfrak{C}^{2}$ (\ref{eq:Kommutatorzeiche C def t.a.d.m.})
(where the square in $\mathsf{C}_{[j]}^{2}$ is just symbolical, as
in $\mathcal{C}{}_{[k]}^{2}$). The left $\mathbf{L}_{[1\cdots k-1]}$
and right $\mathbf{R}_{[k+1\cdots n]}$ can be calculated iteratively
as
\begin{eqnarray}
\mathbf{L}_{[1\cdots j]}^{\bar{\mu}{}_{j}\gamma_{j}\mu{}_{j}} & = & \mathbf{L}_{[1\cdots j-1]}^{\bar{\mu}{}_{j-1}\gamma_{j-1}\mu{}_{j-1}}\nonumber \\
 &  & \cdot\left(\mathsf{M}_{[j]}^{*}\right)_{\bar{\sigma}_{j}}^{\bar{\mu}_{j-1}\bar{\mu}{}_{j}}\left(\mathsf{C}_{[j]}^{2}\right)_{\bar{\sigma}_{j}\sigma_{j}}^{\gamma_{j-1}\gamma_{j}}\mathsf{M}_{[j]\sigma_{j}}^{\mu{}_{j-1}\mu{}_{j}}\nonumber \\
\mathbf{R}_{[j\cdots n]}^{\bar{\mu}{}_{j-1}\gamma_{j-1}\mu{}_{j-1}} & = & \left(\mathsf{M}_{[j]}^{*}\right)_{\bar{\sigma}_{j}}^{\bar{\mu}_{j-1}\bar{\mu}{}_{j}}\left(\mathsf{C}_{[j]}^{2}\right)_{\bar{\sigma}_{j}\sigma_{j}}^{\gamma_{j-1}\gamma_{j}}\mathsf{M}_{[j]\sigma_{j}}^{\mu{}_{j-1}\mu{}_{j}}\nonumber \\
 &  & \cdot\mathbf{R}_{[j+1\cdots n]}^{\bar{\mu}{}_{j}\gamma_{j}\mu{}_{j}}.\label{eq:Iterative L-R Blocks tadm}
\end{eqnarray}
If we perform a gauge transformation as in Eq.~(\ref{eq:Eich matrix TADM}),
we obtain
\begin{eqnarray}
\mathbf{L}_{[1\cdots k-1]}^{\bar{\mu}\gamma\mu} & \rightarrow & \mathbf{L}_{[1\cdots j]}^{\bar{\mu}\gamma\mu}\left(a{}_{[k-1]}^{*}\right)^{\bar{\mu}\bar{\nu}}a{}_{[k-1]}^{\mu\nu}\nonumber \\
\mathbf{R}_{[k+1\cdots n]}^{\bar{\mu}\gamma\mu} & \rightarrow & \mathbf{R}_{[k+1\cdots n]}^{\bar{\mu}\gamma\mu}\left(a{}_{[k]}^{-1\,*}\right)^{\bar{\mu}\bar{\nu}}\left(a_{[k]}^{-1}\right)^{\mu\nu}.\label{eq:Block-Eichung tadm}
\end{eqnarray}
Now, we need to find beneficial gauge matrices $a_{[j]}$. To understand
the procedure, it is helpful to look at the standard norm $\langle M|M\rangle$
without the commutator operator $\mathfrak{C}^{2}$ (Fig.~\ref{fig:EffectiveSimpleNorm tadm}).
Here, the $\gamma$-index in $\mathbf{L}_{[1\cdots k-1]}^{\bar{\mu}\gamma\mu}$
and $\mathbf{R}_{[k+1\cdots n]}^{\bar{\mu}\gamma\mu}$ (\ref{eq:Block-Eichung tadm})
is absent and the remaining $\mathring{\mathbf{L}}_{[1\cdots k-1]}^{\bar{\mu}\mu}$
and $\mathring{\mathbf{R}}_{[k+1\cdots n]}^{\bar{\mu}\mu}$ can be
treated as matrices. Using the trivial fact that $\mathring{\mathbf{L}}=\sqrt{\mathring{\mathbf{L}}}{\mathbbm1}\sqrt{\mathring{\mathbf{L}}}$
with $\sqrt{\mathring{\mathbf{L}}}^{\dagger}=\sqrt{\mathring{\mathbf{L}}}$
and similar for $\mathring{\mathbf{R}}$, we choose the gauge matrices
to be 
\begin{equation}
a_{[k-1]}=\sqrt{\mathring{\mathbf{L}}_{[1\cdots k-1]}^{-1}}\quad{\rm and}\quad a_{[k]}^{-1}=\sqrt{\mathring{\mathbf{R}}_{[k+1\cdots n]}^{-1}}\label{eq:Wurzeleichung tadm}
\end{equation}
and the gauge transformation (\ref{eq:Block-Eichung tadm}) result
in
\begin{alignat}{3}
\mathring{\mathbf{L}}_{[1\cdots k-1]} & \overset{\eqref{eq:Block-Eichung tadm}}{\longrightarrow} & a_{[k-1]}\mathring{\mathbf{L}}_{[1\cdots k-1]}a_{[k-1]} & \overset{\eqref{eq:Wurzeleichung tadm}}{=} & {\mathbbm1}\nonumber \\
\mathring{\mathbf{R}}_{[k+1\cdots n]} & \overset{\eqref{eq:Block-Eichung tadm}}{\longrightarrow} & a_{[k]}^{-1}\mathring{\mathbf{R}}_{[k+1\cdots n]}a_{[k]}^{-1} & \overset{\eqref{eq:Wurzeleichung tadm}}{=} & {\mathbbm1}.\label{eq:Normumgebung wird zur Identity tadm}
\end{alignat}
With these identities, the MPO norm $\left\langle M|M\right\rangle $
reduces to the tensor norm $\langle\mathsf{M}{}_{[k]}|\mathsf{M}{}_{[k]}\rangle$,
as for the canonical form.

Now, we come back to the weighted MPO norm $\left\langle M|\mathfrak{C}^{2}|M\right\rangle $,
where $\mathbf{L}_{[1\cdots k-1]}^{\bar{\mu}\gamma\mu}$ and $\mathbf{R}_{[k+1\cdots n]}^{\bar{\mu}\gamma\mu}$
carry a $\gamma$-index and are given by Eq.~(\ref{eq:Iterative L-R Blocks tadm}).
In this case, the gauge matrices $a_{[k-1]}$ and $a_{[k]}$ do no
longer have the same dimensions as $\mathbf{L}_{[1\cdots k-1]}^{\bar{\mu}\gamma\mu}$
and $\mathbf{R}_{[k+1\cdots n]}^{\bar{\mu}\gamma\mu}$ and hence,
Eq.~(\ref{eq:Wurzeleichung tadm}) can not be used to obtain the
gauge matrices. With the help of the multi-index $\xi=(\bar{\mu},\gamma)$,
$\mathbf{L}_{[1\cdots k-1]}^{\xi\mu}$ and $\mathbf{R}_{[k+1\cdots n]}^{\xi\mu}$
can still be written as matrices and be decomposed by a singular value
decomposition
\begin{eqnarray}
\mathbf{L}_{[1\cdots k-1]}^{\xi\mu} & = & U_{[L,k-1]}^{\xi\nu}D_{[L,k-1]}^{\nu\nu}V_{[L,k-1]}^{\nu\mu}\nonumber \\
\mathbf{R}_{[k+1\cdots n]}^{\xi\mu} & = & U_{[R,k+1]}^{\xi\nu}D_{[R,k+1]}^{\nu\nu}V_{[R,k+1]}^{\nu\mu}.\label{eq:L_R_Block SVD tadm}
\end{eqnarray}
To mimic the effect of the inverse square root in Eq.~(\ref{eq:Wurzeleichung tadm}),
we define the gauge matrices as
\begin{eqnarray}
a_{[k-1]} & = & V_{[L,k-1]}^{\dagger}D_{[L,k-1]}^{-\frac{1}{2}}V_{[L,k-1]}\nonumber \\
a_{[k]}^{-1} & = & V_{[R,k+1]}^{\dagger}D_{[R,k+1]}^{-\frac{1}{2}}V_{[R,k+1]}.\label{eq:Def Eich matrix tadm}
\end{eqnarray}
Inserting these gauge matrices into Eq.~(\ref{eq:Block-Eichung tadm}),
does not turn $\mathbf{L}_{[1\cdots k-1]}^{\bar{\mu}\gamma\mu}$ and
$\mathbf{R}_{[k+1\cdots n]}^{\bar{\mu}\gamma\mu}$ into identities
as it was possible for $\mathring{\mathbf{L}}_{[1\cdots k-1]}^{\bar{\mu}\mu}$
and $\mathring{\mathbf{R}}_{[k+1\cdots n]}^{\bar{\mu}\mu}$ (\ref{eq:Normumgebung wird zur Identity tadm}),
but at least $\mathbf{L}_{[1\cdots k-1]}^{\bar{\mu}\gamma\mu}$ and
$\mathbf{R}_{[k+1\cdots n]}^{\bar{\mu}\gamma\mu}$ get closer to the
identity.

We remark that this procedure is not optimal and could still be improved
by more complicated methods. For example, if we use the gauge matrices
(\ref{eq:Def Eich matrix tadm}) to transform $\mathbf{L}_{[1\cdots k-1]}^{\bar{\mu}\gamma\mu}$
and $\mathbf{R}_{[k+1\cdots n]}^{\bar{\mu}\gamma\mu}$ according to
Eq.~(\ref{eq:Block-Eichung tadm}), we could iterate the procedure
and use the transformed $\mathbf{L}_{[1\cdots k-1]}$ and $\mathbf{R}_{[k+1\cdots n]}$
to obtain further gauge matrices. However, in our applications, a
single gauge transformation turned out to be quite profitable, while
further iterations had no great impact.

\subsection{Decomposing \textmd{\normalsize{}$\mathcal{C}{}_{[k]}^{2}$}}

So far, we just considered the left and right $\mathbf{L}_{[1\cdots k-1]}^{\bar{\mu}\gamma\mu}$
and $\mathbf{R}_{[k+1\cdots n]}^{\bar{\mu}\gamma\mu}$ separately
to obtain the gauge matrices. In case of the standard MPO norm $\left\langle M|M\right\rangle $,
this separated treatment is perfectly justified (see Fig.~\ref{fig:EffectiveSimpleNorm tadm}),
while for the weighted norm $\left\langle M|\mathfrak{C}^{2}|M\right\rangle $,
$\mathbf{L}_{[1\cdots k-1]}^{\bar{\mu}\gamma\mu}$ and $\mathbf{R}_{[k+1\cdots n]}^{\bar{\mu}\gamma\mu}$
are connected by the squared commutator operator $\mathfrak{C}^{2}$
(Fig.~\ref{fig:Effective MPO Norm Operator TADM}). Therefore, the
natural object to consider is the effective tensor operator $\mathcal{C}{}_{[k]}^{2}$.
As already mentioned, calculating the effective tensor operator is
$\mathcal{C}{}_{[k]}^{2}$ is numerically expensive. Fortunately,
it will turn out that we do not require calculating $\mathcal{C}{}_{[k]}^{2}$
explicitly, but for the moment, let us pretend we have done so. We
start by replacing the singular value decompositions in Eq.~(\ref{eq:L_R_Block SVD tadm})
by the more suited decomposition
\begin{eqnarray}
\left(\mathcal{C}{}_{[k]}^{2}\right)^{\xi_{L}\mu_{k-1}} & = & \tilde{U}_{[L,k-1]}^{\xi_{L}\nu}\tilde{D}_{[L,k-1]}^{\nu\nu}\tilde{V}_{[L,k-1]}^{\nu\mu_{k-1}}\nonumber \\
\left(\mathcal{C}{}_{[k]}^{2}\right)^{\xi_{R}\mu_{k}} & = & \tilde{U}_{[R,k+1]}^{\xi_{R}\nu}\tilde{D}_{[R,k+1]}^{\nu\nu}\tilde{V}_{[R,k+1]}^{\nu\mu_{k}},\label{eq:Full C SVD tadm}
\end{eqnarray}
with the multi-indices $\xi_{L}=(\bar{\mu}_{k-1},\bar{\mu}_{k},\mu_{k},\bar{\sigma}_{k},\sigma_{k})$
and $\xi_{R}=(\bar{\mu}_{k-1},\mu_{k-1},\bar{\mu}_{k},\bar{\sigma}_{k},\sigma_{k})$.
These new matrices can now be used in Eq.~(\ref{eq:Def Eich matrix tadm})
to obtain better gauge matrices $a_{[k-1]}=\tilde{V}_{[L,k-1]}^{\dagger}\tilde{D}_{[L,k-1]}^{-0.5}\tilde{V}_{[L,k-1]}$
and $a_{[k]}^{-1}=\tilde{V}_{[R,k+1]}^{\dagger}\tilde{D}_{[R,k+1]}^{-0.5}\tilde{V}_{[R,k+1]}$,
which take the full operator $\mathcal{C}{}_{[k]}^{2}$ into account.

\subsubsection{Successive decomposition of \textup{$\mathcal{C}{}_{[k]}^{2}$}}

Now, we show that we do not require calculating $\mathcal{C}{}_{[k]}^{2}$
explicitly. The needed matrices $\tilde{D}$ and $\tilde{V}$ of Eq.~(\ref{eq:Full C SVD tadm})
can be obtained relatively cheap by successively decomposing the components
$\mathbf{L}_{[1\cdots k-1]}$, $\mathsf{C}_{[k]}^{2}$ and $\mathbf{R}_{[k+1\cdots n]}$
along the indices which connect the components. To demonstrate this
procedure, we assume that we like to calculate the gauge matrix $a_{[k]}^{-1}=\tilde{V}_{[R,k+1]}^{\dagger}\tilde{D}_{[R,k+1]}^{-0.5}\tilde{V}_{[R,k+1]}$.
We start by rewriting $\mathbf{L}_{[1\cdots k-1]}^{\bar{\mu}_{k-1}\gamma_{k-1}\mu_{k-1}}$
as matrix $\mathbf{L}_{[1\cdots k-1]}^{\zeta\gamma_{k-1}}$ with the
multi-index $\zeta=(\bar{\mu}_{k-1},\mu_{k-1})$ and apply a QR-decomposition
\begin{equation}
\mathbf{L}_{[1\cdots k-1]}^{\zeta\gamma_{k-1}}=q_{[L,k-1]}^{\zeta\eta}r_{[L,k-1]}^{\eta\gamma_{k-1}}.\label{eq:L-zerlegung TADm}
\end{equation}
Next, the freshly obtained matrix $r_{[L,k-1]}^{\eta\gamma_{k-1}}$
is multiplied into the MPO tensor $\left(\mathsf{C}_{[k]}^{2}\right)_{\bar{\sigma}_{k}\sigma_{k}}^{\gamma_{k-1}\gamma_{k}}$
\begin{equation}
r_{[L,k-1]}^{\eta\gamma_{k-1}}\left(\mathsf{C}_{[k]}^{2}\right)_{\bar{\sigma}_{k}\sigma_{k}}^{\gamma_{k-1}\gamma_{k}}=:\left(\tilde{\mathsf{C}}_{[k]}^{2}\right)_{\bar{\sigma}_{k}\sigma_{k}}^{\eta\gamma_{k}}.
\end{equation}
Now, we repeat this procedure and rewrite the modified tensor $\tilde{\mathsf{C}}_{[k]}^{2}$
as matrix $\left(\tilde{\mathsf{C}}_{[k]}^{2}\right)^{\kappa\gamma_{k}}$
with $\kappa=(\bar{\sigma}_{k},\sigma_{k},\eta)$ and perform another
QR-decomposition
\begin{equation}
\left(\tilde{\mathsf{C}}_{[k]}^{2}\right)^{\kappa\gamma_{k}}=q_{[C,k]}^{\kappa\eta}r_{[C,k]}^{\eta\gamma_{k}}.\label{eq:C-Zerlegung Tadm}
\end{equation}
Finally, we multiply this $r_{[C,k]}^{\eta\gamma_{k}}$ into $\mathbf{R}_{[k+1\cdots n]}^{\bar{\mu}_{k}\gamma_{k}\mu_{k}}$
\begin{equation}
\tilde{\mathbf{R}}_{[k+1\cdots n]}^{\bar{\mu}_{k}\eta\mu_{k}}=r_{[C,k]}^{\eta\gamma_{k}}\mathbf{R}_{[k+1\cdots n]}^{\bar{\mu}_{k}\gamma_{k}\mu_{k}}.
\end{equation}
This new tensor $\tilde{\mathbf{R}}_{[k+1\cdots n]}$ should now be
used in Eq.~(\ref{eq:L_R_Block SVD tadm}) instead of $\mathbf{R}_{[k+1\cdots n]}$,
i.e.
\begin{equation}
\tilde{\mathbf{R}}_{[k+1\cdots n]}^{\xi\mu_{k}}=\hat{U}_{[R,k+1]}^{\xi\nu}\tilde{D}_{[R,k+1]}^{\nu\nu}\tilde{V}_{[R,k+1]}^{\nu\mu_{k}},\label{eq:UDV successive tadm}
\end{equation}
with $\xi=(\bar{\mu}_{k},\eta)$. Putting these three decompositions
(\ref{eq:L-zerlegung TADm}), (\ref{eq:C-Zerlegung Tadm}), and (\ref{eq:UDV successive tadm})
together, we obtain
\begin{eqnarray}
\mathcal{C}{}_{[k]}^{2} & \overset{\eqref{eq:Effective C as LCR tadm}}{=} & \mathbf{L}_{[1\cdots k-1]}\mathsf{C}_{[k]}^{2}\mathbf{R}_{[k+1\cdots n]}\nonumber \\
 & = & q_{[L,k-1]}q_{[C,k]}\hat{U}_{[R,k+1]}\tilde{D}_{[R,k+1]}\tilde{V}_{[R,k+1]}.\nonumber \\
 &  & \ \label{eq:Product SVD eichung tadm}
\end{eqnarray}
Since $\left(q_{[L,k-1]}q_{[C,k]}\hat{U}_{[R,k+1]}\right)^{\dagger}q_{[L,k-1]}q_{[C,k]}\hat{U}_{[R,k+1]}={\mathbbm1}$
has the same isometric property as $\tilde{U}_{[R,k]}$ in Eq.~(\ref{eq:Full C SVD tadm}),
we find that Eq.~(\ref{eq:Product SVD eichung tadm}) is a correct
singular value decompositions of $\mathcal{C}{}_{[k]}^{2}.$ Hence,
we can use $\tilde{D}_{[R,k+1]}$ and $\tilde{V}_{[R,k+1]}$ of Eq.~(\ref{eq:UDV successive tadm})
for the gauge matrix $a_{[k]}^{-1}=\tilde{V}_{[R,k+1]}^{\dagger}\tilde{D}_{[R,k+1]}^{-0.5}\tilde{V}_{[R,k+1]}$
(\ref{eq:Def Eich matrix tadm}). In order to calculate the gauge
matrix $a_{[k-1]}$, we proceed in the same spirit, only in the opposite
direction (first decomposing $\mathbf{R}_{[k+1\cdots n]}$ then the
modified $\mathsf{C}_{[k]}^{2}$ and finally the modified $\mathbf{L}_{[1\cdots k-1]}$). 

We like to remark that in our algorithm, we optimize the MPO tensors
$\mathsf{M}_{[k]}$ in ascending, as well as in descending sweeps
of the index $k$. In each of these alternating sweep directions,
only one of the two gauge matrices $a_{[k-1]}$, $a_{[k]}^{-1}$ is
calculated while for the other, we assume that the result obtained
in the last sweep in opposite direction is a sufficiently good approximation
(although one could also consider calculating both gauge matrices
anew each sweep). That is, if the next MPO tensor we are going to
optimize is $\mathsf{M}_{[k]}$, the last tensor that has been optimized
should either be $\mathsf{M}_{[k-1]}$ or $\mathsf{M}_{[k+1]}$. In
case $\mathsf{M}_{[k-1]}$ has been optimized last, we use the gauge
matrix $a_{[k-1]}$ to obtain $\mathsf{M}_{[k-1]}\rightarrow\mathsf{M}_{[k-1]}a_{[k-1]}$
and $\mathsf{M}_{[k]}\rightarrow a_{[k-1]}^{-1}\mathsf{M}_{[k]}$,
while after the optimization of $\mathsf{M}_{[k+1]}$, we need the
gauge matrix $a_{[k]}^{-1}$ for the transformation $\mathsf{M}_{[k]}\rightarrow\mathsf{M}_{[k]}a_{[k]}$
and $\mathsf{M}_{[k+1]}\rightarrow a_{[k]}^{-1}\mathsf{M}_{[k+1]}$. 

For the gauging of the MPO tensors, we need to invert matrices respectively
their singular values. Numerically, this procedure might be troublesome.
Therefore, one should regularize the gauge matrices. Further, it might
be helpful to bring the MPO in their canonical form \cite{Schollwoeck2011}
before re-gauging them, since the canonical form is already a good
approximation, which can be obtained without the need of inverting
matrices

\subsection{Physical gauge}

The effective tensor operator $\left(\mathcal{C}{}_{[k]}^{2}\right)_{\bar{\sigma}_{k}\sigma_{k}}^{\bar{\mu}_{k-1}\mu_{k-1}\bar{\mu}_{k}\mu_{k}}$
(\ref{eq:Effective C as LCR tadm}) carries four tensor indices $\bar{\mu}_{k-1},\mu_{k-1},\bar{\mu}_{k},\mu_{k}$
corresponding to the auxiliary bonds and two (multi-)indices $\bar{\sigma}_{k},\sigma_{k}$
corresponding to the physical dimensions. Of these six indices, only
the four auxiliary indices are effected by the MPO tensor gauge. To
obtain an effective $\mathcal{C}{}_{[k]}^{2}$ as close to the identity
as possible, we can also introduce a ``gauge'' for the physical
indices, although this resembles more a transformation than a gauge.
We remark that in our applications, the benefits of this transformation
were far less pronounced than the benefits of the gauging applied
to the auxiliary indices.

The idea is to place a matrix $b_{[k]}$ and its inverse $b_{[k]}^{-1}$
between the physical bonds of the MPO tensor $\mathsf{M}_{[k]}$ and
the tensor $\mathsf{C}_{[k]}^{2}$ respectively $\mathcal{C}\text{\texttwosuperior}_{[k]}$
(\ref{eq:Effective C as LCR tadm})
\begin{eqnarray}
\mathsf{M}_{[k]\sigma_{k}}^{\mu_{k-1}\mu_{k}} & \rightarrow & \mathsf{M}_{[k]\sigma_{k}'}^{\mu_{k-1}\mu_{k}}b_{[k]\sigma_{k}'\sigma_{k}}^{-1}\nonumber \\
\mathcal{C}{}_{[k]\bar{\sigma}_{k}\sigma_{k}}^{2\,\bar{\mu}_{k-1}\mu_{k-1}\bar{\mu}_{k}\mu_{k}} & \rightarrow & \mathcal{C}{}_{[k]\bar{\sigma}'_{k}\sigma'_{k}}^{2\,\bar{\mu}_{k-1}\mu_{k-1}\bar{\mu}_{k}\mu_{k}}\, b_{[k]\,\bar{\sigma}'_{k}\bar{\sigma}_{k}}^{*}b_{[k]\sigma_{k}'\sigma_{k}}.\nonumber \\
 & \,
\end{eqnarray}
Obviously, this transformations keeps the weighted norm $\langle\mathsf{M}_{[k]}|\mathcal{C}{}_{[k]}^{2}|\mathsf{M}_{[k]}\rangle$
unchanged. But we also have to consider that the overlap $\left\langle \varrho_{0}|\mathfrak{C}|M\right\rangle $
(Fig.~(\ref{fig:MPO-Umgebung TADM})) is effected by this transformation
of the physical indices with $b_{[k]}$ and has to be transformed
accordingly.

The procedure to obtain the optimal gauge $b_{[k]}$ for the physical
indices is similar to the procedure used for finding the optimal gauge
$a_{[k]}$ for the auxiliary indices. We start by writing $\mathcal{C}{}_{[k]}^{2}$
as matrix $\left(\mathcal{C}{}_{[k]}^{2}\right)^{\xi\sigma_{k}}$
with the multi-index $\xi=(\bar{\mu}_{k-1},\mu_{k-1},\bar{\mu}_{k},\mu_{k},\bar{\sigma}_{k})$
and the remaining physical index $\sigma_{k}$ and perform a singular
value decomposition 
\begin{equation}
\left(\mathcal{C}{}_{[k]}^{2}\right)^{\xi\sigma_{k}}=U_{[k]}D_{[k]}V_{[k]}.\label{eq:SVD of C for Physical tadm}
\end{equation}
The transformation matrix $b_{[k]}$ is now obtained as
\begin{equation}
b_{[k]}=V_{[k]}^{\dagger}D_{[k]}^{-\frac{1}{2}}V_{[k]}.\label{eq:Transformation matrix phys gauge tadm}
\end{equation}
As before, the singular value decomposition of $\mathcal{C}{}_{[k]}^{2}$
can be obtained by decompositions of its components $\mathbf{L}_{[1\cdots k-1]}$,
$\mathsf{C}_{[k]}^{2}$ and $\mathbf{R}_{[k+1\cdots n]}$ (\ref{eq:Effective C as LCR tadm})
along their connecting indices. That is, we write $\mathbf{L}_{[1\cdots k-1]}^{\bar{\mu}\gamma\mu}$
and $\mathbf{R}_{[k+1\cdots n]}^{\bar{\mu}\gamma\mu}$ as matrices
with the multi-index $\kappa=(\bar{\mu},\mu)$ and perform QR-decompositions
\begin{eqnarray}
\mathbf{L}_{[1\cdots k-1]}^{\kappa\gamma} & \rightarrow & q_{[L,k-1]}r_{[L,k-1]}\nonumber \\
\mathbf{R}_{[k+1\cdots n]}^{\kappa\gamma} & \rightarrow & q_{[R,k+1]}r_{[R,k+1]}.
\end{eqnarray}
Then, the two $r$ matrices are multiplied into $\mathsf{C}_{[k]}^{2}$
\begin{equation}
\left(\tilde{\mathsf{C}}_{[k]}^{2}\right)_{\bar{\sigma}_{k}\sigma_{k}}^{\alpha\beta}:=r_{[L,k-1]}^{\alpha\gamma_{k-1}}\left(\mathsf{C}_{[k]}^{2}\right)_{\bar{\sigma}_{k}\sigma_{k}}^{\gamma_{k-1}\gamma_{k}}r_{[R,k+1]}^{\beta\gamma_{k}}
\end{equation}
Finally, we write $\tilde{\mathsf{C}}_{[k]}^{2}$ as matrix $\left(\tilde{\mathsf{C}}_{[k]}^{2}\right)^{\eta\sigma_{k}}$
with $\eta=(\alpha,\beta,\bar{\sigma}_{k})$ and perform a singular
value decomposition
\begin{equation}
\tilde{\mathsf{C}}_{[k]}^{2}=\hat{U}_{[k]}D_{[k]}V_{[k]},
\end{equation}
which delivers the matrices $D_{[k]}$ and $V_{[k]}$ needed for the
transformation matrix $b_{[k]}=V_{[k]}^{\dagger}D_{[k]}^{-\frac{1}{2}}V_{[k]}$
(\ref{eq:Transformation matrix phys gauge tadm}).

\section{Speeding up convergence by overarching orthonormalization\label{sub:Speeding-up-convergence tadm}}

In this section, we introduce a method which allows for speeding up
the convergence of the algorithm presented in appendix~\ref{sub:Advisable-modification-of the algorithm tadm}.
The key to this method is the insight that essential information gained
in previous optimization sweeps can be passed on to later optimizations
to obtain a faster convergence. 

To understand this new method, we start by reviewing the general MPO
optimization procedure. So far, we mainly discussed how a single MPO
tensor $\mathsf{M}{}_{[k]}$ (\ref{eq:Wiederholung guter Summen Tensor tadm})
is optimized. During the optimization of the tensor $\mathsf{M}{}_{[k]}$,
all other tensors $\mathsf{M}{}_{[j\neq k]}$ are kept constant. Since
these tensors $\mathsf{M}{}_{[j\neq k]}$ are most likely suboptimal,
the optimization of a single tensor $\mathsf{M}{}_{[k]}$ will generally
not allow us to obtain the globally optimal MPO. According to Eq.~(\ref{eq:Wiederholung guter Summen Tensor tadm}),
$\mathsf{M}{}_{[k]}=\sum_{l}\alpha_{l}\cdot\mathsf{M}{}_{[k]}^{(l)}$
is optimized by summing up several weighted basis tensors $\mathsf{M}{}_{[k]}^{(l)}$.
On the one hand, the more $\mathsf{M}{}_{[k]}^{(l)}$ we sum up, the
better the result for the tensor $\mathsf{M}{}_{[k]}$. On the other
hand, an exhaustive optimization of a single tensor $\mathsf{M}{}_{[k]}$
is a waste of computation time if the remaining tensors $\mathsf{M}{}_{[j\neq k]}$
are still far from optimal. Therefore, in practical applications,
one will usually just sum up a few $\mathsf{M}{}_{[k]}^{(l)}$ and
instead perform many repeated sweeps of the index $k$ (which means
we optimize the first to the last tensor and then start over again
and again). During these sweeping cycles, the same tensor $\mathsf{M}{}_{[k]}$
is optimized many times, each time with different environmental tensors
$\mathsf{M}{}_{[j\neq k]}$, which also have been optimized in between.

As a result of the first optimization sweeps, the tensors $\mathsf{M}{}_{[j]}$
are likely to change substantially. But with each completed sweeping
cycle, the modifications of the tensors $\mathsf{M}{}_{[j]}$ should
subside, since the $\mathsf{M}{}_{[j]}$ are converging towards their
optimal value. When the changes have become sufficiently small, the
environmental tensors $\mathsf{M}{}_{[j\neq k]}$ can be considered
to be approximately constant between two optimization cycles of $\mathsf{M}{}_{[k]}$.
This puts us into the position to reuse information won in previous
optimizations of $\mathsf{M}{}_{[k]}$. To see this, we assume for
a moment that the environmental tensors $\mathsf{M}{}_{[j\neq k]}$
do not change at all, and compare the situation where we perform many
short optimizations of the MPO tensor $\mathsf{M}{}_{[k]}$ with the
situation where we do one long optimization. Here, by short optimization,
we mean that we sum up just a few basis tensors $\mathsf{M}{}_{[k]}^{(l)}$
to obtain the new MPO tensor $\mathsf{M}{}_{[k]}^{(l)}$, which corresponds
to the situation in the algorithm.

In order to have a fair comparison, we assume that in total, the many
short optimizations generate the same amount of linearly independent
tensors $\mathsf{M}{}_{[k]}^{(l)}$ (\ref{eq:Wiederholung guter Summen Tensor tadm})
as the one long optimization. The crucial difference is that for the
one long optimization, the tensors $\mathsf{M}{}_{[k]}^{(l)}$ are
not only linearly independent, but also orthonormal (\ref{eq:Weighted norm for HM}).
Solely for orthonormal $\mathsf{M}{}_{[k]}^{(l)}$, Eq.~(\ref{eq:Alpha t.a.d.m})
provides the optimal overlap $\alpha_{l}$ for Eq.~(\ref{eq:Wiederholung guter Summen Tensor tadm}).
In other words: one long optimization is superior to many short optimization.

What we intend to achieve is that the many short optimizations we
use in the algorithm act the same way as one long optimization. That
is, we have to find a way to maintain the orthonormalization of the
basis tensors $\mathsf{M}{}_{[k]}^{(l)}$ over many optimization cycles.
In this context, the observation made in appendix~\ref{sub:Orthogonality f=0000FCr tensoren bei altered method}
is of special relevance that any newly generated basis tensor $\tilde{\mathsf{M}}{}_{[k]}^{(n)}$
(\ref{eq:Tensor M_l>1  tadm}) is already orthonormal to \emph{all}
basis tensors $\mathsf{M}{}_{[k]}^{(l<n)}$ after it has been orthonormalized
against the two basis tensors $\mathsf{M}{}_{[k]}^{(1)}$ and $\mathsf{M}{}_{[k]}^{(n-1)}$.
This limits the amount of information we have to transmit from one
optimization cycle to the next to ensure that all basis tensors $\mathsf{M}{}_{[k]}^{(l)}$
generated in consecutive optimizations cycles are orthonormal. 

The basis tensor $\mathsf{M}{}_{[k]}^{(1)}$ is always given by the
result of the optimization round before (\ref{eq:Alter tensor recyle}).
Since we assume that all tensors $\mathsf{M}{}_{[j]}$ change only
slightly from one short optimization to the next, the first basis
tensor $\mathsf{M}{}_{[k]}^{(1)}$ is roughly the same for the different
optimization rounds. Hence, the orthonormalization against the slightly
different $\mathsf{M}{}_{[k]}^{(1)}$ in the many short optimizations
should approximately have the same effect as the corresponding orthonormalization
against $\mathsf{M}{}_{[k]}^{(1)}$ in the one long optimization,
which we like to mimic. Therefore, the only extra piece of information
we have to transmit from one short optimization to the next is the
lastly generated basis tensor $\mathsf{M}{}_{[k]}^{(n)}$, which we
denote as $\mathsf{M}{}_{[k]}^{(\textrm{last})}$. 

With that, we suggest the following improved rules to generate the
tensors~$\mathsf{M}{}_{[k]}^{(l)}$
\begin{eqnarray}
\tilde{\mathsf{M}}{}_{[k]}^{(1)} & \overset{\eqref{eq:Alter tensor recyle}}{=} & \mathsf{M}^{\textrm{old}}{}_{[k]}\nonumber \\
\tilde{\mathsf{M}}{}_{[k]}^{(2)} & \overset{\textrm{new}}{=} & \mathsf{M}{}_{[k]}^{(\textrm{last})}\nonumber \\
\tilde{\mathsf{M}}{}_{[k]}^{(l>2)} & \overset{\eqref{eq:Tensor M_l>1  tadm}}{=} & \bigl(\prod_{j\neq k}\mathsf{M}_{[j]}\bigr)^{\dagger}\bigl(\mathfrak{C}\varrho_{0}-\sum_{p=1}^{l-1}\alpha_{p}\mathfrak{C}^{2}\mathcal{M}_{p}\bigr).\label{eq:Overarching iteration tadm}
\end{eqnarray}
In comparison to the one long optimization, the concatenated short
optimizations need to do a few extra calculations to patch the different
optimizations together. But the more important comparison is not the
short optimization versus the long one, but the new method presented
in this section versus the old method presented in appendix~\ref{sub:Advisable-modification-of the algorithm tadm}. 

In the worse case scenario, the basis tensor $\tilde{\mathsf{M}}{}_{[k]}^{(2)}=\mathsf{M}{}_{[k]}^{(\textrm{last})}$
has an overlap $\alpha_{2}=0$ (\ref{eq:Alpha t.a.d.m}), i.e.,\ the
basis tensor $\tilde{\mathsf{M}}{}_{[k]}^{(2)}$ of the new method
is completely useless. Under this condition, every further basis tensor
$\tilde{\mathsf{M}}{}_{[k]}^{(l>2)}$ produced by the new method will
be identical to the basis tensor $\tilde{\mathsf{M}}{}_{[k]}^{(l-1>1)}$
in the old method. That is, the maximal ``damage'' in the worse
case scenario is that we have effectively one basis tensor less.

We explained the advantages of the new method for the idealized scenario
that the tensors $\mathsf{M}{}_{[j]}$ do not change at all, but it
should be clear that also for slightly varying $\mathsf{M}{}_{[j]}$,
a positive residual effect remains. The less the tensors $\mathsf{M}{}_{[j]}$
change from one optimization cycle to the next, the better are the
results we can expect. Therefore, we might use the old method described
in appendix~\ref{sub:Advisable-modification-of the algorithm tadm}
as long as we detect strong changes in the $\mathsf{M}{}_{[j]}$.
When these changes drop below a preset threshold value, we might change
to the new method presented here.

So far, our main argument for the new iteration Eq.~(\ref{eq:Overarching iteration tadm})
has been the overarching orthonormalization. A much more trivial point
might also be that the old iteration schema without the new definition
for $\tilde{\mathsf{M}}{}_{[k]}^{(2)}$ (\ref{eq:Overarching iteration tadm})
runs a certain risk generating each optimization cycle some basis
tensors $\mathsf{M}{}_{[k]}^{(l)}$ which are very much alike the
$\mathsf{M}{}_{[k]}^{(l)}$ of the round before. This is likely to
happen if the changes in the tensor $\mathsf{M}{}_{[k]}$ per optimization
cycle are only small compared to the changes which are necessary to
reach the optimal tensor $\mathsf{M}_{[k]}^{\textrm{optimal}}$. That
is, especially when the basis tensors $\mathsf{M}{}_{[k]}^{(l)}$
prove to be badly chosen, the probability is high that these bad basis
tensors are reproduced to a great part in the next round. The new
definition for $\tilde{\mathsf{M}}{}_{[k]}^{(2)}$ breaks this vicious
cycle.

We like to finish with a practical advice: The prerequisite for the
overarching orthonormalization to work is that all tensors only change
slightly from one optimization sweep to another. We also have to take
care not to introduce any changes gauging the MPO, as described in
appendix~\ref{sub:Gauging-the-MPO tadm}. That is, we have to use
the adequate gauge for $\mathsf{M}{}_{[k]}^{(\textrm{last})}$, as
well. Now, we find that after several optimization rounds the gauge
becomes approximately statical, as well and only changes slightly
from sweep to sweep. Therefore, one might also keep the old gauge
of $\mathsf{M}{}_{[k]}^{(\textrm{last})}$ \--- at least theoretically.
Unfortunately, we learned that for the QR-decomposition, some software
libraries take care that the diagonal elements of the upper triangle
matrix $R$ has only positive diagonal elements, while other libraries
do not. In case of combined ascending and descending optimization
sweeps, gauging with negative diagonal elements can induce alternating
signs of some tensor elements, wrecking the entire procedure, if the
gauge for $\mathsf{M}{}_{[k]}^{(\textrm{last})}$ is not adapted.

\subsection{Comparison with other problems}

At the beginning of appendix~\ref{sub:Advisable-modification-of the algorithm tadm},
we shortly compared the approach for the time averaged density matrices
with the Krylov subspace based MPO optimization for ground states
(\ref{eq:Beispiel Grund state tadm}). Here, we refer again to the
example of the ground state search to obtain a better understanding
of the ingredients which are necessary for a successful application
of the overarching orthonormalization method.

Usually, the MPO ground state search consists of many optimization
sweeps, where for each tensor optimization, we build up a small Krylov
subspace (\ref{eq:Krylov f=0000FCr Grundzustand tadm}), as well.
Hence, we can also aim for an orthonormalization which overarches
many optimization cycles. For the ground state search, this can be
obtained with slight modifications, i.e., we need to transmit the
last two Krylov subspace basis (see appendix~\ref{sub:Orthogonality of the M matricen tadm}).
But unfortunately, this will not help us to improve the algorithm.

The important difference between the optimization of the ground state
and the TADM is founded in the way the equation $\mathsf{M}{}_{[k]}^{\textrm{optimized}}=\sum_{l}\alpha_{l}\mathsf{M}{}_{[k]}^{(l)}$
(\ref{eq:MPO tensors locally bad optimized tadm}) is executed. Except
for the choice of the symbols and their interpretation, the ground
state search uses the same type of equation to find the optimal solution.
The crucial point is that for the ground state search, the calculation
of the optimal coefficients $\alpha_{l}$ (\ref{eq:Alpha t.a.d.m})
and the execution of the summation can only be done at the very end,
when all $\mathsf{M}{}_{[k]}^{(l)}$ are known. Another way to say
this is to state that the optimal value of $\alpha_{l}$ might depend
on some $\mathsf{M}{}_{[k]}^{(p)}$ which are calculated much later.
For the TADM on the other hand, the optimal $\alpha_{l}$ can be calculated
directly after a new $\mathsf{M}{}_{[k]}^{(l)}$ has been generated.
This allows us to sum up the $\alpha_{l}\mathsf{M}{}_{[k]}^{(l)}$
components to a partial sum, immediately after they have been computed.
That is, soon after the $\mathsf{M}{}_{[k]}^{(l)}$ have been generated,
we can forget them completely.

Principally, it is a solvable problem to memorize all the $\mathsf{M}{}_{[k]}^{(l)}$
for the ground state search. But then we also need a strategy which
takes into account that the environmental tensors $\mathsf{M}{}_{[j\neq k]}$
are not really constant. As a consequence, calculations done with
the $\mathsf{M}{}_{[k]}^{(l)}$ become increasingly imprecise when
they get older. Without going into further details, we state that
for the ground state search, these problems seem to eat up most of
the advantages one could hope to gain. 

In conclusion, we find that the overarching orthonormalization method
appears to be quite specific for the problem at hand, i.e.,\ a linear
problem with a bilinear side condition (weighted norm (\ref{eq:Weighted norm for HM})).

\section{Mapping Hermitian matrices onto real matrices in the MPO framework\label{sec:Mapping-hermitian-matrices onto real tADM}}

Any density matrix is Hermitian, which entails a redundant encoding.
In this section, we show how this redundancy can be exploited by mapping
the complex density matrix onto a real matrix, which contains the
same amount of information but without the Hermitian redundancy. We
remark that this mapping is not suitable for the double MPS ansatz
(appendix~\ref{sec:Double-MPS tadm}), but in the MPO framework,
this mapping can be performed efficiently and allows us to obtain
an algorithm which is entirely based on real numbers and hence, runs
faster. 

Exploiting symmetries to obtain a faster algorithm is a quite common
approach. Taking advantage of the Hermitian symmetry is nonetheless
unusual, since most symmetries are based on special properties of
the physical system, while the Hermitian symmetry is universal and
based on the mathematical formalism of quantum mechanics. We are not
aware if a similar approach for MPO has ever been presented in the
literature. Since the symmetry is universal, the corresponding algorithm
can be applied to \emph{all }physical systems.

Hermitian matrices are ubiquitous in quantum mechanics and it might
surprise that they are not exploited more often in numerical algorithms.
One reason why it is difficult to take advantage of the Hermitian
symmetry is that there are no universal matrices $A$, $B$ which
could turn each Hermitian matrix $M$ with the correct dimensions
reversible into a real matrix $M_{\textrm{real}}$ 
\begin{equation}
AMB\overset{?}{=}M_{\textrm{real}}.
\end{equation}
For a matrix $M=\sum_{jk}m_{jk}|j\rangle\langle k|$, hermiticity
$m_{jk}=m_{kj}^{*}$ is a combined property of the bra and ket vector
$|j\rangle$ and $\langle k|$, while the matrices $A$, $B$ each
only ``know'' either of them, i.e.\ bra or ket. To turn a Hermitian
matrix into a real matrix, we need a linear map $\mathfrak{U}$ which
receives the combined information of $|j\rangle\langle k|$ as input.
In this context, it is helpful to vectorize all matrices 
\begin{equation}
M=\sum_{jk}m_{jk}|j\rangle\langle k|\rightarrow\sum_{j,k}m_{jk}|j,k\rangle,
\end{equation}
which in turn allows to write any linear map in form of a matrix $\sum_{jklm}s_{\left(jk\right)\left(lm\right)}|j,k\rangle\langle l,m|$.
Now, a suitable map $\mathfrak{U}$ to turn a Hermitian matrix into
a real matrix is given by 
\begin{eqnarray}
\mathfrak{U} & = & \sum_{j}|j,j\rangle\langle j,j|\nonumber \\
 &  & +\frac{1}{\sqrt{2}}\sum_{j>k}\Bigl[|j,k\rangle\bigl(\langle j,k|+\langle k,j|\bigr)\nonumber \\
 &  & \qquad\quad\;+i|k,j\rangle\bigl(\langle j,k|-\langle k,j|\bigr)\Bigr],\label{eq:Def U hermit real tadm}
\end{eqnarray}
with $i=\sqrt{-1}.$ The factor $\frac{1}{\sqrt{2}}$ was inserted
to ensure 
\begin{equation}
\mathfrak{U}^{\dagger}\mathfrak{U}={\mathbbm1}:=\sum_{j,k}|j,k\rangle\langle j,k|,\label{eq:Identity UU hermit real tadm}
\end{equation}
where $\mathfrak{U}^{\dagger}$ is given by $\left(u_{\left(jk\right)\left(lm\right)}|j,k\rangle\langle l,m|\right)^{\dagger}=u_{\left(jk\right)\left(lm\right)}^{*}|l,m\rangle\langle j,k|$,
i.e.,\ to obtain the Hermitian conjugate, $\mathfrak{U}$ is treated
as a matrix.

Up to now, we just remarked that any density matrix is Hermitian.
Since we search for a $M$ with $\bar{\varrho}=\varrho_{0}-c\mathfrak{C}M$
(\ref{eq:=0000DCbersichtsformel Time averaged d.m.}), the term $c\mathfrak{C}M$
has to be Hermitian, as well. Any phase factor $e^{i\phi}$ in $c=|c|e^{i\phi}$
can be absorbed into $M$, which allows us to demand that $c\in\mathbb{R}$.
With that, $M$ has to be antihermitian $M=-M^{\dagger}$ to have
a Hermitian $c\mathfrak{C}M=\left(c\mathfrak{C}M\right)^{\dagger}$.
Since we prefer $M$ to be Hermitian, we include an extra factor $i=\sqrt{-1}$
into Eq.~(\ref{eq:=0000DCbersichtsformel Time averaged d.m.}), i.e.,
we now use the approach 
\begin{equation}
\bar{\varrho}=\varrho_{0}-ci\mathfrak{C}M.\label{eq:Imag Formel tadm}
\end{equation}
Multiplying this equation from the left with $\mathfrak{U}$ (\ref{eq:Def U hermit real tadm})
and inserting the identity $\mathfrak{U}^{\dagger}\mathfrak{U}={\mathbbm1}$
(\ref{eq:Identity UU hermit real tadm}), we obtain the real equation
\begin{eqnarray}
\underbrace{\mathfrak{U}\bar{\varrho}}_{\bar{\varrho}^{\textrm{real}}} & = & \underbrace{\mathfrak{U}\varrho_{0}}_{\varrho_{0}^{\textrm{real}}}-c\underbrace{\mathfrak{U}i\mathfrak{C}\mathfrak{U}^{\dagger}}_{\mathfrak{C}^{\textrm{real}}}\underbrace{\mathfrak{U}M}_{M^{\textrm{real}}}\nonumber \\
\bar{\varrho}^{\textrm{real}} & = & \varrho_{0}^{\textrm{real}}-c\mathfrak{C}^{\textrm{real}}M^{\textrm{real}},\label{eq:Reele Formel tadm}
\end{eqnarray}
with $c=\left\langle i\mathfrak{C}M|\varrho_{0}\right\rangle =\left\langle \mathfrak{C}^{\textrm{real}}M^{\textrm{real}}|\varrho_{0}^{\textrm{real}}\right\rangle \in\mathbb{R}$
and the side condition
\begin{equation}
\bigl\langle\mathfrak{C}^{\textrm{real}}M^{\textrm{real}}\big|\mathfrak{C}^{\textrm{real}}M^{\textrm{real}}\bigr\rangle=1.
\end{equation}
Since $\bigl\langle\mathfrak{C}^{\textrm{real}}M^{\textrm{real}}\big|\mathfrak{C}^{\textrm{real}}M^{\textrm{real}}\bigr\rangle=\bigl\langle\mathfrak{C}M\big|\mathfrak{C}M\bigr\rangle,$
this is exactly the same side condition we used all the time $\eqref{eq:Norm Dichte Matrix}$. 

So far, we just denoted $\mathfrak{C}^{\textrm{real}}$ as a real-valued
map, but have not proved it. The map $\mathfrak{U}$ (\ref{eq:Def U hermit real tadm})
was constructed such that it maps Hermitian matrices onto real matrices,
but it is not evident that this entails $\mathfrak{C}^{\textrm{real}}=\mathfrak{U}i\mathfrak{C}\mathfrak{U}^{\dagger}$
to be real, as well. One could confirm this either via a detailed
component by component check or simply by noticing that $\mathfrak{C}^{\textrm{real}}$
maps \emph{arbitrary }real matrices onto real matrices and hence cannot
contain any imaginary elements. Still, we have to come back to this
point in appendix~\ref{sub:Real-MPO-with complex tensors tadm},
when we look at the MPO structure of $\mathfrak{C}^{\textrm{real}}$.

Now, we like to have a look at what we have found. The main idea of
the entire transformation was to have a faster algorithm. Any (linear)
map acting on $n\times n$-matrices corresponds to a $n\times n\times n\times n$-tensor.
If we were relying on standard matrix and tensor multiplication, using
such a huge tensor would be highly questionable. For MPO calculations
on the other hand, the physical dimensions are often of secondary
importance. The decisive characteristic is the bond dimension. In
this context, it is relevant to note that for maps $\mathfrak{U}_{i}$
which map Hermitian matrices onto real matrices, the outer product
\begin{equation}
\mbox{\ensuremath{\mathfrak{U}}}_{\otimes}=\bigotimes_{i=1}^{n}\mbox{\ensuremath{\mathfrak{U}}}_{i}=\mbox{\ensuremath{\mathfrak{U}}}_{1}\otimes\mbox{\ensuremath{\mathfrak{U}}}_{2}\otimes\mbox{\ensuremath{\mathfrak{U}}}_{3}\otimes\ldots\otimes\mbox{\ensuremath{\mathfrak{U}}}_{n}\label{eq:MPO hermit to real tadm}
\end{equation}
describes a mapping from Hermitian matrices to real matrices, as well,
with $\mbox{\ensuremath{\mathfrak{U}}}_{\otimes}^{\dagger}\mbox{\ensuremath{\mathfrak{U}}}_{\otimes}={\mathbbm1}$
(\ref{eq:Identity UU hermit real tadm}). This is easily checked applying
$\mbox{\ensuremath{\mathfrak{U}}}_{\otimes}$ to a suitable base consisting
of outer products $\bigotimes_{i}H_{i}$ of Hermitian matrices $H_{i}$.
The structure of $\mbox{\ensuremath{\mathfrak{U}}}_{\otimes}$ corresponds
to a trivial MPO with the bond dimension being one.

Further, we need to provide a single MPO representing the commutator
operator $\mathfrak{C}$ to perform the mapping $\mathfrak{C}^{\textrm{real}}=\mathfrak{U}i\mathfrak{C}\mathfrak{U}^{\dagger}$.
We cannot use the definition of the commutator operator $\mathfrak{C}=[H,\ldots]$
(\ref{eq:Kommutatorzeiche C def t.a.d.m.}), since the mapping cannot
be decomposed accordingly. To see this, remember that for a mapping
like $H_{\textrm{real}}=\mathfrak{U}H$, the matrix $H$ has to be
vectorized, i.e.,  $\mathfrak{U}$ acts on the bra \emph{and} ket
side. Vectorized matrices $\mathfrak{U}H$ and $\mathfrak{U}M$ do
not allow a standard matrix multiplication $\mathfrak{U}H\mathfrak{U}M$.
If we rewrite $\mathfrak{U}H$ and $\mathfrak{U}M$ as matrices, the
resulting matrix product is no longer the correct multiplication needed
for $\mathfrak{U}^{\dagger}\mathfrak{U}={\mathbbm1}$ to hold.

Many frequently used Hamiltonians $H$  possess relatively simple
MPO descriptions, with bond dimensions which are small compared to
the bond dimensions needed to obtain suitable MPO descriptions for
$M$ (\ref{eq:Imag Formel tadm}). In this case, it is reasonable
to write $\mathfrak{C}$ as a single MPO  as described in appendix~\ref{sec:Constructing-a-MPO for C TADM}
instead of using the definition $\mathfrak{C}M=HM-MH$ (\ref{eq:Kommutatorzeiche C def t.a.d.m.}).
As a bonus, this enables us to use a MPO compression algorithm to
pre-compute $\mathfrak{C}^{2}$, which is needed to calculate$\left\langle \mathfrak{C}\mathcal{M}_{j}|\mathfrak{C}\mathcal{M}_{k}\right\rangle $
(\ref{eq:Weighted norm for HM}). Compared to the explicit use of
$\left\langle \left[H,\mathcal{M}_{j}\right]|\left[H,\mathcal{M}_{k}\right]\right\rangle $,
this often entails a speed up.

Since the Hermitian to real mapping MPO $\mbox{\ensuremath{\mathfrak{U}}}_{\otimes}$
(\ref{eq:MPO hermit to real tadm}) has the trivial bond dimension
one, the real-valued MPO $\varrho_{0}^{\textrm{real}}=\mbox{\ensuremath{\mathfrak{U}}}_{\otimes}\varrho_{0}$,
$\mathfrak{C}^{\textrm{real}}=\mbox{\ensuremath{\mathfrak{U}}}_{\otimes}i\mathfrak{C}\mbox{\ensuremath{\mathfrak{U}}}_{\otimes}^{\dagger}$
and $M^{\textrm{real}}=\mbox{\ensuremath{\mathfrak{U}}}_{\otimes}M$
(\ref{eq:Reele Formel tadm}) have the same bond dimensions as their
Hermitian counterparts. Further, we remark that the mappings $\varrho_{0}\rightarrow\varrho_{0}^{\textrm{real}}$
and $\mathfrak{C}\rightarrow\mathfrak{C}^{\textrm{real}}$ only have
to be applied once, at the beginning. Afterwards, we can compute $M^{\textrm{real}}$
with the same algorithm we would have used to obtain the Hermitian
$M$. At the very end, when $M^{\textrm{real}}$ is calculated, one
 final mapping gives us $M=\mbox{\ensuremath{\mathfrak{U}}}_{\otimes}^{\dagger}M^{\textrm{real}}$.

\subsection{Real MPO with complex MPO tensors\label{sub:Real-MPO-with complex tensors tadm}}

When we stated that $\varrho_{0}^{\textrm{real}}$ and $\mathfrak{C}^{\textrm{real}}$
are real-valued objects, we indirectly included the assumption that
they are described by a single matrix or tensor. If we represent $\varrho_{0}^{\textrm{real}}$
and $\mathfrak{C}^{\textrm{real}}$ as MPO, they are decomposed into
a product of tensors. These MPO tensors no longer have to be real.
Usually, the simple transformation $\mathfrak{C}^{\textrm{real}}=\mbox{\ensuremath{\mathfrak{U}}}_{\otimes}i\mathfrak{C}\mbox{\ensuremath{\mathfrak{U}}}_{\otimes}^{\dagger}$
(\ref{eq:Reele Formel tadm}) produces complex-valued MPO tensors.
In this subsection, we describe a procedure to turn the complex-valued
MPO tensors of $\varrho_{0}^{\textrm{real}}$ and $\mathfrak{C}^{\textrm{real}}$
into real-valued tensors.

In the following, we assume that we deal with open boundary conditions
for the MPO. For most physical systems of interest, it should be no
problem to find MPO with open boundary conditions for $\varrho_{0}^{\textrm{real}}$
and $\mathfrak{C}^{\textrm{real}}$. Under certain circumstances,
periodic boundary conditions might be advisable for the MPO $M^{\textrm{real}}$
(\ref{eq:Reele Formel tadm}). But $M^{\textrm{real}}$ is generated
by the algorithm and not the result of a transformation. Therefore,
the MPO tensors of $M^{\textrm{real}}$ are real by construction. 

Let us look at an arbitrary real operator $\hat{O}$ represented as
MPO in its left-canonical form 
\begin{equation}
\hat{O}=\sum_{\alpha_{1}\ldots\alpha_{n-1}}\mathsf{U}_{[1]\sigma_{1}}^{\alpha_{1}}\mathsf{U}_{[2]\sigma_{2}}^{\alpha_{1}\alpha_{2}}\cdot\ldots\cdot\mathsf{U}_{[n-1]\sigma_{n-1}}^{\alpha_{n-2}\alpha_{n-1}}\mathsf{R}_{\sigma_{n}}^{\alpha_{n-1}}.\label{eq:Beispiel Left canonical MPO tadm}
\end{equation}
Here, the $\sigma_{j}$ are multi-indices comprising all physical
indices of the MPO tensors (i.e.,  in case of the map $\mathfrak{C}^{\textrm{real}}$,
the MPO tensors have four physical indices). For an MPO in a left-canonical
form, all MPO tensors $\mathsf{U}_{[j]}$ are left-normalized except
for the rightmost tensor $\mathsf{R}$, i.e.,
\begin{equation}
\sum_{\sigma_{j},\alpha_{j-1}}\bigl(\mathsf{U}_{[j]\sigma_{j}}^{\alpha_{j-1}\alpha_{j}}\bigr)^{*}\mathsf{U}_{[j]\sigma_{j}}^{\alpha_{j-1}\alpha'_{j}}={\mathbbm1}^{\alpha_{j}\alpha'_{j}},\label{eq:Left canonical in Beweis tadm}
\end{equation}
respectively $\sum_{\sigma_{1}}\bigl(\mathsf{U}_{[1]\sigma_{1}}^{\alpha_{1}}\bigr)^{*}\mathsf{U}_{[1]\sigma_{1}}^{\alpha'_{1}}={\mathbbm1}^{\alpha_{1}\alpha'_{1}}$.
Any MPO can be brought into the left-canonical form via a repeated
application of a singular value or QR decomposition, starting with
the leftmost tensor. For details, see e.g.\ Sec.~4.4 of Ref.~\cite{Schollwoeck2011}.

Using unitary matrices $V_{[j]}$, we can construct new tensors $\mathsf{O}_{[j]}$
\begin{equation}
\mathsf{O}_{[j]\sigma_{j}}^{\alpha{}_{j-1}\alpha_{j}}=\sum_{\beta\gamma}\left(V_{[j-1]}^{\beta\alpha{}_{j-1}}\right)^{*}\mathsf{U}_{[j]\sigma_{j}}^{\beta\gamma}V_{[j]}^{\gamma\alpha_{j}},\label{eq:U-O-V relation tasdm}
\end{equation}
respectively $\mathsf{O}_{[1]\sigma_{1}}^{\alpha{}_{1}}=\sum_{\beta}\mathsf{U}_{[1]\sigma_{1}}^{\beta}V_{[1]}^{\beta\alpha_{1}}$
and $\mathsf{P}_{\sigma_{n}}^{\alpha{}_{n-1}}=\sum_{\beta}\left(V_{[n-1]}^{\beta\alpha{}_{n-1}}\right)^{*}\mathsf{R}_{\sigma_{n}}^{\beta}$.
With these new tensors, an alternative MPO representation for the
operator $\hat{O}$ is given by
\begin{equation}
\hat{O}=\sum_{\alpha_{1}\ldots\alpha_{n-1}}\mathsf{O}_{[1]\sigma_{1}}^{\alpha_{1}}\mathsf{O}_{[2]\sigma_{2}}^{\alpha_{1}\alpha_{2}}\cdot\ldots\cdot\mathsf{O}_{[n-1]\sigma_{n-1}}^{\alpha_{n-2}\alpha_{n-1}}\mathsf{P}_{\sigma_{n}}^{\alpha_{n-1}}.
\end{equation}
In appendix~\ref{sub:Existens-of-the matricees V tadm}, we will
prove the existence of unitary matrices $V_{[j]}$ such that all MPO
tensors $\mathsf{O}_{[j]}$ and $\mathsf{P}$ are real valued. Interestingly,
we have to demand that the MPO representation (\ref{eq:Beispiel Left canonical MPO tadm})
is maximally compressed to ensure the existence of suitable $V_{[j]}$.
That is, we do not allow MPO dimensions which belong to vanishing
singular values.

\subsubsection{Finding the gauge matrices $V_{[j]}$\label{sub:Finding-the-matrices V_j tadm}}

Once the existence of the unitary matrices $V_{[j]}$ is guaranteed,
calculating them is relatively easy. We start with $V_{[1]}$ and
note that due to the unitarity of the matrices $V_{[j]}$, we have
\begin{equation}
\mathsf{O}_{[1]\sigma_{1}}^{\alpha{}_{1}}\overset{\eqref{eq:U-O-V relation tasdm}}{=}\sum_{\beta}\mathsf{U}_{[1]\sigma_{1}}^{\beta}V_{[1]}^{\beta\alpha_{1}}\Leftrightarrow\mathsf{U}_{[1]\sigma_{1}}^{\alpha{}_{1}}=\sum_{\beta}\mathsf{O}_{[1]\sigma_{1}}^{\beta}V_{[1]}^{*\ \alpha_{1}\beta}.\label{eq:O1 spezial TADM}
\end{equation}
With this, we find 
\begin{align}
\sum_{\alpha_{1}}\mathsf{U}_{[1]\sigma_{1}}^{\alpha{}_{1}}\mathsf{U}_{[1]\sigma'_{1}}^{*\ \alpha{}_{1}} & =\sum_{\alpha_{1},\beta,\gamma}\mathsf{O}_{[1]\sigma_{1}}^{\beta}\underbrace{V_{[1]}^{*\ \alpha_{1}\beta}V_{[1]}^{\alpha_{1}\gamma}}_{\delta^{\beta\gamma}}\mathsf{O}_{[1]\sigma'_{1}}^{*\ \gamma}\nonumber \\
 & =\sum_{\beta}\mathsf{O}_{[1]\sigma_{1}}^{\beta}\mathsf{O}_{[1]\sigma'_{1}}^{*\ \beta}\nonumber \\
 & =:\mathsf{W}_{[1]\sigma_{1}\sigma_{1}'}.\label{eq:Def W=00003DOO tadm}
\end{align}
We remark that $\sum_{\sigma}\mathsf{O}_{[1]\sigma}^{\beta}\mathsf{O}_{[1]\sigma}^{*\ \gamma}={\mathbbm1}^{\beta\gamma},$
while $\mathsf{W}_{[1]\sigma\sigma'}=\sum_{\beta}\mathsf{O}_{[1]\sigma}^{\beta}\mathsf{O}_{[1]\sigma'}^{*\ \beta}$
only equals ${\mathbbm1}_{\sigma\sigma'}$ iff $\dim\left(\sigma\right)=\dim\left(\beta\right)$.
For real objects as $\varrho_{0}^{\textrm{real}}$ and $\mathfrak{C}^{\textrm{real}}$,
we know that a decomposition into real MPO tensors exists. In this
case, $\mathsf{W}_{[1]\sigma\sigma'}$ has to be real, as well. 

Remember that the tensor $\mathsf{O}_{[1]s}^{\beta}$ is still unknown,
while $\mathsf{W}_{[1]\sigma\sigma'}=\sum_{\alpha}\mathsf{U}_{[1]\sigma}^{\alpha}\mathsf{U}_{[1]\sigma'}^{*\ \alpha}$
can be calculated. Since $\mathsf{W}_{[1]}$ (\ref{eq:Def W=00003DOO tadm})
can be written as matrix equation 
\begin{equation}
\mathsf{W}_{[1]}=\mathsf{O}_{[1]}{\mathbbm1}\mathsf{O}_{[1]}^{\dagger},\label{eq:W-SVD tadm}
\end{equation}
we can obtain $\mathsf{O}_{[1]}$ as the eigenvectors of $\mathsf{W}_{[1]}$
(with all eigenvalues being one) or alternatively, via a singular
value decomposition. The matrix $\mathsf{O}_{[1]}$ is not unique,
but the important part is that it is always real-valued, in case $\mathsf{W}_{[1]\sigma\sigma'}$
is real, as it is the case for $\varrho_{0}^{\textrm{real}}$ and
$\mathfrak{C}^{\textrm{real}}$.

Having $\mathsf{O}_{[1]}$, we can calculate the matrix $V_{[1]}$
via 
\begin{eqnarray}
\sum_{s}\mathsf{O}_{[1]\sigma}^{*\ \alpha}\mathsf{U}_{[1]\sigma}^{\beta} & \overset{\eqref{eq:U-O-V relation tasdm}}{=} & \sum_{\beta,s}\underbrace{\mathsf{O}_{[1]\sigma}^{*\ \alpha}\mathsf{O}_{[1]\sigma}^{\gamma}}_{\delta^{\alpha\gamma}}V_{[1]}^{*\ \beta\gamma}\nonumber \\
 & = & V_{[1]}^{*\ \beta\alpha}.\label{eq:Eich-V aus OU tadm}
\end{eqnarray}
With some slight adjustments, we can use the same technique to calculate
the matrix $V_{[2]}$ and successively all following matrices $V_{[j]}$.
Instead of Eq.~(\ref{eq:O1 spezial TADM}), we now have
\begin{equation}
\sum_{\alpha}V_{[j-1]}^{*\ \alpha\beta}\mathsf{U}_{[j]\sigma}^{\alpha\gamma}\overset{\eqref{eq:U-O-V relation tasdm}}{=}\sum_{\delta}\mathsf{O}_{[j]\sigma}^{\beta\delta}V_{[j]}^{*\ \gamma\delta},\label{eq:U-O-Beziehung tadm}
\end{equation}
where we assume that $V_{[j-1]}$ is already known. To facilitate
the notation, we introduce 
\begin{equation}
\mathsf{Q}_{[j]\mathfrak{S}}^{\gamma}=\mathsf{Q}_{[j](\sigma,\beta)}^{\gamma}:=\sum_{\alpha}V_{[j-1]}^{*\ \alpha\beta}\mathsf{U}_{[j]\sigma}^{\alpha\gamma},
\end{equation}
with the multi-index $\mathfrak{S}=(\sigma,\beta)$. Replacing $\mathsf{U}_{[1]\sigma}^{\gamma}$
by $\mathsf{Q}_{[j]\mathfrak{S}}^{\gamma}$, we can repeat all the
steps above to obtain $V_{[j]}$. In short, we calculate $\mathsf{W}_{[j]\mathfrak{S}\mathfrak{S}'}=\sum_{\gamma}\mathsf{Q}_{[j]\mathfrak{S}}^{\gamma}\mathsf{Q}_{[j]\mathfrak{S}'}^{\gamma}$
(\ref{eq:Def W=00003DOO tadm}) and via a singular value decomposition
of $\mathsf{W}_{[j]\mathfrak{S}\mathfrak{S}'}$, we obtain $\mathsf{O}_{[j]\mathfrak{S}}^{\alpha}$
(\ref{eq:W-SVD tadm}), which leads to $V_{[j]}^{*\ \beta\alpha}=\sum_{\mathfrak{S}}\mathsf{O}_{[j]\mathfrak{S}}^{*\ \alpha}\mathsf{Q}_{[j]\mathfrak{S}}^{\beta}$
(\ref{eq:Eich-V aus OU tadm}).

\section{Existence of the gauge matrices $V_{[j]}$\label{sub:Existens-of-the matricees V tadm}}

In this section, we prove the existent of the unitary matrices $V_{[j]}$,
which we used in the last section (appendix~\ref{sub:Finding-the-matrices V_j tadm})
to transform the complex-valued tensors $\mathsf{U}_{[j]}$ into real-valued
tensors $\mathsf{O}_{[j]}$ (\ref{eq:U-O-Beziehung tadm}). This proof
is added for formal reasons only and is of no importance for the practical
application of the algorithm.

MPO tensors are not uniquely defined. We look at the case where we
have two different MPO which represent the same object $\hat{O}$.
\begin{eqnarray}
\hat{O} & = & \sum_{\alpha_{1}\ldots\alpha_{n-1}}\mathsf{U}_{[1]\sigma_{1}}^{\alpha_{1}}\cdot\ldots\cdot\mathsf{U}_{[n-1]\sigma_{n-1}}^{\alpha_{n-2}\alpha_{n-1}}\mathsf{R}_{\sigma_{n}}^{\alpha_{n-1}}\nonumber \\
 & = & \sum_{\alpha_{1}\ldots\alpha_{n-1}}\mathsf{O}_{[1]\sigma_{1}}^{\alpha_{1}}\cdot\ldots\cdot\mathsf{O}_{[n-1]\sigma_{n-1}}^{\alpha_{n-2}\alpha_{n-1}}\mathsf{P}_{\sigma_{n}}^{\alpha_{n-1}}.\label{eq:Zwei gleiche MPO tadm}
\end{eqnarray}
Both MPO are supposed to be maximally compressed and in the left-canonical
form (\ref{eq:Left canonical in Beweis tadm}). We like to show that
for these two MPO, all tensors $\mathsf{U}_{[j]}$ and $\mathsf{O}_{[j]}$
(respectively $\mathsf{R}$ and $\mathsf{P}$) can always be related
by unitary matrices $V_{[k]}$, as in Eq.~(\ref{eq:U-O-V relation tasdm}). 

For the upcoming proof, we need to shorten the notation. To this end,
we use the Einstein summation convention, i.e., we imply summation
over identical indices. Further, we introduce the two matrices $\boldsymbol{U}_{[j]}^{\mathfrak{S}_{j}\alpha_{j}}$
and $\boldsymbol{O}_{[j]}^{\mathfrak{S}_{j}\alpha_{j}}$, given as
tensor products of the first $j$ MPO tensors $\mathsf{U}_{[k]}$
respectively $\mathsf{O}_{[k]}$ 
\begin{eqnarray}
\boldsymbol{U}_{[j]}^{\mathfrak{S}_{j}\alpha_{j}} & = & \boldsymbol{U}_{[j-1]}^{\mathfrak{S}_{j-1}\alpha_{j-1}}\mathsf{U}_{[j]\sigma_{j}}^{\alpha_{j-1}\alpha_{j}}\nonumber \\
 & = & \mathsf{U}_{[1]\sigma_{1}}^{\alpha_{1}}\cdot\ldots\cdot\mathsf{U}_{[j]\sigma_{j}}^{\alpha_{j-1}\alpha_{j}}\label{eq:Def U short tadm}
\end{eqnarray}
and
\begin{eqnarray}
\boldsymbol{O}_{[j]}^{\mathfrak{S}_{j}\alpha_{j}} & = & \boldsymbol{O}_{[j-1]}^{\mathfrak{S}_{j-1}\alpha_{j-1}}\mathsf{O}_{[j]\sigma_{j}}^{\alpha_{j-1}\alpha_{j}}\nonumber \\
 & = & \mathsf{O}_{[1]\sigma_{1}}^{\alpha_{1}}\cdot\ldots\cdot\mathsf{O}_{[j]\sigma_{j}}^{\alpha_{j-1}\alpha_{j}},\label{eq:Def O short tadm}
\end{eqnarray}
with the physical multi-index $\mathfrak{S}_{j}=(\sigma_{1}\ldots\sigma_{j})$.
Since the two MPO (\ref{eq:Zwei gleiche MPO tadm}) are in left-canonical
form (\ref{eq:Left canonical in Beweis tadm}), we find 
\begin{equation}
\boldsymbol{U}_{[j]}^{\dagger}\boldsymbol{U}_{[j]}={\mathbbm1}=\boldsymbol{O}_{[j]}^{\dagger}\boldsymbol{O}_{[j]},\label{eq:UU OO eins tadm}
\end{equation}
while generally $\boldsymbol{U}_{[j]}\boldsymbol{U}_{[j]}^{\dagger}\neq{\mathbbm1}\neq\boldsymbol{O}_{[j]}\boldsymbol{O}_{[j]}^{\dagger}$.
Still, $\boldsymbol{U}_{[j]}\boldsymbol{U}_{[j]}^{\dagger}$ acts
like an identity, when applied to $\boldsymbol{U}_{[j]}$ 
\begin{equation}
\left(\boldsymbol{U}_{[j]}\boldsymbol{U}_{[j]}^{\dagger}\right)\boldsymbol{U}_{[j]}=\boldsymbol{U}_{[j]}\underbrace{\left(\boldsymbol{U}_{[j]}^{\dagger}\boldsymbol{U}_{[j]}\right)}_{{\mathbbm1}}=\boldsymbol{U}_{[j]}.
\end{equation}
With that, $\boldsymbol{U}_{[j]}\boldsymbol{U}_{[j]}^{\dagger}$ also
acts like an identity when applied to the MPO $\hat{O}$\ (\ref{eq:Zwei gleiche MPO tadm})
\begin{equation}
\boldsymbol{U}_{[j]}\boldsymbol{U}_{[j]}^{\dagger}\hat{O}=\hat{O},
\end{equation}
which is easily seen when we use the MPO representation of $\hat{O}$
based on the tensors $\mathsf{U}_{[j]}$. On the other hand, when
we apply $\boldsymbol{U}_{[j]}\boldsymbol{U}_{[j]}^{\dagger}$ to
the MPO $\hat{O}$ represented as
\begin{equation}
\hat{O}\overset{\eqref{eq:Zwei gleiche MPO tadm},\eqref{eq:Def O short tadm}}{=}\underbrace{\boldsymbol{O}_{[j]}^{\mathfrak{S}_{j}\alpha_{j}}}_{\eqref{eq:Def O short tadm}}\mathsf{O}_{[j+1]\sigma_{j+1}}^{\alpha_{j}\alpha_{j+1}}\cdot\ldots\cdot\mathsf{O}_{[n-1]\sigma_{n-1}}^{\alpha_{n-2}\alpha_{n-1}}\mathsf{P}_{\sigma_{n}}^{\alpha_{n-1}},\label{eq:OO-MPO tadm}
\end{equation}
we find
\begin{eqnarray}
\hat{O} & = & \boldsymbol{U}_{[j]}\boldsymbol{U}_{[j]}^{\dagger}\hat{O}\nonumber \\
 & \overset{\eqref{eq:OO-MPO tadm}}{=} & \boldsymbol{U}_{[j]}^{\mathfrak{S}'_{j}\gamma}\underbrace{\boldsymbol{U}_{[j]}^{*\ \mathfrak{S}_{j}\gamma}\boldsymbol{O}_{[j]}^{\mathfrak{S}_{j}\alpha_{j}}}_{=:W_{[j]}^{\gamma\alpha_{j}}}\mathsf{O}_{[j+1]\sigma_{j+1}}^{\alpha_{j}\alpha_{j+1}}\cdot\ldots\cdot\mathsf{P}_{\sigma_{n}}^{\alpha_{n-1}}\nonumber \\
 & = & \boldsymbol{U}_{[j]}^{\mathfrak{S}'_{j}\gamma}W_{[j]}^{\gamma\alpha_{j}}\mathsf{O}_{[j+1]\sigma_{j+1}}^{\alpha_{j}\alpha_{j+1}}\cdot\ldots\cdot\mathsf{O}_{[n-1]\sigma_{n-1}}^{\alpha_{n-2}\alpha_{n-1}}\mathsf{P}_{\sigma_{n}}^{\alpha_{n-1}}.\nonumber \\
 &  & \ \label{eq:Mixt-MPO-form tadm}
\end{eqnarray}
We demanded that the MPO $\hat{O}$ is maximally compressed, i.e.,
it contains no vanishing singular values. Hence, the expression $\mathcal{R}=\mathsf{O}_{[j+1]\sigma_{j+1}}^{\alpha_{j}\alpha_{j+1}}\cdot\ldots\cdot\mathsf{O}_{[n-1]\sigma_{n-1}}^{\alpha_{n-2}\alpha_{n-1}}\mathsf{P}_{\sigma_{n}}^{\alpha_{n-1}}$
built from the right-hand tensors of the MPO is invertible. Applying
this inverse $\mathcal{R}^{-1}$ to the MPO $\hat{O}$ in the form
of the last line of Eq.~(\ref{eq:Mixt-MPO-form tadm}) as well as
to the representation in Eq.~(\ref{eq:OO-MPO tadm}), we find
\begin{eqnarray}
\underbrace{\hat{O}}_{\eqref{eq:OO-MPO tadm}}\mathcal{R}^{-1} & = & \underbrace{\hat{O}}_{\eqref{eq:Mixt-MPO-form tadm}}\mathcal{R}^{-1}\nonumber \\
\boldsymbol{O}_{[j]} & = & \boldsymbol{U}_{[j]}W_{[j]}\label{eq:O gleich UW tadm}
\end{eqnarray}
Multiplying this equation with $\boldsymbol{O}_{[j]}^{\dagger}$ we
find
\begin{eqnarray}
\underbrace{\boldsymbol{O}_{[j]}^{\dagger}\boldsymbol{O}_{[j]}}_{{\mathbbm1}\ \eqref{eq:UU OO eins tadm}} & = & \underbrace{\boldsymbol{O}_{[j]}^{\dagger}\boldsymbol{U}_{[j]}}_{W_{[j]}^{\dagger}\ \eqref{eq:Mixt-MPO-form tadm}}W_{[j]}\nonumber \\
{\mathbbm1} & = & W_{[j]}^{\dagger}W_{[j]}.
\end{eqnarray}
Repeating the some line of argumentation for $\boldsymbol{O}_{[j]}\boldsymbol{O}_{[j]}^{\dagger}\hat{O}$
as we used for $\boldsymbol{U}_{[j]}\boldsymbol{U}_{[j]}^{\dagger}\hat{O}$,
we arrive at the conclusion that
\begin{eqnarray}
\boldsymbol{U}_{[j]} & = & \boldsymbol{O}_{[j]}W_{[j]}^{\dagger}\nonumber \\
W_{[j]}W_{[j]}^{\dagger} & = & {\mathbbm1}.\label{eq:U gleich OW tadm}
\end{eqnarray}
Putting all together, we find
\begin{eqnarray}
\boldsymbol{O}_{[j]}^{\mathfrak{S}_{j}\gamma} & \overset{\eqref{eq:O gleich UW tadm}}{=} & \boldsymbol{U}_{[j]}^{\mathfrak{S}_{j}\alpha_{j}}W_{[j]}^{\alpha_{j}\gamma}\nonumber \\
 & \overset{\eqref{eq:Def U short tadm}}{=} & \boldsymbol{U}_{[j-1]}^{\mathfrak{S}_{j-1}\alpha_{j-1}}\mathsf{U}_{[j]\sigma_{j}}^{\alpha_{j-1}\alpha_{j}}W_{[j]}^{\alpha_{j}\gamma}\nonumber \\
 & \overset{\eqref{eq:U gleich OW tadm}}{=} & \boldsymbol{O}_{[j-1]}^{\mathfrak{S}_{j-1}\beta}W_{[j-1]}^{*\ \alpha_{j-1}\beta}\mathsf{U}_{[j]\sigma_{j}}^{\alpha_{j-1}\alpha_{j}}W_{[j]}^{\alpha_{j}\gamma}.
\end{eqnarray}
Multiplying this equation with $\boldsymbol{O}_{[j-1]}^{*\ \mathfrak{S}_{j-1}\beta}$
and using the identity (\ref{eq:UU OO eins tadm}), we finally obtain
\begin{eqnarray}
\boldsymbol{O}_{[j-1]}^{*\ \mathfrak{S}_{j-1}\beta}\boldsymbol{O}_{[j]}^{\mathfrak{S}_{j}\gamma} & = & W_{[j-1]}^{*\ \alpha_{j-1}\beta}\mathsf{U}_{[j]\sigma_{j}}^{\alpha_{j-1}\alpha_{j}}W_{[j]}^{\alpha_{j}\gamma}\nonumber \\
\boldsymbol{O}_{[j-1]}^{*\ \mathfrak{S}_{j-1}\beta}\boldsymbol{O}_{[j-1]}^{\mathfrak{S}_{j-1}\delta}\mathsf{O}_{[j]\sigma_{j}}^{\delta\gamma} & \overset{\eqref{eq:Def O short tadm}}{=} & W_{[j-1]}^{*\ \alpha_{j-1}\beta}\mathsf{U}_{[j]\sigma_{j}}^{\alpha_{j-1}\alpha_{j}}W_{[j]}^{\alpha_{j}\gamma}\nonumber \\
\mathsf{O}_{[j]\sigma_{j}}^{\beta\gamma} & \overset{\eqref{eq:UU OO eins tadm}}{=} & W_{[j-1]}^{*\ \alpha_{j-1}\beta}\mathsf{U}_{[j]\sigma_{j}}^{\alpha_{j-1}\alpha_{j}}W_{[j]}^{\alpha_{j}\gamma}.\nonumber \\
 &  & \ \label{eq:Fertig Beweis 1 tadm}
\end{eqnarray}
In the same way, it is easily shown that
\begin{eqnarray}
\mathsf{O}_{[1]\sigma_{1}}^{\gamma} & = & \mathsf{U}_{[1]\sigma_{1}}^{\alpha_{1}}W_{[1]}^{\alpha_{1}\gamma}\nonumber \\
\mathsf{P}_{\sigma_{n}}^{\beta} & = & W_{[n-1]}^{*\ \alpha_{n-1}\beta}\mathsf{R}_{\sigma_{n}}^{\alpha_{n-1}}.\label{eq:Fertig Beweis 2 tadm}
\end{eqnarray}
Since we can be sure that for a real-valued operator $\hat{O}$ an
MPO based on real-valued tensors $\mathsf{O}_{[j]}$ and $\mathsf{P}$
exist, we can also deduce the existence of some gauge matrices $V_{[j]}=W_{[j]}$
with the help of Eq.~(\ref{eq:Fertig Beweis 1 tadm}) and $\eqref{eq:Fertig Beweis 2 tadm}$.

\section{Constructing a MPO for the commutator operator~$\mathfrak{C}$\label{sec:Constructing-a-MPO for C TADM}}

To construct a MPO representation for the commutator operator $\mathfrak{C}$,
first, we need to construct a MPO representation of the Hamilton operator
$H$. This is e.g.\ described in Ref.~\cite{Crosswhite2008_Automata,Froewis2010}. 

The commutator operator $\mathfrak{C}$ acts on the vector space of
linear operators with $\mathfrak{C}A=HA-AH$. Evidently, this can
also be written as 
\[
\mathfrak{C}A=HA{\mathbbm1}-{\mathbbm1}AH
\]
Now, let us rewrite the commutator operator symbolically as
\begin{equation}
\mathfrak{C}=H\otimes{\mathbbm1}-{\mathbbm1}\otimes H,\label{eq:C als H1-1H tadm}
\end{equation}
which is to be understood as $\left(H\otimes{\mathbbm1}\right)A=HA{\mathbbm1}$
and $\left({\mathbbm1}\otimes H\right)A={\mathbbm1}AH$. Knowing a
MPO representation of the Hamiltonian $H=\prod_{j}\mathsf{H}_{[j]}$,
we immediately obtain
\begin{eqnarray}
H\otimes{\mathbbm1} & = & \prod_{j}\mathsf{H}_{[j]\bar{s}_{j}s_{j}}^{\alpha_{j-1}\alpha_{j}}\otimes{\mathbbm1}_{[j]\bar{s}_{j}'s_{j}'}\nonumber \\
{\mathbbm1}\otimes H & = & \prod_{j}{\mathbbm1}_{[j]\bar{s}_{j}s_{j}}\otimes\mathsf{H}_{[j]\bar{s}_{j}'s_{j}'}^{\beta_{j-1}\beta_{j}},
\end{eqnarray}
where $|s_{j}\rangle$ and $\langle s'_{j}|$ are the ket and bra
components of the operator $A$.

To take care of the minus sign in the commutator, we multiply the
MPO tensor ${\mathbbm1}_{[1]}\otimes\mathsf{H}_{[1]}$ with $-1$.
Then, we simply have to add the two MPO $H\otimes{\mathbbm1}$ and
$-{\mathbbm1}\otimes H$. Adding two MPO is a standard procedure,
which is e.g.\ explained in Sec.~4.3 and 5.2 of Ref.~\cite{Schollwoeck2011}.

\section{Double MPS\label{sec:Double-MPS tadm}}

So far, we presented a general computation method for the time averaged
density matrix (TADM) and explained in detail, how this method can
be adapted for matrix product operators (MPO). MPO are just one example
for tensor networks. Here, we discuss another (non-standard) type
of tensor networks, where the TADM is obtained as a double sized matrix
product states (MPS), which we dubbed double MPS. 

\begin{figure}
\includegraphics[width=1\columnwidth]{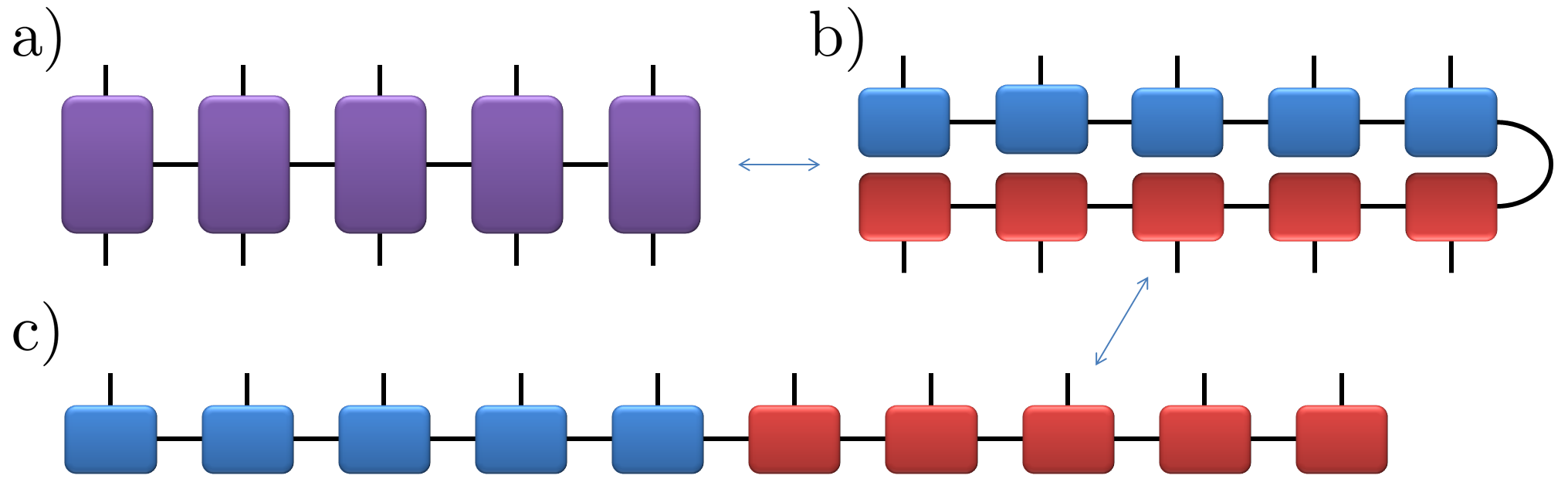}

\protect\caption{\label{fig:Double MPS tadm}a) Graphical representation of a finite
MPO with open boundary conditions. b) Any finite operator $\hat{O}=\sum_{ij}\lambda_{ij}|u_{i}\rangle\langle v_{j}|$
can be decomposed into two connected MPS, where the connecting bond
between the MPS corresponds to $\lambda_{ij}$. Due to the connecting
bond, the two MPS actually correspond to two collections of several
MPS $|u_{i}\rangle$ and $\langle v_{j}|$. c) Formally, any finite
MPO can be represented as MPS of twice the size.}
\end{figure}

Formally, a double MPS is a MPS with twice as many sites as the physical
system has components. Hereby, the first part of the double MPS represents
the ket-states $|u_{i}\rangle$ of the TADM or any other matrix $M=\sum_{ij}\lambda_{ij}\cdot|u_{i}\rangle\langle v_{j}|$,
while the second part of the double MPS represents the bra-states
$\langle v_{j}|$. The matrix $\lambda_{ij}$ is encoded into the
MPS-bond which connects the two parts, see also Fig.~\ref{fig:Double MPS tadm}.
If the double MPS is brought into a suitable canonical form \cite{Schollwoeck2011},
the basis states $|u_{i}\rangle$ and $|v_{i}\rangle$ encoded in
$M$ are orthogonal (i.e.,  $\langle u_{i}|u_{j\neq i}\rangle=0=\langle v_{i}|v_{j\neq i}\rangle$)
and we can extract the matrix $\lambda_{ij}$ from the double MPS.
This allows e.g.\ to check whether or not $M$ is a positive matrix.
Assuming that the double MPS represents a positive Hermitian matrix,
its entanglement entropy of bipartion for the half chain corresponds
to the entropy of the entire matrix $M$.

We emphasize that the need for doubling the number of tensors to accommodate
bra- and ket-vectors in a double MPS arises from our special ansatz
taking advantage of the commutator, which needs to operate on the
bra- and ket-vectors at the same time. As a consequence of this doubling,
the bra- and ket-part are treated independently in a numerical algorithm
which optimizes tensor by tensor. Therefore, the resulting operator
is not forcedly Hermitian by construction 
\begin{equation}
\sum_{ij}\lambda_{ij}|u_{i}\rangle\langle v_{j}|\overset{?}{=}\sum_{ij}\left(\lambda^{\dagger}\right)_{ji}|v_{j}\rangle\langle u_{i}|.
\end{equation}
Still, since we intend to express the Hermitian TADM as double MPS,
the optimization objective forces the algorithm to come up with a
solution which is very close to Hermitian. At the end, for most applications,
a lack of Hermiticity should not be more severe than any other numerical
imprecision. If Hermiticity is of importance, we can still resort
to $M'=\tfrac{1}{2}(M+M^{\dagger})$. 

Comparing Fig.~\ref{fig:Double MPS tadm} b) and c), we see that
due to the unfolding process ${\rm b)\rightarrow c)}$, the order
of sites in the second part of the double MPS is inverted. That is,
in a double MPS, the tensors $\mathsf{M}_{[1]}\cdots\mathsf{M}_{[2n]}$
correspond to the physical sites $1,2,\cdots n-1,n,n,n-1,\cdots,2,1$.
This ordering should be kept, since it is very convenient if we like
to calculate expectation values, where we need to fold the double
MPS as in Fig.~\ref{fig:Double MPS tadm} b).

\subsection{Implementation}

One of the great advantages of the double MPS is that with marginal
adaptations, all algorithms we have developed so far for the MPO TADM
can be reused, except the Hermite to real mapping explained in appendix~\ref{sec:Mapping-hermitian-matrices onto real tADM}. 

The MPO based algorithm is operating with three different MPO, representing
the original density matrix $\varrho_{0}$, the commutator operator
$\mathfrak{C}$ and the matrix $M$, which we optimize. All three
have to be replaced by double MPS (where $\mathfrak{C}$ actually
corresponds to a double MPO). First, we observe that throughout the
algorithm, a single multi-index $\sigma_{j}=(s_{j},s_{j}')$ is used
for the two physical indices $s_{j}$ and $s_{j}'$ corresponding
to the bra- and ket-index of the MPO tensors. Therefore, it is straight
forward to replace the MPO structure in the algorithm by a (double)
MPS structure. Of course, this is just a formal argument and we have
to ensure the correct correspondence between MPO and double MPS. 

The tensors $\mathsf{M}_{[j]}$ of the double MPS which represents
$M$ are determined by the algorithm. For us, it remains to find the
correct double MPS representation for $\varrho_{0}$ and $\mathfrak{C}$.
For many interesting cases, the initial state is a pure state $\varrho_{0}=|\Psi_{0}\rangle\langle\Psi_{0}|$.
In this case, if $|\Psi_{0}\rangle$ can be represented as MPS, the
construction of the double MPS is trivial. On the other hand, if we
are not interested in the TADM $\bar{\varrho}$ but in the time average
of an operator $O_{0}$, a double MPS is generally not a suitable
choice. For commonly used operators (as e.g.\  a Pauli matrix acting
on the $j$th site $\boldsymbol{\alpha}\cdot\boldsymbol{\sigma}^{(j)}\equiv{\mathbbm1}^{(1\dots j-1)}\otimes\boldsymbol{\alpha}\cdot\boldsymbol{\sigma}^{(j)}\otimes{\mathbbm1}^{(j+1\dots n)}$),
the needed bond dimension for a double MPS scales exponentially with
the number of sites. 

Finally, we need to construct the commutator operator $\mathfrak{C}$,
which has formally the appearance of a double MPO, where each tensor
carries two physical indices. In appendix~\ref{sec:Constructing-a-MPO for C TADM},
we briefly outline the construction of $\mathfrak{C}$ for the MPO
based algorithm, where the commutator operator is symbolically written
as 
\begin{equation}
\mathfrak{C}=H\otimes{\mathbbm1}-{\mathbbm1}\otimes H,
\end{equation}
see Eq.~(\ref{eq:C als H1-1H tadm}). For the double MPS based algorithm,
this symbolical form can be directly translated into a double MPO.
That is, $\mathfrak{C}$ is the difference of two double MPO, where
one part of each double MPO represents the Hamiltonian and the other
part the identity. It is easy to verify that for arbitrary double
MPS $A$ and $B$, this construction fulfills the property $\left\langle \mathfrak{C}A|B\right\rangle =\left\langle A|\mathfrak{C}B\right\rangle $,
as it should (\ref{eq:Selbstadjungierter kommutator t.a.d.m.}).

\section{Numerical aspects\label{sec:Numerical-aspects TADM}}

In regard to numerical aspects, the result section focused strongly
on the dependence of the results on the bond dimension. Here, we add
a few comments concerning convergence properties and numerical precision.

\subsection{Convergence}

Tensor networks are usually optimized by successive local optimizations
of one or two tensors at a time. Although it is well known that locally
optimizing algorithms often run the risk of getting stuck in a local
extremum, we find e.g.\  that matrix product state (MPS) based ground
state search algorithms seem to be widely immune against this problem.
They exhibits superb convergence properties for many physical systems
of interest. Can we hope that this is true for the optimization of
the time averaged density matrix (TADM), as well? 

An important difference between these two algorithms is that many
commonly used Hamiltonians are sums of local operators only, while
the squared commutator operator $\mathfrak{C}^{2}$ used in the TADM
algorithm (\ref{eq:Optimization Objective tadm}) is a highly non-local
object. If we follow the alternative optimization strategy of appendix~\ref{sub:Alternative-optimization-ansatz tadm},
we even have to use the third and fourth power of $\mathfrak{C}$.
In this case, we occasionally observed strong difficulties in finding
the optimal solution if we started out with a randomized initial state.
In case of the standard algorithm based on $\mathfrak{C}^{2}$, we
noticed convergence into local extrema, as well, but the observed
deviations were only marginal.

\subsection{Precision}

A well known source for losses in the numerical precision are differences
of big numbers which differ only by a very small number. To soften
this effect for the commutator, we recommend to gauge the Hamiltonian
such that ${\rm tr(H\varrho_{0})=0}$, i.e 
\begin{equation}
H\rightarrow H-\textrm{tr}(H\varrho_{0}).
\end{equation}
Still, certain losses in the precision are inevitable. Especially
the $T+$ method (\ref{eq:T+ und T- TADM}) discussed in appendix~\ref{sub:Alternative-optimization-ansatz tadm}
is prone to numerical imprecision, since it employs the third and
fourth power of $\mathfrak{C}$. In this algorithm, the value of $\varepsilon=\left\Vert \mathfrak{C}\bar{\varrho}_{{\rm approx}}\right\Vert ^{2}$
is minimized and should become zero for a perfect $\bar{\varrho}_{{\rm approx}}=\bar{\varrho}$.
Now, we have to see that the value of $\varepsilon$ is not just limited
by the achievable numerical precision, but during the optimization
of $\varepsilon$, the average improvement per optimization step should
also exceed the achievable numerical precision. Further, we are actually
interested in the square root of $\varepsilon$ respectively in the
value $q=\tfrac{\left\Vert \mathfrak{C}\varrho_{0}\right\Vert }{\left\Vert \mathfrak{C}\bar{\varrho}_{{\rm approx}}\right\Vert }\propto\sqrt{\varepsilon^{-1}}$
(\ref{eq:Guete again TADM}). Especially the combination double MPS
(appendix~\ref{sec:Double-MPS tadm}) and $T+$ method seems to be
quite vulnerable. In some of our numerical simulations, the maximal
reliable value of $q$ was limited around $10^{3}\dots10^{4}$ due
to numerical imprecision. An example for this effect can be seen in
Fig.~\ref{fig:AbleitungRhoAllTyps TADM}~b), where the precision
of the double MPS $T+$ method saturates already for a bond dimension
$D=128$. Of course, we can always resort to a more precise floatingpoint
operation, but this is usually not supported by the hardware and hence,
needs a software emulation, which is significantly slower.
\end{document}